# The Strategy-Proofness Landscape of Merging

**Patricia Everaere**                                        EVERAERE@CRIL.FR
**Sébastien Konieczny**                                     KONIECZNY@CRIL.FR
**Pierre Marquis**                                           MARQUIS@CRIL.FR
*CRIL – CNRS*
*Faculté des Sciences, Université d'Artois*
*62300 Lens, France*

## Abstract

Merging operators aim at defining the beliefs/goals of a group of agents from the beliefs/goals of each member of the group. Whenever an agent of the group has preferences over the possible results of the merging process (i.e., the possible merged bases), she can try to rig the merging process by lying on her true beliefs/goals if this leads to a better merged base according to her point of view. Obviously, strategy-proof operators are highly desirable in order to guarantee equity among agents even when some of them are not sincere. In this paper, we draw the strategy-proof landscape for many merging operators from the literature, including model-based ones and formula-based ones. Both the general case and several restrictions on the merging process are considered.

## 1. Introduction

Merging operators aim at defining the beliefs/goals of a group of agents from the beliefs/goals of each member of the group. Though beliefs and goals are distinct notions, merging operators can typically be used for merging either beliefs or goals. Thus, most of the logical properties from the literature (Revesz, 1993, 1997; Konieczny & Pino Pérez, 1998, 2002) for characterizing rational belief merging operators can be used for characterizing as well rational goal merging operators.

Whatever beliefs or goals are to be merged, there are numerous situations where agents have preferences on the possible results of the merging process (i.e., the merged bases). As far as goals are concerned, an agent is surely satisfied when her individual goals are chosen as the goals of the group. In the case of belief merging, an agent can be interested in imposing her beliefs to the group (i.e., "convincing" the other agents), especially because the result of a further decision stage at the group level may depend on the beliefs of the group.

So, as soon as an agent participates to a merging process, the *strategy-proofness* problem has to be considered. The question is: is it possible for a given agent to improve the result of the merging process with respect to her own point of view by lying on her true beliefs/goals, given that she knows (or at least she assumes) the beliefs/goals of each agent of the group and the way beliefs/goals are merged?

As an illustration, let us consider the following scenario of goal merging (that will be used as a running example in the rest of the paper):

**Example 1** *Three friends, Marie, Alain and Pierre want to spend their summer holidays together. They have to determine whether they will go to the seaside and/or to the mountains, or to stay at home, and also to determine whether they will take a long period of vacations or not. The goals of Marie are to go to the seaside and to the mountains if it is for a long period; otherwise she wants to go to the mountains, only, or to stay at home. The goals of Alain are to go to the seaside if it is for a long period; or to go to the mountains if it is for a short period. Finally, Pierre is only interested in going to the seaside for a long period, otherwise he prefers to stay at home. If one uses a common merging operator for defining the choice of the group,[1] then the goals of the group will be either to go to the seaside for a long period, or to go to the mountains or to stay at home for a short period. Accordingly, the group may choose to go to the seaside, only, for a long period, which is not among the goals of Marie. However, if Marie lies and claims that, for a short period, she wants to go to the mountains only, or to stay at home, then the result of the merging process will be different. Indeed, in this case, the goals of the group will be to go to the mountains for a short period, or to stay at home, which corresponds to the goals of Marie.*

Similarly, the strategy-proofness issue has to be considered in many belief merging scenarios, just because rational decision making typically takes account for the "true" state of the world. When agents have conflicting beliefs about it, belief merging can be used to determine what is the "true" state of the world for the group; manipulating the belief merging process is a way for an agent to change the resulting beliefs at the group level so as to make them close to her own beliefs. As a consequence, the decisions made by the group may also change and become closer to those the agent would made alone. For instance, assume that the three friends agree that the mountains must be avoided when the weather is bad. If the beliefs of the group is that the weather is bad, then the decision to go to the mountains will be given up. If Pierre believes that the weather is bad, then he may be tempted to make *the weather is bad* accepted at the group level. Therefore, a collective decision will be not to go to the mountains.

There are several multi-agent settings in which some agents exchange information and must make individual decisions based on their beliefs. In many scenarios, agents are tempted to get an informational advantage over other agents, which can be achieved by gathering as much information as possible and by hiding their own ones. Indeed, being better informed may help an agent making better decisions than the other agents of the group. For instance, Shoham and Tennenholtz (2005) investigate non-cooperative computation: each agent delivers some piece of information (truthfully or not), all such pieces are used to compute the value of a (commonly-known) function, and this value is given back to the agents; the aim of each agent is to get the true value of the function, and if possible to be the only one to get it. In the work of Shoham and Tennenholtz (2005), information is considered at an abstract level; assuming that information represent beliefs and the function is a belief merging operator, each agent wants to get the true merged base and possibly to be the only one to get it. Contrastingly, in other scenarios, where decisions have to be made collectively and are based on the beliefs of the group, agents are satisfied if the beliefs of the group

---

1. Formally, the model-based operator $\triangle_\mu^{d_H, \Sigma}$, using the Hamming distance and the sum aggregation function, defined in Section 3.1.

are close to their own beliefs. In this paper, we focus on such an issue, which has to be addressed in many everyday life situations. Let us illustrate this on an example:

**Example 2** *There is a position available in some university. The committee in charge of the recruitment has to determine the right profile for this position. Four criteria are considered: research skills, teaching skills, relationship skills, and the past positions of the candidate. Suppose that a member of the committee believes that the important criteria for the job are research skills and relationship skills, and that it is better to recruit a candidate who got a good position in the past. She will be pleased by the recruitment if the group shares her beliefs about the right profile. So she can be tempted to manipulate the merging process in order to achieve such a situation.*

Determining whether a belief/goal merging operator is strategy-proof, and in the negative case, identifying restrictions under which strategy-proofness can be ensured is thus an important issue. Indeed, merging operators are intended to characterize the beliefs/goals of a group of agents, from the beliefs/goals of each agent from the group; obviously, this objective cannot be reached if the agents do not report their "true" beliefs/goals, which can easily be the case when manipulable merging operators are used (since agents will be tempted to manipulate the process in such a case).

Since merging operators are typically used in artificial systems, one can wonder whether strategy-proofness is really a relevant issue in this context. The answer actually depends on the sophistication of the agents under consideration. Thus, in a distributed database setting, the (low-level) agents (i.e., the databases) have typically no evaluation/preference on the merged base, and the strategy-proofness issue does not make sense. It is not the same story when the agents have goals and reasoning capacities. In this case, it cannot be discarded that some agents will be able to foresee weaknesses in the aggregation process and to exploit them for their own benefits. When high-level artificial agents are involved, the strategy-proofness problem is even more stricking than in the case of human agents because of the superior computational abilities of artificial systems.

The strategy-proofness issue has been studied for years in the domain of Social Choice Theory (Arrow, Sen, & Suzumura, 2002). An important objective is to design preference aggregation procedures (and, in particular, voting procedures) which are strategy-proof. A very famous result, known as Gibbard-Sattherwaite theorem, is that this objective cannot be reached in an absolute manner: under a number of sensible requirements, no strategy-proof voting procedure may exist (Gibbard, 1973; Satterthwaite, 1975). Strategy-proofness can only be achieved by relaxing some of those requirements, which is enough to escape from Gibbard-Sattherwaite theorem. We shall return to this topic in Section 7 where the connections between belief merging and preference aggregation will be considered in more depth.

The very objective of this paper is to draw the strategy-proofness landscape for many merging operators from the literature, including model-based ones and formula-based ones. We focused on operators for merging bases that are sets of propositional formulas, where no priorities between the bases are available. The (classical) propositional logic framework can be argued as a representation setting expressive enough for many AI scenarios; furthermore, it is natural to investigate first the key problems raised by aggregation and manipulation in this simple setting, before considering more sophisticated logics. For each operator under consideration, we aim at determining whether it is strategy-proof in the general case, and under some restrictions on the merging process (including the number of agents and the presence of integrity constraints) and on the set of available strategies for the agents.

The rest of the paper is organized as follows. In Section 2, some formal preliminaries are provided. In Section 3, the definitions of the main propositional merging operators from the literature are recalled. Several definitions of strategy-proofness based on a general notion of satisfaction index are given in Section 4 and our strategy-proofness results are reported in Section 5. They are discussed in Section 6. Then, connections with Social Choice Theory and other related works are pointed out in Section 7, just before the conclusion. Proofs are reported at the end of the paper.

## 2. Formal Preliminaries

We consider a propositional language $\mathcal{L}$ defined from a finite (and non-empty) set of propositional variables $\mathcal{P}$ and the standard connectives, including $\top$, the Boolean constant always true, and $\bot$, the Boolean constant always false.

An interpretation (or world) $\omega$ is a total function from $\mathcal{P}$ to $\{0, 1\}$, denoted by a bit vector whenever a strict total order on $\mathcal{P}$ is specified. The set of all interpretations is noted $\mathcal{W}$. An interpretation $\omega$ is a model of a formula $\phi \in \mathcal{L}$ if and only if it makes it true in the usual truth functional way.

$[\phi]$ denotes the set of models of formula $\phi$, i.e., $[\phi] = \{\omega \in \mathcal{W} \mid \omega \models \phi\}$. In order to avoid too heavy notations, we identify each interpretation $\omega$ with the canonical term on $\mathcal{P}$ which has $\omega$ as its unique model. For instance, if $\mathcal{P} = \{a, b\}$ and $\omega(a) = 1$, $\omega(b) = 0$, $\omega$ is identified with the term $a \wedge \neg b$.

A formula $\phi$ from $\mathcal{L}$ is consistent if and only if $[\phi] \neq \emptyset$. $\phi$ is a logical consequence of a formula $\psi$, noted $\psi \models \phi$ if and only if $[\psi] \subseteq [\phi]$. Two formulas are logically equivalent ($\equiv$) if and only if they share the same models.

A belief/goal *base* $K$ denotes the set of beliefs/goals of an agent. It is a finite and consistent set of propositional formulas, interpreted conjunctively. When $K$ is a belief/goal base, $\widehat{K}$ denotes the singleton base containing the conjunction $\bigwedge K$ of all formulas from $K$. A base is said to be complete if and only if it has exactly one model. Each belief/goal base $K$ characterizes a bipartition of the set of all interpretations: the models of $K$ are the interpretations which are acceptable for the agent, and the countermodels are not. When $K$ is a belief base, an interpretation $\omega$ is acceptable when there is enough evidence that $\omega$ is the "true" state of the world; when $K$ is a goal base, $\omega$ is acceptable when it is sufficiently desired. Such a bipartition can be considered as an approximation of the full belief/goal preference structure of the corresponding agent: in the belief case, $\omega$ is at least as preferred as $\omega'$ means that the fact that $\omega$ is the "true" state of the world is at least as plausible as the fact that $\omega'$ is the "true" state of the world; in the goal case, $\omega$ is at least as preferred as $\omega'$ means that the fact that $\omega$ would be the "true" state of the world is at least as desired as the fact that $\omega'$ would be the "true" state of the world.

A belief/goal *profile* $E$ is associated with the group of $n$ agents involved in the merging process. It is a non-empty multi-set (bag) of belief/goal bases $E = \{K_1, \ldots, K_n\}$ (hence different agents are allowed to exhibit identical bases). Note that profiles are *non-ordered* (multi-)sets; thus, the profile representation of groups of agents induces an anonymity prop-erty: each agent has the same importance as the other agents of the group and the result of the merging process only depends on the bases themselves (i.e., exchanging the bases of two agents gives the same profile, hence the same merged base).

We denote by $\bigwedge E$ the conjunction of bases of $E = \{K_1, \ldots, K_n\}$, i.e., $\bigwedge E = (\bigwedge K_1) \wedge \ldots \wedge (\bigwedge K_n)$, and we denote by $\bigvee E$ the disjunction of bases of $E$, i.e., $\bigvee E = (\bigwedge K_1) \vee \ldots \vee (\bigwedge K_n)$.

A profile $E$ is said to be consistent if and only if $\bigwedge E$ is consistent. The multi-set union is noted $\sqcup$ and the multi-set containment relation is noted $\sqsubseteq$. The cardinal of a finite set (or a finite multi-set) $A$ is noted $\#(A)$. $\subseteq$ will denote set containment and $\subset$ strict set containment, i.e., $A \subset B$ if and only if $A \subseteq B$ and $A \neq B$.

If $\leq_E$ denotes a pre-order on $\mathcal{W}$ (i.e., a reflexive and transitive relation), then $<_E$ denotes the associated strict ordering defined by $\forall \omega, \omega' \in \mathcal{W}$, $\omega <_E \omega'$ if and only if $\omega \leq_E \omega'$ and $\omega' \not\leq_E \omega$.

The result of the merging of (the bases from) a profile $E$ with the merging operator $\triangle$, under the integrity constraints $\mu$ is the base, noted $\triangle_\mu(E)$, called the *merged base*. The *integrity constraints* consist of a consistent formula (or, equivalently, a (finite) consistent conjunction of formulas) the merged base has to satisfy (it may represent some physical laws, some norms, etc.); in other words, models of the merged base are models of the integrity constraints.

## 3. Merging Operators

We recall in this section the two main families of merging operators from the literature. The first family is defined by a selection of some interpretations (model-based operators). The second family is defined by a selection of some formulas in the union of the bases (formula-based operators). For more details on those two families, see for example (Konieczny, Lang, & Marquis, 2004).

### 3.1 Model-Based Operators

The first family is based on the selection of some models of the integrity constraints, the "closest" ones to the profile. Closeness is usually defined from a notion of distance and an aggregation function (Revesz, 1997; Konieczny & Pino Pérez, 1998, 1999; Lin & Mendelzon, 1999; Liberatore & Schaerf, 1998).

### Definition 1 (pseudo-distances)

- A pseudo-distance *between interpretations is a total function* $d : \mathcal{W} \times \mathcal{W} \mapsto \mathbb{R}^+$ *s.t. for any* $\omega$, $\omega'$, $\omega'' \in \mathcal{W}$: $d(\omega, \omega') = d(\omega', \omega)$, *and* $d(\omega, \omega') = 0$ *if and only if* $\omega = \omega'$.

- A distance *between interpretations is a pseudo-distance that satisfies the triangular inequality:* $\forall \omega, \omega', \omega'' \in \mathcal{W}$, $d(\omega, \omega') \leq d(\omega, \omega'') + d(\omega'', \omega')$.

Two widely used distances between interpretations are Dalal distance (Dalal, 1988), denoted $d_H$, that is the Hamming distance between interpretations (the number of propositional atoms on which the two interpretations differ); and the drastic distance, denoted $d_D$, that is one of the simplest pseudo-distances one can define: it gives 0 if the two interpretations are the same one, and 1 otherwise.

**Definition 2 (aggregation functions)** *An aggregation function $f$ is a total function associating a nonnegative real number to every finite tuple of nonnegative real numbers s.t. for any $x_1, \ldots, x_n, x, y \in \mathbb{R}^+$:*

- *if $x \leq y$, then $f(x_1, \ldots, x, \ldots, x_n) \leq f(x_1, \ldots, y, \ldots, x_n)$.*  *(non-decreasingness)*

- *$f(x_1, \ldots, x_n) = 0$ if and only if $x_1 = \ldots = x_n = 0$.*  *(minimality)*

- *$f(x) = x$.*  *(identity)*

Widely used functions are the max (Revesz, 1997; Konieczny & Pino Pérez, 2002), the sum $\Sigma$ (Revesz, 1997; Lin & Mendelzon, 1999; Konieczny & Pino Pérez, 1999), or the leximax $GMax$ (Konieczny & Pino Pérez, 1999, 2002).

The chosen distance between interpretations induces a "distance"[2] between an interpretation and a base, which in turn gives a "distance" between an interpretation and a profile, using the aggregation function. This latter distance gives the needed notion of closeness represented by a pre-order on $\mathcal{W}$ induced by $E$, noted $\leq_E$. Such a pre-order can be interpreted as a plausibility ordering associated with the merged base.

**Definition 3 (distance-based merging operators)** *Let $d$ be a pseudo-distance between interpretations and $f$ be an aggregation function. The result $\triangle_\mu^{d,f}(E)$ of the merging of $E$ given the integrity constraints $\mu$ is defined by:*

$$[\triangle_\mu^{d,f}(E)] = \min([\mu], \leq_E) = \{\omega \in [\mu] \mid \nexists \omega' \in [\mu], \ \omega' <_E \omega\}$$

*where the pre-order $\leq_E$ on $\mathcal{W}$ induced by $E$ is defined by:*

- *$\omega \leq_E \omega'$ if and only if $d(\omega, E) \leq d(\omega', E)$, where*

- *$d(\omega, E) = f_{K \in E}(d(\omega, K))$, where*

- *$d(\omega, K) = min_{\omega' \models K} d(\omega, \omega')$.*

Observe that $d^{d,f}(\omega, E)$ would be a more correct notation than $d(\omega, E)$; however, since there is no ambiguity in the choice of the function $f$ and the distance between interpretations $d$ in the following, we prefer the lighter notation $d(\omega, E)$.

Let us step back to the example given in the introduction in order to illustrate model-based merging operators:

**Example 3** *Consider the set $\mathcal{P}$ with three propositional variables l(long period), s(easide) and m(ountains), taken in this order. The goals of the three agents are then given by the following bases: $[K_1] = \{000, 001, 111\}$ (Marie's wishes), $[K_2] = \{001, 110\}$ (Alain's wishes) and $[K_3] = \{000, 110\}$ (Pierre's wishes). There is no integrity constraint ($\mu \equiv \top$).*

---

2. Abusing words since it is not a distance from the mathematical standpoint.

We have $[\Delta_\mu^{d_H, \Sigma}(\{K_1, K_2, K_3\})] = \{000, 001, 110\}$. *Table 1 gives some details of the computation. The first column gives all possible words. The $K_i$ ($i = 1 \ldots 3$) columns give for each interpretation $\omega$ the value $d_H(\omega, K_i)$. Finally, the rightmost column gives for each interpretation $\omega$ the value of $d_H(\omega, \{K_1, K_2, K_3\})$. The interpretations $\omega$ for which $d_H(\omega, \{K_1, K_2, K_3\})$ is minimal (in bold) are the models of the merged base $\Delta_\mu^{d_H, \Sigma}(\{K_1, K_2, K_3\})$.*

| $\omega$ | $K_1$ | $K_2$ | $K_3$ | $\Delta_\mu^{d_H, \Sigma}(\{K_1, K_2, K_3\})$ |
|---|---|---|---|---|
| 000 | 0 | 1 | 0 | **1** |
| 001 | 0 | 0 | 1 | **1** |
| 010 | 1 | 1 | 1 | 3 |
| 011 | 1 | 1 | 2 | 4 |
| 100 | 1 | 1 | 1 | 3 |
| 101 | 1 | 1 | 2 | 4 |
| 110 | 1 | 0 | 0 | **1** |
| 111 | 0 | 1 | 1 | 2 |

Table 1: Merging with $\Delta_\mu^{d_H, \Sigma}$.

### 3.2 Formula-Based Operators

The other main family of merging operators gather the so-called "formula-based operators" or "syntax-based operators". Formula-based operators are based on the selection of consistent subsets of formulas in the union of the bases of the profile $E$. Several operators are obtained by letting vary the selection criterion. The result of the merging process is the set of consequences that can be inferred from all selected subsets (Baral, Kraus, Minker, & Subrahmanian, 1992; Rescher & Manor, 1970; Konieczny, 2000). For these operators, the syntactic form of the bases may easily influence the result of the merging process: replacing a base $K = \{\varphi_1, \ldots, \varphi_n\}$ by the (logically equivalent) base $\widehat{K} = \{\varphi_1 \wedge \ldots \wedge \varphi_n\}$ may lead to change the corresponding merged base (while this is not the case for model-based operators).

**Definition 4 (maximal consistent subsets)** *Let $K$ be a base and $\mu$ be an integrity constraint.* MAXCONS$(K, \mu)$ *is the set of all the maximal (w.r.t. set inclusion) consistent subsets (maxcons for short) of $K \cup \{\mu\}$ that contains $\mu$, i.e.,* MAXCONS$(K, \mu)$ *is the set of all consistent $M$ that satisfy:*

- $M \subseteq K \cup \{\mu\}$, *and*

- $\mu \in M$, *and*

- *If $M \subset M' \subseteq K \cup \{\mu\}$, then $M'$ is not consistent.*

*When maximality must be taken with respect to cardinality (instead of set inclusion), we shall use the notation* MAXCONS$_{card}(K, \mu)$.

Now, for any profile $E$ and integrity constraint $\mu$, we set

$$\text{MAXCONS}(E, \mu) = \text{MAXCONS}\left(\bigcup_{K_i \in E} K_i, \mu\right)$$

Observe that set-theoretic union (and not multi-set union) is used here.

The following operators have been defined so far (Baral, Kraus, & Minker, 1991; Baral et al., 1992; Konieczny, 2000):

**Definition 5 (formula-based merging operators)** *Let $E$ be a profile and let $\mu$ be an integrity constraint:*

- $\triangle_\mu^{C1}(E) \equiv \bigvee_{M \in \text{MAXCONS}(E,\mu)}(\bigwedge M)$.

- $\triangle_\mu^{C3}(E) \equiv \bigvee_{M|M \in \text{MAXCONS}(E,\top) \text{ and } M \cup \{\mu\} \text{ consistent}}(\bigwedge M)$.

- $\triangle_\mu^{C4}(E) \equiv \bigvee_{M \in \text{MAXCONS}_{card}(E,\mu)}(\bigwedge M)$.

- $\triangle_\mu^{C5}(E) \equiv \begin{cases} \bigvee_{M \in \text{MAXCONS}(E,\top) \text{ and } M \cup \{\mu\} \text{ consistent}}(\bigwedge\{M \cup \{\mu\}\}) & \\ & \text{if consistent,} \\ \mu \text{ otherwise.} \end{cases}$

Those operators clearly select as much formulas as they can from the union of the bases, under the consistency requirement. The differences between them lie in the meaning of "as much". Those operators were defined by Baral et al. (1992), except $\triangle^{C5}$ which is a modification of $\triangle^{C3}$ that ensures consistency. Indeed, unlike the other operators listed here, $\triangle_\mu^{C3}$ may generate inconsistent merged bases (as the empty disjunction). A $\triangle^{C2}$ operator has also been introduced by Baral et al. (1992), and shown equivalent to $\triangle^{C1}$ (this is why it is not listed above). An important drawback of those operators is that they do not take account for the sources from which the formulas are issued.[3] Nevertheless, they have an appealing, simple definition.

To illustrate the behaviour of formula-based operators, let us step back to the example given in the introduction. Since the absence of constraints makes the operators $\triangle^{C1}$, $\triangle^{C3}$ and $\triangle^{C5}$ coincide, we shall add the following constraint: $\mu = l \wedge \neg s$, i.e., it turns out that the group will have to take holidays for a long period and that they cannot go to the seaside.

**Example 4** *Suppose that Marie, Alain and Pierre's goals are encoded by the following bases: $K_1 = \{l \leftrightarrow s, l \rightarrow m\}$, $K_2 = \{l \leftrightarrow s, s \leftrightarrow \neg m\}$, and $K_3 = \{l \leftrightarrow s, \neg m\}$. With the integrity constraints $\mu = l \wedge \neg s$, $\text{MAXCONS}(E, \mu)$ contains two sets: $\{l \rightarrow m, s \leftrightarrow \neg m\}$ and $\{\neg m\}$. We get $\triangle_\mu^{C1}(E) \equiv \mu$, $\triangle_\mu^{C3}(E) \equiv \bot$, $\triangle_\mu^{C4}(E) \equiv l \wedge \neg s \wedge m$, and $\triangle_\mu^{C5}(E) \equiv \mu$.*

The "syntax sensitivity" of those operators is due to the fact that the comma symbol "," which appears in the expression of the bases, is a specific, yet not truth-functional, connective (Konieczny, Lang, & Marquis, 2005) that is usually not equivalent to standard conjunction in the formula-based framework. Such operators may easily lead to merged bases which differ from their counterparts where commas are replaced by conjunctions in the input bases. For example, with $\mu = \top, K_1 = \{a \wedge b\}, K_2 = \{\neg(a \wedge b)\}$ and $K_1' = \{a, b\}$, the fact that $K_1' \equiv K_1$ does not imply that $\triangle_\mu^{C1}(\{K_1, K_2\}) \equiv \triangle_\mu^{C1}(\{K_1', K_2\})$, since $\triangle_\mu^{C1}(\{K_1, K_2\}) \equiv \top$ and $\triangle_\mu^{C1}(\{K_1', K_2\}) \equiv a \vee b$. See (Konieczny et al., 2005) for a more detailed discussion on the meaning of the comma connective in frameworks for reasoning under inconsistency.

---

3. It is possible to avoid it by taking advantage of a further selection function (Konieczny, 2000).

Clearly enough, if one replaces each base $K$ by $\widehat{K}$, the singleton base containing the conjunction of its elements before making the union, the resulting operators, noted $\triangle_\mu^{\widehat{C}}$, are not any longer sensitive to the syntactic presentation of the bases (replacing every base by a logically equivalent one leads to the same merged base). Formally, we have:

**Definition 6 (other formula-based merging operators)** *Let $E = \{K_1, \ldots, K_n\}$ be a profile and let $\mu$ be an integrity constraint:*

$$\triangle_\mu^{\widehat{Ci}}(E) = \triangle_\mu^{Ci}(\{\widehat{K_1}, \ldots, \widehat{K_n}\}).$$

**Remark 1** *Observe that $\triangle^{\widehat{C4}}$ is equivalent to the model-based operator $\Delta^{d_D, \Sigma} = \Delta^{d_D, GMax}$. Indeed, $\triangle^{\widehat{C4}}$ returns the disjunction of the maximal (for cardinality) consistent subsets of the profile under the constraints. This is exactly what the operator $\Delta^{d_D, \Sigma} = \Delta^{d_D, GMax}$ achieves since it amounts to define the set of models of the merged base as the set of interpretations that satisfy the constraints and a maximal number of bases of the profile, i.e., the interpretations that are models of a maximal (for cardinality) consistent subset of the profile under the constraints.*

## 4. Strategy-Proofness

The strategy-proofness issue for a merging operator can be stated as follows: is it possible for a given agent to improve the result of the merging process with respect to her own point of view by lying on her true beliefs/goals, given that she knows the beliefs/goals of each agent of the group and the way beliefs/goals are merged? If this question can be answered positively, then the operator is not strategy-proof (the agent may benefit from being untruthful). Thus, a merging operator is not strategy-proof if one can find a profile $E = \{K_1, \ldots, K_n\}$ which represents the bases of the other agents, an integrity constraint $\mu$, and two bases $K$ and $K'$ s.t. the result of the merging of $E$ and $K'$ is better for the agent than the result of the merging of $E$ with her true base $K$ (called the initial base).

**Definition 7 (strategy-proofness)** *Let $i$ be a satisfaction index, i.e., a total function from $\mathcal{L} \times \mathcal{L}$ to $\mathbb{R}$.*

- *A profile $E$ is said to be* manipulable *by a base $K$ for $i$ given a merging operator $\Delta$ and an integrity constraint $\mu$ if and only if there exists a base $K'$ such that $i(K, \Delta_\mu(E \sqcup \{K'\})) > i(K, \Delta_\mu(E \sqcup \{K\}))$.*

- *A merging operator $\Delta$ is* strategy-proof *for $i$ if and only if there are no integrity constraint $\mu$ and profile $E = \{K_1, \ldots, K_n\}$ such that $E$ is manipulable for $i$.*

Given two bases (interpreted conjunctively) $K$, $K_\Delta$, the value of $i(K, K_\Delta)$ is intended to indicate how much a base $K$ is close to the merged base $K_\Delta$. The need for such satisfaction indexes comes from the fact that the *only information* given by each agent is her own base $K$: if the full preference structure over sets of interpretations were available for each agent as an additional input (e.g., under the form of a utility function), then one could use it to define strategy-proofness directly for a given agent (as it is done in Social Choice Theory, Arrow et al., 2002) and not in a uniform way for all agents. This explains why we call $i$ a satisfaction index and not a utility function.

Clearly, there are many different ways to define the satisfaction of an agent given a merged base. While many *ad hoc* definitions can be considered, we consider the following three indexes which are, according to us, the most meaningful ones when no additional information about the agents are available. As far as we know, this is the first time such indexes are considered in the context of pure propositional merging.

The first two indexes are drastic ones: they range to $\{0, 1\}$, so the agent is either fully satisfied or not satisfied at all.

**Definition 8 (weak drastic index)**

$$i_{dw}(K, K_\Delta) = \left\{ \begin{array}{ll} 1 & \text{if } K \wedge K_\Delta \text{ is consistent,} \\ 0 & \text{otherwise.} \end{array} \right.$$

This index takes value 1 if the result of the merging process (noted $K_\Delta$ in the definition) is consistent with the agent's base ($K$), and 0 otherwise. It means that the agent is considered fully satisfied as soon as its beliefs/goals are consistent with the merged base.

**Definition 9 (strong drastic index)**

$$i_{ds}(K, K_\Delta) = \left\{ \begin{array}{ll} 1 & \text{if } K_\Delta \models K, \\ 0 & \text{otherwise.} \end{array} \right.$$

This index takes value 1 if the agent's base is a logical consequence of the result of the merging process, and 0 otherwise. In order to be fully satisfied, the agent must impose her beliefs/goals to the whole group.

The last index is not a Boolean one, leading to a more gradual notion of satisfaction. The more compatible the merged base with the agent's base the more satisfied the agent. The compatibility degree of $K$ with $K_\Delta$ is the (normalized) number of models of $K$ that are models of $K_\Delta$ as well:

**Definition 10 (probabilistic index)** *If $\#([K_\Delta]) = 0$, then $i_p(K, K_\Delta) = 0$, otherwise:*

$$i_p(K, K_\Delta) = \frac{\#([K] \cap [K_\Delta])}{\#([K_\Delta])}.$$

$i_p(K, K_\Delta)$ is the probability to get a model of $K$ from a uniform sampling in the models of $K_\Delta$. This index takes its minimal value when no model of $K$ is in the models of the merged base $K_\Delta$, and its maximal value when each model of the merged base is a model of $K$.

Strategy-proofness for these three indexes are not independent notions:

**Theorem 1**

1. *If a merging operator is strategy-proof for $i_p$, then it is strategy-proof for $i_{dw}$.*

2. *Consider a merging operator $\Delta$ that generates only consistent bases.[4] If it is strategy-proof for $i_p$, then it is strategy-proof for $i_{ds}$.*

On the other hand, it is easy to prove that strategy-proofness for $i_{dw}$ and strategy-proofness for $i_{ds}$ are logically independent in the general case (an operator can be strategy-proof for one of them without being strategy-proof for the other, and it can be strategy-proof for both of them or for neither).

Let us conclude this section with our running example, and give formal arguments explaining how Marie can manipulate the merging process:

**Example 5** *We consider three bases $[K_1] = \{000, 001, 111\}$ (Marie's wishes), $[K_2] = \{110, 001\}$ (Alain's wishes) and $[K_3] = \{110, 000\}$ (Pierre's wishes). There is no constraint $(\mu \equiv \top)$.*
*$[\Delta_\top^{d_H, \Sigma}(\{K_1, K_2, K_3\})] = \{000, 001, 110\}$ and $i_{ds}(K_1, \Delta_\top^{d_H, \Sigma}(\{K_1, K_2, K_3\})) = 0$.*

*If Marie reports $[K_1'] = \{000, 001\}$ instead of $K_1$, then $[\Delta_\top^{d_H, \Sigma}(\{K_1', K_2, K_3\})] = \{000, 001\}$ and $i_{ds}(K_1, \Delta_\top^{d_H, \Sigma}(\{K_1', K_2, K_3\})) = 1$.*

*See Table 2 for details of the computations.*

| $\omega$ | $K_1$ | $K_1'$ | $K_2$ | $K_3$ | $\Delta_\top^{d_H, \Sigma}(\{K_1, K_2, K_3\})$ | $\Delta_\top^{d_H, \Sigma}(\{K_1', K_2, K_3\})$ |
|---|---|---|---|---|---|---|
| 000 | 0 | 0 | 1 | 0 | **1** | **1** |
| 001 | 0 | 0 | 0 | 1 | **1** | **1** |
| 010 | 1 | 1 | 1 | 1 | 3 | 3 |
| 011 | 1 | 1 | 2 | 2 | 4 | 4 |
| 100 | 1 | 1 | 1 | 1 | 3 | 3 |
| 101 | 1 | 1 | 2 | 2 | 4 | 4 |
| 110 | 1 | 2 | 0 | 0 | **1** | 2 |
| 111 | 0 | 2 | 1 | 1 | 2 | 4 |

Table 2: $\Delta^{d_H, \Sigma}$ is not strategy-proof for $i_{ds}$.

In the rest of this paper, we shall focus on those three indexes, $i_{dw}$, $i_{ds}$ and $i_p$. Note that investigating other indexes can be interesting. In particular, the probabilistic index can be viewed as a rough measure of similarity between the bases. Could more complex similarity measures between sets (see e.g., Tversky, 2003) also prove useful to define other sensible indexes is an interesting question that we let for further research.

## 5. Strategy-Proofness Results

In the general case, both the family of model-based operators and the family of formula-based operators are not strategy-proof for the three indexes we consider. This means that there are operators from those families which are not strategy-proof. However, imposing further restrictions may lead to some strategy-proofness results. Considering them in a systematic way allows us to draw the strategy-proofness landscape for both families.

In the following, we consider four natural restrictions for the merging process, as listed below:

---

4. I.e., $\Delta_\mu(E)$ is always consistent.

- A first restriction concerns the number of bases to be merged. The question is the following: does the number of bases involved in the merging process have an influence on the strategy-proofness of an operator? In general, we can answer positively to this question. More precisely, there is a distinction between the cases $\#(E) = 2$ and $\#(E) > 2$. In some situations, no manipulation is possible with two bases, while with more bases, it becomes possible. Since a base $\{\top\}$ typically plays the role of a "neutral element" for all the operators we consider (i.e., $\triangle_\mu(E) \equiv \triangle_\mu(E \sqcup \{\top\})$), if an operator is not strategy-proof for profiles with $n$ bases, then it is not strategy-proof for profiles with $m > n$ bases.

- A second parameter is the completeness of the beliefs/goals of the agent who aims at manipulating. In some cases, having such strong beliefs/goals renders any manipulation impossible. Working with complete bases (i.e., singleton sets of models) makes the merging process close to a uninominal vote, i.e., a vote for a unique interpretation.

- A third significant parameter is the presence of integrity constraints. On the one hand, having nontrivial integrity constraints ($\mu \not\equiv \top$) can make a manipulation possible, while it is not the case when no integrity constraints are considered, and the converse also holds.

- Another restriction bears on the available manipulations. In the general case the untruthful agent is free to reporting any base, even if it is "quite far" from her true base. However, there are numerous situations where the other agents participating to the merging process have some information about her true base. In these cases, the agents have to report some bases close to the real ones. Two particular manipulations will be studied: erosion manipulation when the agent pretends to believe/desire more that she does (the agent gives only some parts of its models); and dilatation manipulation when the agent pretends to believe/desire less that she does (the agent gives only parts of its countermodels).

## 5.1 Model-Based Operators

The first result is that there is no general strategy-proofness result (i.e., for any aggregation function and any distance) for model-based operators. This is not very surprising when one reminds the existence of Gibbard-Satterthwaite theorem, which states that there is no "good" strategy-proof preference aggregation method (see Section 7).

However, some quite general strategy-proofness results can be obtained. The following theorem gives some of them, organized from the most general one (for any aggregation function when the drastic distance $d_D$ is considered), to more specific ones (for any distance $d$ when the aggregation function is $\Sigma$):

## Theorem 2

- *Let $f$ be any aggregation function. $\Delta_\mu^{d_D, f}$ is strategy-proof for $i_p$, $i_{dw}$ and $i_{ds}$.*

- *Let $d$ be any distance. Provided that only two bases are to be merged, $\Delta_\top^{d, \Sigma}$ is strategy-proof for the indexes $i_{dw}$ and $i_{ds}$.*

- *For any distance $d$, $\Delta_\mu^{d,\Sigma}$ is strategy-proof for the indexes $i_p$, $i_{dw}$ and $i_{ds}$ when the initial base $K$ is complete.*

It is interesting to note that $\Delta_\mu^{d_D,\Sigma}$ (which coincides with $\Delta_\mu^{d_D,GMax}$) is close to a voting procedure called approval voting (Brams & Fishburn, 1983), where each agent can vote for (approve) as many candidates as she wants, and the elected candidates are the ones who get the greatest number of votes.

As shown by the running example, the family of merging operators $\Delta_\mu^{d_H,f}$ obtained by considering the Hamming distance and letting the aggregation function $f$ vary is not strategy-proof. Let us now focus on this family, and consider successively the two operators obtained by considering $\Sigma$ and $GMax$ as aggregation functions.

As to $\Delta_\mu^{d_H,\Sigma}$, the number of bases, the presence of integrity constraints and the completness of the bases are significant. For this operator, the next theorem makes precise the boundaries between strategy-proofness and manipulation (in the following properties, $K$ represents the initial base and $\#(E)$ the number of bases in the profile $E$):

**Theorem 3**

- *$\Delta_\mu^{d_H,\Sigma}$ is strategy-proof for $i_{dw}$ or $i_{ds}$ if and only if ($\mu \equiv \top$ and $\#(E) = 2$) or $K$ is complete.*

- *$\Delta_\mu^{d_H,\Sigma}$ is strategy-proof for $i_p$ if and only if $K$ is complete.*

In contrast to $\Delta_\mu^{d_H,\Sigma}$, $\Delta_\mu^{d_H,GMax}$ is not strategy-proof even in very restricted situations:

**Theorem 4**

- *$\Delta_\mu^{d_H,GMax}$ is not strategy-proof for the satisfaction indexes $i_{dw}$ and $i_p$ (even if $\mu \equiv \top$, $K$ is complete and $\#(E) = 2$).*

- *$\Delta_\mu^{d_H,GMax}$ is strategy-proof for the satisfaction index $i_{ds}$ if and only if $\mu \equiv \top$, $K$ is complete and $\#(E) = 2$.*

## 5.2 Formula-Based Operators

For the probabilistic index, none of the formula-based operators $\triangle_\mu^{C1}$, $\triangle_\mu^{C3}$, $\triangle_\mu^{C4}$, and $\triangle_\mu^{C5}$ is strategy-proof. But, for the two drastic indexes, there are some situations where strategy-proofness can be ensured:

**Theorem 5**

- *$\triangle_\mu^{C1}$, $\triangle_\mu^{C3}$, $\triangle_\mu^{C4}$, and $\triangle_\mu^{C5}$ are not strategy-proof for $i_p$ (even if $\mu \equiv \top$, $K$ is complete and $\#(E) = 2$).*

- *$\triangle_\mu^{C1}$ is strategy-proof for $i_{dw}$ and $i_{ds}$.*

- *$\triangle_\mu^{C3}$ is strategy-proof for $i_{dw}$ and $i_{ds}$ if and only if $\mu \equiv \top$.*

- *$\triangle_\mu^{C4}$ is not strategy-proof for $i_{dw}$ and $i_{ds}$ (even if $\mu \equiv \top$, $K$ is complete and $\#(E) = 2$).*

- $\triangle_\mu^{C5}$ is strategy-proof for $i_{dw}$ if and only if $\mu \equiv \top$ or $K$ is complete, and is strategy-proof for $i_{ds}$ if and only if $\mu \equiv \top$.

For the other formula-based merging operators $\triangle_\mu^{\widehat{C1}}$, $\triangle_\mu^{\widehat{C3}}$, $\triangle_\mu^{\widehat{C4}}$, $\triangle_\mu^{\widehat{C5}}$, the results are more balanced, with more strategy-proofness results:

**Theorem 6**

- $\triangle_\mu^{\widehat{C1}}$ is strategy-proof for $i_{dw}$ and $i_{ds}$, and is strategy-proof for $i_p$ if and only if $\#(E) = 2$.

- $\triangle_\mu^{\widehat{C3}}$ is strategy-proof for $i_{dw}$ and $i_{ds}$ if and only if $\mu \equiv \top$, and is strategy-proof for $i_p$ if and only if $\#(E) = 2$ and $\mu \equiv \top$.

- $\triangle_\mu^{\widehat{C4}}$ is strategy-proof for $i_p$, $i_{dw}$ and $i_{ds}$.

- $\triangle_\mu^{\widehat{C5}}$ is strategy-proof for $i_{dw}$ if and only if $\#(E) = 2$ or $\mu \equiv \top$ or $K$ is complete. $\triangle_\mu^{\widehat{C5}}$ is strategy-proof for $i_{ds}$ if and only if $\#(E) = 2$ or $\mu \equiv \top$. Finally, $\triangle_\mu^{\widehat{C5}}$ is strategy-proof for $i_p$ if and only if $\#(E) = 2$.

## 5.3 Ensuring Strategy-Proofness: The Case of Complete Bases

Let us now focus on a very specific case: the situation where every base is complete. While this situation is rather infrequent when dealing with usual belief bases, it can be imposed in a goal merging setting, especially if it guarantees strategy-proofness. This explains why we consider such a case in this paper. As said above, it is also interesting because of the relationship with uninominal voting systems if one interprets each complete base as a vote for the corresponding interpretation.

**Theorem 7** *The strategy-proofness results reported in Table 3 hold, under the restriction that each base is complete ($f$ stands for any aggregation function, and $d$ for any distance).* **sp** *means "strategy-proof",* $\overline{\textbf{sp}}$ *means "non strategy-proof" even if $\#(E) = 2$ and $\mu \equiv \top$,* $\overline{\textbf{sp}}^*$ *means "non strategy-proof" even if either $\#(E) = 2$ or $\mu \equiv \top$, but "strategy-proof" if both $\#(E) = 2$ and $\mu \equiv \top$. Finally,* $\overline{\textbf{sp}}^\top$ *means "non strategy-proof" even if $\#(E) = 2$, but "strategy-proof" whenever $\mu \equiv \top$.*

As Theorem 7 shows, no operator among the $\Delta_\mu^{d_H, GMax}$ and the $\triangle_\mu^C$ ones ensures full strategy-proofness in the restricted case where two complete bases are merged and no integrity constraint is considered. Contrastingly, all the other operators offer strategy-proofness for the three indexes whenever every base is complete.

## 5.4 Dalal Index

As explained before, the fact that $i_p$ is based on model counting allows some form of graduality in the corresponding notion of satisfaction, and this contrasts with the drastic indexes. Actually, other non drastic indexes can be defined. In particular, in cases where the agent knows that the result could not fit her beliefs/goals (e.g., if her beliefs/goals are

| $\Delta$ | $i_p$ | $i_{dw}$ | $i_{ds}$ |
|---|---|---|---|
| $\Delta_\mu^{d_D,J}$ | **sp** | **sp** | **sp** |
| $\Delta_\mu^{d,\Sigma}$ | **sp** | **sp** | **sp** |
| $\Delta_\mu^{d_H,GMax}$ | $\overline{sp}$ | $\overline{sp}$ | $\overline{sp}^*$ |
| $\triangle_\mu^{C1}$ | $\overline{sp}$ | **sp** | **sp** |
| $\triangle_\mu^{C3}$ | $\overline{sp}$ | $\overline{sp}^\dagger$ | $\overline{sp}^\dagger$ |
| $\triangle_\mu^{C4}$ | $\overline{sp}$ | $\overline{sp}$ | $\overline{sp}$ |
| $\triangle_\mu^{C5}$ | $\overline{sp}$ | **sp** | $\overline{sp}^\dagger$ |
| $\triangle_\mu^{\widehat{C1}}$ | **sp** | **sp** | **sp** |
| $\triangle_\mu^{\widehat{C3}}$ | **sp** | **sp** | **sp** |
| $\triangle_\mu^{\widehat{C4}}$ | **sp** | **sp** | **sp** |
| $\triangle_\mu^{\widehat{C5}}$ | **sp** | **sp** | **sp** |

Table 3: Strategy-proofness: complete bases

.

not consistent with the integrity constraints), she still can be interested in achieving a result that is close to her beliefs/goals. Closeness can be captured by a notion of distance, and a possible satisfaction index is the following "Dalal index":

**Definition 11 (Dalal index)** $i_{Dalal}(K, K_\Delta) = 1 - \frac{min(\{d_{Hal}(\omega, K_\Delta) \mid \omega \models K\})}{\#(\mathcal{P})}$.

As far as we know, this index has never been carried out. For the sake of homogeneity with previous indexes, the greater $i_{Dalal}(K, K_\Delta)$ the more satisfied the agent associated with $K$. $i_{Dalal}$ grows antimonotonically with the Hamming distance between the two bases under consideration, i.e., the minimal distance between a model of the first base and a model of the second one. Thus, this index takes its minimal value when every variable must be flipped to obtain a model of $K_\Delta$ from a model of $K$, while it takes its maximal value whenever $K$ is consistent with $K_\Delta$ (no flip is required).

A direct observation is that $i_{Dalal}(K, K_\Delta) \geq i_{dw}(K, K_\Delta)$, whatever the bases $K$ and $K_\Delta$. Investigating the strategy-proofness of a profile $E$ for the Dalal index given a merging operator $\Delta$ and an integrity constraint $\mu$ makes sense only in the situation $K \wedge \Delta_\mu(E)$ is inconsistent. Indeed, in the remaining case, $i_{Dalal}(K, \Delta_\mu(E))$ takes the maximal value 1 and no manipulation is possible.

In contrast to the three previous indexes we have considered, merging operators are not strategy-proof for $i_{Dalal}$, even in very restricted situations.

**Theorem 8** *None of* $\Delta_\mu^{d_D,\Sigma}$, $\Delta_\mu^{d_D,GMax}$, $\Delta_\mu^{d_H,\Sigma}$ *or* $\Delta_\mu^{d_H,GMax}$ *is strategy-proof for* $i_{Dalal}$, *even in the restricted case where* $E$ *consists of two complete bases.*

**Theorem 9** *None of the* $\triangle_\mu^{\widehat{C}}$ *operators (hence, none of the* $\triangle_\mu^C$ *operators) is strategy-proof for* $i_{Dalal}$, *even in the restricted case where* $E$ *consists of two complete bases.*

## 5.5 Restricted Strategies

There are situations where the other agents participating to the merging process have some information about the bases of the other agents. For instance, in cooperative problem solving, it can be decided that whenever an agent is able to answer a query within a limited amount of time, she has to communicate it to the other agents. Contrastingly, the communication protocol may force an agent to inform the other agents that she is definitely not able to answer the query. Such information exchanges allow the other agents to get a partial view of the models or the countermodels of the true beliefs/goals of the agent, and if this conflicts with the reported beliefs/goals, the untruthful agent can be unmasked. That is clearly a wrong thing for the untruthful agent since her opinion could then be ignored; she can even be punished for her guilty behaviour.

We consider here two restrictions on available manipulations (and the corresponding notions of strategy-proofness): erosion manipulation is when the agent pretends to believe/desire more that she does (the agent gives only some parts of its models); and dilatation manipulation is when the agent pretends to believe/desire less that she does (the agent gives only parts of its countermodels).

## Definition 12 (erosion and dilation)

- *Erosion manipulation holds when the reported base $K'$ is logically stronger than the true one $K$: $K' \models K$*

- *Dilation manipulation holds when the reported base $K'$ is logically weaker than the true one $K$: $K \models K'$.*

Erosion (resp. dilation) manipulation is safe for the untruthful agent when the other agents may only have access to a subset of the countermodels (resp. models) of her true beliefs/goals, while it is unsafe in general when the other agents may have access to a subset of the models (resp. countermodels).

The next theorem gives the dilation strategy-proofness of model-based operators:

**Theorem 10** *Let $d$ be a pseudo-distance and let $f$ be an aggregation function. $\Delta_\mu^{d,f}$ is dilation strategy-proof for the indexes $i_p$, $i_{dw}$ and $i_{ds}$.*

This result has to be compared with the ones in the unrestricted case (previous sections), where most of the operators are not strategy-proof.

It is not the same story for erosion. One can easily find profiles that can be manipulated using erosion manipulation (see the running example). Interestingly, focusing on erosion strategy-proofness proves sufficient in some situations. Indeed, when $d$ is any distance, $\Sigma$ is the aggregation function and any drastic index $i_d$ are considered, $\Delta_\mu^{d,\Sigma}$ is strategy-proof for $i_d$ if and only if it is erosion strategy-proof for $i_d$:

**Theorem 11** *Let $d$ be any distance. If $\Delta_\mu^{d,\Sigma}$ is not strategy-proof for the index $i_{dw}$ (resp. $i_{ds}$), then it is not erosion strategy-proof for $i_{dw}$ (resp. $i_{ds}$).*

This result has a corollary, showing that it it enough to focus on each complete base that implies $K$ to determine whether a profile $E$ is manipulable by a base $K$ for $i_{dw}$:

**Corollary 12** *A profile $E$ is manipulable by $K$ for $i_{dw}$ (resp. $i_{ds}$) given $\Delta_\mu^{d,\Sigma}$ and $\mu$ if and only if the manipulation is possible using a complete base $K_\omega \models K$, i.e., there exists $K_\omega \models K$ s.t. $i_{dw}(K, \Delta_\mu^{d,\Sigma}(E \sqcup \{K_\omega\})) > i_{dw}(K, \Delta_\mu^{d,\Sigma}(E \sqcup \{K\}))$ (resp. $i_{ds}(K, \Delta_\mu^{d,\Sigma}(E \sqcup \{K_\omega\})) > i_{ds}(K, \Delta_\mu^{d,\Sigma}(E \sqcup \{K\})))$.*

## 6. Discussion

In this paper, we have drawn the strategy-proofness landscape for many merging operators, including model-based ones and formula-based ones. While both families are not strategy-proof in the general case, we have shown that several restrictions on the merging framework or on the available strategies may lead to strategy-proofness.

As to model-based operators, the choice of a "right" distance appears crucial. Thus, model-based operators are strategy-proof when based on the drastic distance, while they are typically not strategy-proof when based on Dalal distance.

Among formula-based merging operators $\triangle_\mu^{C1}$ achieves the highest degree of strategy-proofness in the sense that it is strategy-proof for the drastic indexes.

Most of the results are summed up in Table 4 (**sp** means "strategy-proof" and $\overline{sp}$ means "not strategy-proof"). For space reasons, the results on restricted strategies are not reported there (see Section 5.5), as well as the ones concerning complete bases (see Table 3).

From the derived results, it appears that strategy-proofness is easier to achieve with formula-based operators than with model-based ones, especially when the bases are singletons (i.e., with the $\triangle_\mu^{\widehat{Ci}}$ operators). This could be explained by the fact that the latter operators obey an all-or-nothing principle – a base is either selected as a whole (and included as such in a maxcons) or it is not selected at all – and this may forbid some subtle manipulations.

We have also exhibited some restricted strategies that constrain the agent who wants to manipulate. For example, all model-based operators are strategy-proof for dilation.

Most of the results of this paper are based on three satisfaction indexes, that are, according to us, the most natural ones when no additional information about the merging process is available. These three indexes share the property that if two bases are jointly inconsistent, then the satisfaction is minimal. We call it the consistency property. In order to handle scenarios where the consistency property is not discriminant enough, we introduced the Dalal index. It turns out that none of the operators considered in the paper is strategy-proof for this index.

The choice of a satisfaction index has (by definition) a major impact on the existence of manipulation. As explained before, if the full preferences of all the agents over sets of interpretations were available, strategy-proofness could be easily defined as in Social Choice Theory: a manipulation occurs if and only if the merged base obtained when the agent lies is strictly preferred (w.r.t. her own preference) to the merged base obtained when she reports her "true" base. When such information are not available, several choices for an index are possible, capturing different intuitions.

When beliefs are to be merged, indexes satisfying the consistency property seem more suited than the remaining ones; indeed, for the latter indexes, even if a merged base is "close" to the agent's beliefs, it is still not compatible with them. Drawing such a conclusion is not so easy when goals are to be merged, since for instance, it can be the case that for some agents, the more goals satisfied, the better. Nevertheless, our three satisfaction indexes $i_{dw}$, $i_{ds}$ and $i_p$ are still meaningful in such scenarios, since an agent is typically more satisfied if the goals of the group are compatible with her own ones than if it is not the case.

| $i$ | $\#(E)$ | $K$ | $\mu$ | $\Delta_\mu^{d_D,f}$ | $\Delta_\mu^{d_H,\Sigma}$ | $\Delta_\mu^{d_H,Gmax}$ | $\triangle_\mu^{C1}$ | $\triangle_\mu^{C3}$ | $\triangle_\mu^{C4}$ | $\triangle_\mu^{C5}$ | $\triangle_\mu^{\widehat{C1}}$ | $\triangle_\mu^{\widehat{C3}}$ | $\triangle_\mu^{\widehat{C4}}$ | $\triangle_\mu^{\widehat{C5}}$ |
|---|---|---|---|---|---|---|---|---|---|---|---|---|---|---|
| $i_p$ | $=2$ | complete | $\equiv\top$ | **sp** | **sp** | $\overline{sp}$ | $\overline{sp}$ | $\overline{sp}$ | $\overline{sp}$ | $\overline{sp}$ | **sp** | **sp** | **sp** | **sp** |
| | | | $\not\equiv\top$ | **sp** | **sp** | $\overline{sp}$ | $\overline{sp}$ | $\overline{sp}$ | $\overline{sp}$ | $\overline{sp}$ | **sp** | $\overline{sp}$ | **sp** | **sp** |
| | | not complete | $\equiv\top$ | **sp** | $\overline{sp}$ | $\overline{sp}$ | $\overline{sp}$ | $\overline{sp}$ | $\overline{sp}$ | $\overline{sp}$ | **sp** | **sp** | **sp** | **sp** |
| | | | $\not\equiv\top$ | **sp** | $\overline{sp}$ | $\overline{sp}$ | $\overline{sp}$ | $\overline{sp}$ | $\overline{sp}$ | $\overline{sp}$ | **sp** | $\overline{sp}$ | **sp** | **sp** |
| | $>2$ | complete | $\equiv\top$ | **sp** | **sp** | $\overline{sp}$ | $\overline{sp}$ | $\overline{sp}$ | $\overline{sp}$ | $\overline{sp}$ | $\overline{sp}$ | $\overline{sp}$ | **sp** | $\overline{sp}$ |
| | | | $\not\equiv\top$ | **sp** | **sp** | $\overline{sp}$ | $\overline{sp}$ | $\overline{sp}$ | $\overline{sp}$ | $\overline{sp}$ | $\overline{sp}$ | $\overline{sp}$ | **sp** | $\overline{sp}$ |
| | | not complete | $\equiv\top$ | **sp** | $\overline{sp}$ | $\overline{sp}$ | $\overline{sp}$ | $\overline{sp}$ | $\overline{sp}$ | $\overline{sp}$ | $\overline{sp}$ | $\overline{sp}$ | **sp** | $\overline{sp}$ |
| | | | $\not\equiv\top$ | **sp** | $\overline{sp}$ | $\overline{sp}$ | $\overline{sp}$ | $\overline{sp}$ | $\overline{sp}$ | $\overline{sp}$ | $\overline{sp}$ | $\overline{sp}$ | **sp** | $\overline{sp}$ |
| $i_{dw}$ | $=2$ | complete | $\equiv\top$ | **sp** | **sp** | $\overline{sp}$ | **sp** | **sp** | $\overline{sp}$ | **sp** | **sp** | **sp** | **sp** | **sp** |
| | | | $\not\equiv\top$ | **sp** | **sp** | $\overline{sp}$ | **sp** | $\overline{sp}$ | $\overline{sp}$ | **sp** | **sp** | $\overline{sp}$ | **sp** | **sp** |
| | | not complete | $\equiv\top$ | **sp** | **sp** | $\overline{sp}$ | **sp** | **sp** | $\overline{sp}$ | **sp** | **sp** | **sp** | **sp** | **sp** |
| | | | $\not\equiv\top$ | **sp** | $\overline{sp}$ | $\overline{sp}$ | **sp** | $\overline{sp}$ | $\overline{sp}$ | $\overline{sp}$ | **sp** | $\overline{sp}$ | **sp** | **sp** |
| | $>2$ | complete | $\equiv\top$ | **sp** | **sp** | $\overline{sp}$ | **sp** | **sp** | $\overline{sp}$ | **sp** | **sp** | **sp** | **sp** | **sp** |
| | | | $\not\equiv\top$ | **sp** | **sp** | $\overline{sp}$ | **sp** | $\overline{sp}$ | $\overline{sp}$ | **sp** | **sp** | $\overline{sp}$ | **sp** | **sp** |
| | | not complete | $\equiv\top$ | **sp** | $\overline{sp}$ | $\overline{sp}$ | **sp** | **sp** | $\overline{sp}$ | **sp** | **sp** | **sp** | **sp** | **sp** |
| | | | $\not\equiv\top$ | **sp** | $\overline{sp}$ | $\overline{sp}$ | **sp** | $\overline{sp}$ | $\overline{sp}$ | $\overline{sp}$ | **sp** | $\overline{sp}$ | **sp** | $\overline{sp}$ |
| $i_{ds}$ | $=2$ | complete | $\equiv\top$ | **sp** | **sp** | **sp** | **sp** | **sp** | $\overline{sp}$ | **sp** | **sp** | **sp** | **sp** | **sp** |
| | | | $\not\equiv\top$ | **sp** | **sp** | $\overline{sp}$ | **sp** | $\overline{sp}$ | $\overline{sp}$ | $\overline{sp}$ | **sp** | $\overline{sp}$ | **sp** | **sp** |
| | | not complete | $\equiv\top$ | **sp** | **sp** | $\overline{sp}$ | **sp** | **sp** | $\overline{sp}$ | **sp** | **sp** | **sp** | **sp** | **sp** |
| | | | $\not\equiv\top$ | **sp** | $\overline{sp}$ | $\overline{sp}$ | **sp** | $\overline{sp}$ | $\overline{sp}$ | $\overline{sp}$ | **sp** | $\overline{sp}$ | **sp** | **sp** |
| | $>2$ | complete | $\equiv\top$ | **sp** | **sp** | $\overline{sp}$ | **sp** | **sp** | $\overline{sp}$ | **sp** | **sp** | **sp** | **sp** | **sp** |
| | | | $\not\equiv\top$ | **sp** | **sp** | $\overline{sp}$ | **sp** | $\overline{sp}$ | $\overline{sp}$ | $\overline{sp}$ | **sp** | $\overline{sp}$ | **sp** | $\overline{sp}$ |
| | | not complete | $\equiv\top$ | **sp** | $\overline{sp}$ | $\overline{sp}$ | **sp** | **sp** | $\overline{sp}$ | **sp** | **sp** | **sp** | **sp** | **sp** |
| | | | $\not\equiv\top$ | **sp** | $\overline{sp}$ | $\overline{sp}$ | **sp** | $\overline{sp}$ | $\overline{sp}$ | $\overline{sp}$ | **sp** | $\overline{sp}$ | **sp** | $\overline{sp}$ |

Table 4: Synthesis of the results.

In light of our study, strategy-proofness appears as a property independent of the computational complexity of query answering from a merged base (see Konieczny et al., 2004). It means that having a low/high complexity does not prevent/imply strategy-proofness.

Strategy-proofness appears also independent of the fact that the operator satisfies the rationality postulates given by Konieczny and Pino Pérez (1999, 2002). Indeed, as a direct consequence of Theorems 2 and 3, we have that some majority (resp. arbitration) merging operators (Konieczny & Pino Pérez, 2002) are strategy-proof, while others are not. Thus, satisfying the rationality postulates for merging proves not sufficient to ensure strategy-proofness or manipulability. Nevertheless, we can note that arbitration operators, like $\Delta^{d,GMax}$, are more sensitive to manipulation than majority operators, like $\Delta^{d,\Sigma}$. This is easily explained by the fact that arbitration operators are egalitarist ones: they aim at giving a result that is close to each base of the profile. Intuitively, a small change in a base can heavily change the whole result. Contrastingly, majority operators, that listen to majority wishes for defining the merged base, often do not take into account bases that are "far" from the majority. So, when using majority operators, it is likely that a small change in a base has no impact on the merged base.

Thus, strategy-proofness can be viewed as a further dimension that can be used to compare merging operators, besides the computational complexity and rationality criteria. Its independence to the latter criteria may also explain why no strategy-proofness results for wide families of operators seem to exist.

## 7. Related Work

As explained through this paper, the manipulation problem has been studied extensively in Social Choice Theory for years. In the next subsection we will relate our work to this stream of research. In a second subsection we mention some other related work concerning the strategy-proofness issue for weighted bases.

### 7.1 Social Choice Theory

In the propositional merging framework considered in the paper, the beliefs/goals $K$ of each agent induce a two-strata partition of the interpretations: all the models of $K$ are equally preferred, and strictly preferred to its countermodels, which are equally disliked. When agents report full preference relations (that can be encoded in various ways, e.g., explicitly, or by a prioritized base, an ordinal conditional function, etc.), the aggregation problem consists in defining a global preference relation from individual preference relations. This problem has been addressed for centuries in Social Choice Theory. This can be traced back at least to Condorcet (1785) and Borda (1781).

In Social Choice Theory (Arrow et al., 2002), the strategy-proofness problem has received great attention. In this framework, an agent $A$ is untruthful when she reports a preference relation (a complete pre-order over the set of alternatives) that is not the true one. A social choice function (associating an alternative to a profile of such preference relations) is not strategy-proof when the alternative chosen by the function when $A$ lies is ranked higher for $A$ than the alternative chosen when she reports her true preferences. One of the most famous result in Social Choice Theory is that there is no "good" strategy-proof preference aggregation procedure. This result is known as Gibbard-Satterthwaite impossibility theorem (Gibbard, 1973; Satterthwaite, 1975; Moulin, 1988).

Formally, consider a set of agents (*individuals*) $N = \{1, \ldots, n\}$, and a set of *alternatives* $\mathcal{A} = \{a, b, \ldots\}$. Each agent $i$ has a *preference relation* on those alternatives, that is supposed to be a complete, reflexive and transitive binary relation, noted $\geq_i$. A *preference profile* $P = (\geq_1, \ldots, \geq_n)$ assigns a preference relation to each agent. Let us note $\mathfrak{P}$ the set of all possible preference profiles. A given preference profile can be noted $P = (P_{-i}, \geq_i)$, $i \in N$, where $P_{-i}$ denotes the profile $P$ without (the preferences of) individual $i$. A *social choice function* $f$ is a mapping from $\mathfrak{P}$ to $\mathcal{A}$. A social choice function is *manipulable* if there is an individual $i \in N$, a preference relation $\geq'$, and a preference profile $P$ such that $f(P_{-i}, \geq'_i) >_i f(P)$, i.e., when there is an agent $i$ that is best satisfied with the result when she claims her preferences are $\geq'_i$ instead of her true preference $\geq_i$. If a social choice function is not manipulable, it is said to be *strategy-proof*. A social choice function is *dictatorial* if there is an individual $i \in N$ (the dictator), such that $f(P) \geq_i a$ for all $a \in \mathcal{A}$ and for all $P \in \mathfrak{P}$. A social choice function is *onto* if for each alternative $a \in \mathcal{A}$ there is a preference profile $P \in \mathfrak{P}$ such that $f(P) = a$. Then Gibbard-Satterthwaite theorem (Gibbard, 1973; Satterthwaite, 1975) can be stated as:

**Theorem 13** *(Gibbard, 1973; Satterthwaite, 1975) If $\mathcal{A}$ contains at least three alternatives, then there is no social choice function $f$ that is onto, strategy-proof and non-dictatorial.*

Since this result has been stated, there has been a lot of work for deriving strategy-proofness results under some restrictions (see Kelly, 1988; Arrow et al., 2002). In some sense, our work is relevant to such approaches. Nonetheless, our work is original - as far as we know - from two points of view: on the one hand, the preference relations considered here are two-strata total pre-orders, and not arbitrary pre-orders; on the other hand, the result of a merging process is usually not a single interpretation but still a two-strata total pre-order (and the number of models of the merged base is not constrained *a priori*). This leads to more complex notions of strategy-proofness where different definitions are possible, depending on an index which formalizes one of the various intuitive notions of "how satisfied an agent is by the result of the merging process".

In social choice theory, there are also some works on social choice correspondences, that are mapping from $\mathfrak{P}$ to $2^{\mathcal{A}}$, and that are closer to our framework than social choice functions. The data coming from individuals are preference relations on $\mathcal{A}$, and the problem is to shift them to preference relations on $2^{\mathcal{A}}$. The standard way to achieve it is to consider that even if a set is chosen by the correspondence, then ultimately only one alternative will be realized, and to suppose that each individual has a subjective probability measure on this realization. Then a social choice correspondance is strategy-proof if it is not possible for an individual to increase her expected utility of the result. In this case, results similar to Gibbard-Satterthwaite theorems can be derived (see e.g., Barberà, Dutta, & Sen, 2001; Chin & Zhou, 2002; Duggan & Schwartz, 2000). These works are related to the one conducted in this paper, but they all suppose that each agent makes available not only its full preference relation on alternatives under the form of a utility function – typically not reducible to a two-strata complete pre-order – but also its subjective probability measure on alternatives. Contrastingly, in this work, the only information coming from each agent is the corresponding base, which typically approximates her full preference relation, and is of pure ordinal nature.

## 7.2 Strategy-Proofness for Weighted-Bases Merging

A study of strategy-proofness of some merging operators has been carried out by Meyer, Ghose, and Chopra (2001). The framework they consider is distinct from the one used in our work. On the one hand, agents may report full preference relations (encoded as ordinal conditional functions, also called $\kappa$-functions, see Spohn, 1987). On the other hand, the merging operators under consideration escape Gibbard-Satterthwaite theorem since Meyer et al. (2001) make a commensurability assumption between the agents' preference relations (the same remark applies also to possibilistic base merging as defined by Benferhat, Dubois, Kaci, & Prade, 2002). Roughly, this commensurability assumption amounts to consider that the weights (or the levels) associated to formulas have the same meaning for all the agents, i.e., that a weight 3 for agent 1 is the same that a weight 3 for agent 2. The commensurability assumption is sensible in many situations, but when dealing with agents' preferences, commensurability must be used carefully. For human agents, it is commonly accepted in Social Choice Theory that this assumption is very strong. Arrow (1963) illustrates this idea by quoting Bentham:

> " This is vain to talk of adding quantities which after the addition will continue distinct as they were before, one man's happiness will never be another man's happiness: a gain to one man is no gain to another; you might as well pretend to add 20 apples to 20 pears..."

The notion of strategy-proofness and the merging operators of Meyer et al. (2001) and Chopra, Ghose, and Meyer (2006) are defined in the framework of ordinal conditional functions. In this section, we study the corresponding operators in the pure propositional framework, i.e., when the profile contains "flat" belief/goal bases, in order to compare them with our approach.

An ordinal conditional function (OCF) $\kappa$ is a total function from the set of interpretations $\mathcal{W}$ to the set of non-negative integers (originally, an OCF maps an interpretation to the class of ordinals, and is such that at least one interpretation is mapped to zero, but considering integers is sufficient here). Intuitively, the greater the number, the less credible the interpretation. To each OCF $\kappa$ one can associate a base $Bel(\kappa)$ defined as $[Bel(\kappa)] = \{\omega \in \mathcal{W} \mid \kappa(\omega) = min_{\omega' \in \mathcal{W}}(\kappa(\omega'))\}$. The aim of OCF merging operators is, from a profile of OCFs $E = \{\kappa_1, \ldots, \kappa_n\}$ to define an OCF $\kappa_\Delta(E)$ that best represents the profile. The operators studied by Meyer et al. (2001) are the following ones:

- $\kappa_{\Delta_{\max}}(E)(\omega) = \max_{\kappa_i \in E} \kappa_i(\omega)$,

- $\kappa_{\Delta_{\min_1}}(E)(\omega) = \begin{cases} 2\kappa_1(\omega) \text{ if } \kappa_i(\omega) = \kappa_j(\omega) \text{ for all } \kappa_i, \kappa_j \in E \\ 2\min_{\kappa_i \in E} \kappa_i(\omega) + 1 \text{ otherwise,} \end{cases}$

- $\kappa_{\Delta_{\min_2}}(E)(\omega) = \begin{cases} \kappa_1(\omega) \text{ if } \kappa_i(\omega) = \kappa_j(\omega) \text{ for all } \kappa_i, \kappa_j \in E \\ \min_{\kappa_i \in E} \kappa_i(\omega) + 1 \text{ otherwise,} \end{cases}$

- $\kappa_{\Delta_\Sigma}(E)(\omega) = \sum_{\kappa_i \in E} \kappa_i(\omega)$.

The straightforward way to translate the framework of propositional merging into ordinal conditional functions is to consider a propositional base as a special case of OCF: a propositional base is a two-strata OCF, with the models of the bases having rank 0 and the countermodels having rank 1. If we consider only two-strata OCFs $\kappa_i$ and note $K_i = Bel(\kappa_i)$ and $\Delta = Bel(\kappa_\Delta)$, the previous definitions of merging operators give:

- $\Delta_{max}(E) \equiv \Delta_{min_2}(E) \equiv \bigwedge E$ if consistent and $\Delta_{max}(E) \equiv \Delta_{min_2}(E) \equiv \top$ otherwise.

- $\Delta_{min_1}(E) \equiv \bigwedge E$ if consistent and $\Delta_{min_1}(E) \equiv \bigvee E$ otherwise.

- $\Delta_\Sigma(E) \equiv \Delta^{d_D, \Sigma}(E)$.

The resulting propositional merging operators $\Delta_{max}$, $\Delta_{min_1}$, $\Delta_{min_2}$, and $\Delta_\Sigma$ are quite simple and well-known. $\Delta_{max}$ (or equivalently $\Delta_{min_2}$) is the so-called basic merging operator (in absence of integrity constraints) (Konieczny & Pino Pérez, 1999). $\Delta_{min_1}$ is the 1-quota operator defined by Everaere, Konieczny, and Marquis (2005) (without integrity constraints). $\Delta_\Sigma$ corresponds to the intersection operator defined by Konieczny (2000).

All those operators are strategy-proof for our indexes:

**Theorem 14** $\Delta_{max}$, $\Delta_{min_1}$, $\Delta_{min_2}$, and $\Delta_\Sigma$ are strategy-proof for $i_{dw}$, $i_{ds}$ and $i_p$.

Besides those operators, Meyer, Chopra and Ghose also proposed general definitions of strategy-proofness for OCF merging. More precisely, they have studied two properties. The first one is the **(IP)** property (Meyer et al., 2001):

**Definition 13 (IP)** *An OCF merging operator $\kappa_\Delta$ satisfies the **(IP)** property if and only if for every OCF profile $E$, for every agent $i$, we have whatever the OCF $\kappa$*

$$\forall \omega \in \mathcal{W}, |\kappa_\Delta(E)(\omega) - \kappa_i(\omega)| \leq |\kappa_\Delta(rep(E, \{i\}, \kappa)(\omega) - \kappa_i(\omega)|$$

*where $rep(E, \{i\}, \kappa)$ is the profile identical to $E$ except that the OCF $\kappa_i$ is replaced by $\kappa$.*

Focusing on two-strata OCFs, we say that a merging operator $\Delta(= Bel(\kappa_\Delta))$ is strategy-proof for **(IP)** if and only if $\kappa_\Delta$ satisfies the **(IP)** property for every agent any profile. We have obtained the following characterization:

**Theorem 15** $\Delta$ *is strategy-proof for **(IP)** if and only if for every profile $E$ and every pair of bases $K$ and $K'$:*

- $K \wedge \neg\Delta(E \sqcup \{K\}) \models \neg\Delta(E \sqcup \{K'\})$, *and*

- $\neg K \wedge \Delta(E \sqcup \{K\}) \models \Delta(E \sqcup \{K'\})$.

The second strategy-proofness property that Meyer, Chopra and Ghose have investigated is **(WIP)**:

**Definition 14 (WIP)** *An OCF merging operator $\kappa_\Delta$ satisfies the* **(WIP)** *property if and only if for every profile $E$, for every agent $i$, we have whatever the OCF $\kappa$:*

$$\Sigma_{\omega \in \mathcal{W}} |\kappa_\Delta(E)(\omega) - \kappa_i(\omega)| \leq \Sigma_{\omega \in \mathcal{W}} |\kappa_\Delta(rep(E, \{i\}, \kappa)(\omega) - \kappa_i(\omega)|.$$

**(WIP)** is weaker than **(IP)** in the sense that if an OCF merging operator $\kappa_\Delta$ satisfies **(IP)** for an agent $i$, then $\kappa_\Delta$ satisfies **(WIP)** for $i$ (but the converse does not always hold).

Again, focusing on two-strata OCFs, we say that a merging operator $\Delta(= Bel(\kappa_\Delta))$ is strategy-proof for **(WIP)** if and only if $\kappa_\Delta$ satisfies the **(WIP)** property for every agent given any profile.

Let us note $\oplus$ the *exclusive or* operator, i.e., $K \oplus K' = (\bigwedge K \wedge \neg \bigwedge K') \vee (\neg \bigwedge K \wedge \bigwedge K')$. Then **(WIP)** can be characterized in our framework by :

**Theorem 16** *Let $i_{wip}(K, K_\Delta) = \frac{1}{\#([K \oplus K_\Delta]) + 1}$. $\Delta$ satisfies the* **(WIP)** *property if and only if it is strategy-proof for $i_{wip}$.*

Note that the "wip index" $i_{wip}$ is very close to the probabilistic index $i_p$. The probabilistic index measures the closeness of the merged base to the agent base, whereas the "wip index" measures the difference between the merged base and the agent base.

However, the corresponding notion of strategy-proofness (and a fortiori the one induced by **(IP)**) appears too strong in the pure propositional setting. Consider the following belief merging scenario:

**Example 6** *Consider $K = a$ and $K_1 = b$. We have $\Delta^{d_H, \Sigma}(\{K, K_1\}) \equiv a \wedge b$. If the agent gives $K' = \{a \wedge \neg b\}$ instead of $K$, then the merged base is $\Delta^{d_H, \Sigma}(\{K', K_1\}) \equiv a$. Accordingly, this an example of manipulation for* **(WIP)** *($i_{wip}(\Delta^{d_H, \Sigma}(\{K, K_1\})) = \frac{1}{2} < i_{wip}(\Delta^{d_H, \Sigma}(\{K', K_1\})) = 1$).*

In this example, the untruthful agent actually manages to change the merged base to one which is more similar to her initial base (with respect to $i_{wip}$). This is because she is not fully satisfied by the merged base equivalent to $a \wedge b$ but still strictly prefers her initial base $\{a\}$, despite the fact that $a \wedge b$ *refines* her own beliefs. Accordingly, such an agent wants to preserve both her beliefs and *her ignorance*. In many scenarios when an agent participates to a merging process in order to get new information, this is counter-intuitive. Chopra et al. (2006) give more general definitions of strategy-proofness, by considering other similarity relations. In the propositional case, they all suffer from the same above-mentioned drawback. This explains why we did not investigate the strategy-proofness of the purely propositional model-based operators and formula-based operators for criteria like **(WIP)** or **(IP)**.

## 8. Conclusion

Investigating the strategy-proofness of merging operators is important from a multi-agent perspective whenever some agents can get the information conveyed by the other agents participating to the merging process. When strategy-proofness is not guaranteed, it may be questioned whether the result of the merging process actually represents the beliefs/goals of the group.

In this paper, we have drawn the strategy-proofness landscape for many existing merging operators, including model-based ones and formula-based ones, both in the general case and under several natural restrictions. Strategy-proofness appears as independent of complexity and rationality aspects, and can be used as such, as a further criterion to evaluate merging operators. All those results have been discussed in Section 6.

This work calls for a number of perspectives. A first perspective is to identify the complexity of determining whether a profile can be manipulated by a base given an operator. Indeed, using a merging operator that is not strategy-proof is not necessarily harmful if finding out a strategy is computationally hard. Such a complexity issue has been investigated for voting schemes (Conitzer & Sandholm, 2003; Conitzer, Lang, & Sandholm, 2003; Conitzer & Sandholm, 2002a, 2002b) when individual preferences are given explicitly (which is not the case in our framework). A first result follows easily from Theorem 11: if the distance $d$ between interpretations can be computed in polynomial time in the input size (which is not a strong assumption), determining whether a given profile can be manipulated by a base for a drastic index given $\Delta_\mu^{d,\Sigma}$ and $\mu$ is in $\Sigma_2^p$.

In Social Choice Theory, the Gibbard-Sattertwhaite theorem states that every "sensible" social choice function is manipulable. Taking into account the fact that the agents are tempted to manipulate transforms the aggregation process into a game between agents. For ensuring strategy-proofness, it can prove sufficient to build a game where telling the truth is an optimal strategy for each agent. How to achieve it is the aim of implementation theory (also called mechanism design), see e.g., Maskin & Sjostrom, 2002. A perspective is to determine whether building such mechanisms is possible in a belief merging setting in order to force the agents to tell the truth. Most of the work on mechanism design assume transferable utility, and use payments as part as the process. Importing such ideas in a fully qualitative framework surely is a hard task.

Another interesting perspective is to study the strategy-proofness problem when coalitions are allowed. Instead of considering manipulation by single agents, one can be interested in manipulation by coalition of agents who coordinate to improve the result for the coalition. See (Meyer et al., 2001; Chopra et al., 2006) for such a definition in a different framework. Since manipulation by a single agent is a particular case of manipulation by a coalition, and since we have seen that many operators are not strategy-proof for a single agent, it is clear that strategy-proofness results for coalitions will be very hard to achieve.

## Acknowledgements

The authors would like to thank the anonymous referees for their thoughful comments which helped us a lot to improve the paper. The authors have been supported by the Université d'Artois, the Région Nord/Pas-de-Calais, the IRCICA Consortium, and by the European Community FEDER Program.

## Appendix A. Proofs

**Theorem 1**

1. *If a merging operator is strategy-proof for $i_p$, then it is strategy-proof for $i_{dw}$.*

2. *Consider a merging operator $\Delta$ that generates only consistent bases.[5] If it is strategy-proof for $i_p$, then it is strategy-proof for $i_{ds}$.*

**Proof:**

1. Assume that $\Delta_\mu$ is not strategy-proof for $i_{dw}$. Then there exists a profile $E$, a base $K$, a base $K'$ and an integrity constraint $\mu$ s.t. (1) $\Delta_\mu(E \sqcup \{K\}) \wedge K$ is inconsistent, and (2) $\Delta_\mu(E \sqcup \{K'\}) \wedge K$ is consistent. (1) implies that $\frac{\#([K] \cap [\Delta_\mu(E \sqcup \{K\})])}{\#([\Delta_\mu(E \sqcup \{K\})])} = 0$. (2) implies that $\frac{\#([K] \cap [\Delta_\mu(E \sqcup \{K'\})])}{\#([\Delta_\mu(E \sqcup \{K'\})])} > 0$. Hence, $\Delta_\mu$ is not strategy-proof for $i_p$.

2. Assume that $\Delta_\mu$ is not strategy-proof for $i_{ds}$. Then there exists a profile $E$, a base $K$, a base $K'$ and an integrity constraint $\mu$ s.t. (1) $\Delta_\mu(E \sqcup \{K\}) \not\models K$, and (2) $\Delta_\mu(E \sqcup \{K'\}) \models K$. (1) implies that $\frac{\#([K] \cap [\Delta_\mu(E \sqcup \{K\})])}{\#([\Delta_\mu(E \sqcup \{K\})])} \neq 1$. (2) implies that $\frac{\#([K] \cap [\Delta_\mu(E \sqcup \{K'\})])}{\#([\Delta_\mu(E \sqcup \{K'\})])} = 1$ if $\Delta_\mu(E \sqcup \{K\})$ is consistent. Hence, $\Delta_\mu$ is not strategy-proof for $i_p$.

$\square$

**Theorem 2**

- *Let $f$ be any aggregation function. $\Delta_\mu^{d_D,f}$ is strategy-proof for $i_p$, $i_{dw}$ and $i_{ds}$.*

- *Let $d$ be any distance. Provided that only two bases are to be merged, $\Delta_\top^{d,\Sigma}$ is strategy-proof for the indexes $i_{dw}$ and $i_{ds}$.*

- *For any distance $d$, $\Delta_\mu^{d,\Sigma}$ is strategy-proof for the indexes $i_p$, $i_{dw}$ and $i_{ds}$ when the initial base $K$ is complete.*

**Proof:**

- Let $f$ be any aggregation function. $\Delta_\mu^{d_D,f}$ is strategy-proof for $i_p$, $i_{dw}$ and $i_{ds}$.

  *The proof is organized in three steps: by reduction ad absurdum, we show that the minimal drastic distance between a model of $\mu$ and $E \sqcup \{K\}$ is equal to the minimal drastic distance between a model of $\mu$ and $E \sqcup \{K'\}$. Then, it is easy to show that the number of $K$'s models is greater in $E \sqcup \{K\}$ than in $E \sqcup \{K'\}$. Finally, we prove that the number of countermodels of $K$ is greater in $E \sqcup \{K'\}$ than in $E \sqcup \{K\}$, which entails a contradiction.*

---

5. I.e., $\Delta_\mu(E)$ is always consistent.

From Theorem 1, we know that if any operator $\Delta_\mu^{d_D,f}$ is strategy-proof for $i_p$, it is also strategy-proof for both $i_{dw}$ and $i_{ds}$ (indeed, $\Delta_\mu^{d_D,f}(E)$ is always consistent). So it is sufficient to prove the strategy-proofness of $\Delta_\mu^{d_D,f}$ for $i_p$ to prove its strategy-proofness for the three indexes.

Let us prove it by *reductio ad absurdum*: assume that there is an operator $\Delta_\mu^{d_D,f}$, where $d_D$ is the drastic distance and $f$ is any aggregation function, that is not strategy-proof for $i_p$. Then there exist an integrity constraint $\mu$, a profile $E$, and two bases $K$ and $K'$ s.t. $i_p(K, \Delta_\mu^{d_D,f}(\{K\} \sqcup E)) < i_p(K, \Delta_\mu^{d_D,f}(\{K'\} \sqcup E))$, which is equivalent to

$$\frac{\#([K] \cap [E \triangle_\mu^{d_D,f} K])}{\#([E \triangle_\mu^{d_D,f} K])} < \frac{\#([K] \cap [E \triangle_\mu^{d_D,f} K'])}{\#([E \triangle_\mu^{d_D,f} K'])}$$

where $E \triangle_\mu^{d_D,f} K$ is a light notation for $\Delta_\mu^{d_D,f}(\{K\} \sqcup E))$. Let us note $d_{min}(E \sqcup_\mu^{d_D,f} \{K\}) = min(\{d_D(\omega, E \sqcup \{K\}) \mid \omega \models \mu\}, \leq)$. We now show that $d_{min}(E \sqcup_\mu^{d_D,f} \{K\}) = d_{min}(E \sqcup_\mu^{d_D,f} \{K'\})$:

- Let us first notice that we have $i_p(K, E \triangle_\mu^{d_D,f} K) \neq 1$: if $i_p(K, E \triangle_\mu^{d_D,f} K) = 1$, then the probabilistic satisfaction index takes its maximal value, so it is impossible to increase it.
  Since $i_p(K, E \triangle_\mu^{d_D,f} K) < 1$, we have that $\#([K] \cap [E \triangle_\mu^{d_D,f} K]) < \#([E \triangle_\mu^{d_D,f} K])$, so at least one model of $E \triangle_\mu^{d_D,f} K$ does not belong to $K$:

  $$\exists \omega_1 \models (\neg K) \wedge \mu, d_D(\omega_1, E \sqcup \{K\}) = d_{min}(E \sqcup_\mu^{d_D,f} \{K\}).$$

  Since $\omega_1 \models (\neg K) \wedge \mu$, we have $d_D(\omega_1, K) = 1$ and this distance is maximal (because we use the drastic distance). We get immediately that $d_D(\omega_1, K) \geq d_D(\omega_1, K')$. Hence $d_D(\omega_1, E \sqcup \{K\}) \geq d_D(\omega_1, E \sqcup \{K'\})$ (because the aggregation function $f$ satisfy non-decreasingness). Since $d_D(\omega_1, E \sqcup \{K\}) = d_{min}(E \sqcup_\mu^{d_D,f} \{K\})$, we get $d_{min}(E \sqcup_\mu^{d_D,f} \{K\}) \geq d_D(\omega_1, E \sqcup \{K'\})$. Since $d_D(\omega_1, E \sqcup \{K'\}) \geq d_{min}(E \sqcup_\mu^{d_D,f} \{K'\})$ by definition of *min* and since $\omega_1 \models \mu$, we have

  $$d_{min}(E \sqcup_\mu^{d_D,f} \{K\}) \geq d_{min}(E \sqcup_\mu^{d_D,f} \{K'\}) \ \ (*).$$

- We can also conclude that $i_p(K, E \triangle_\mu^{d_D,f} K') \neq 0$: if $i_p(K, E \triangle_\mu^{d_D,f} K') = 0$, then $i_p(K, E \triangle_\mu^{d_D,f} K')$ is minimal, so the value taken by $i_p$ has not increased, and this contradicts the assumption (manipulation).
  If $i_p(K, E \triangle_\mu^{d_D,f} K') \neq 0$, then we can find at least one model of $K \wedge \mu$ in $E \triangle_\mu^{d_D,f} K'$: $\exists \omega_1 \models K \wedge \mu, d_D(\omega_1, E \sqcup \{K'\}) = d_{min}(E \sqcup_\mu^{d_D,f} \{K'\})$. Since $\omega_1 \models K$, we have $d_D(\omega_1, K) = 0$ and since this distance is minimal, we get $d_D(\omega_1, E \sqcup \{K\}) \leq d_D(\omega_1, E \sqcup \{K'\})$, and then $d_D(\omega_1, E \sqcup \{K\}) \leq d_{min}(E \sqcup_\mu^{d_D,f} \{K'\})$, because $d_D(\omega_1, E \sqcup \{K'\}) = d_{min}(E \sqcup_\mu^{d_D,f} \{K'\})$. Furthermore, since $d_D(\omega_1, E \sqcup \{K\}) \geq d_{min}(E \sqcup_\mu^{d_D,f} \{K\})$ by definition of *min* and because $\omega_1 \models \mu$, we have:

  $$d_{min}(E \sqcup_\mu^{d_D,f} \{K\}) \leq d_{min}(E \sqcup_\mu^{d_D,f} \{K'\}) \ \ (**).$$

From the inequations (*) and (**), we get:

$$d_{min}(E \sqcup_{\mu}^{d_D,f} \{K\}) = d_{min}(E \sqcup_{\mu}^{d_D,f} \{K'\}). \tag{1}$$

Let us show now that we can only increase the number of countermodels of $K$ in $E \bigtriangleup_{\mu}^{d_D,f} K'$, and decrease the number of models of $K$ in $E \bigtriangleup_{\mu}^{d_D,f} K'$.

– Let $\omega$ be a countermodel of $K$ which is a model of $E \bigtriangleup_{\mu}^{d_D,f} K$: $\omega \models (\neg K) \wedge (E \bigtriangleup_{\mu}^{d_D,f} K)$.

Since $\omega \models \neg K$, we have $d_D(\omega, K) = 1$ and this distance is maximal. Hence $d_D(\omega, K) \geq d_D(\omega, K')$. So:

$$d_D(\omega, E \sqcup \{K\}) \geq d_D(\omega, E \sqcup \{K'\}) \tag{2}$$

because the aggregation function $f$ satisfies non-decreasingness.

Since $\omega \models E \bigtriangleup_{\mu}^{d_D,f} K$, we have $d_D(\omega, E \sqcup \{K\}) = d_{min}(E \sqcup_{\mu}^{d_D,f} \{K\})$. With (2), we get $d_{min}(E \sqcup_{\mu}^{d_D,f} \{K\}) \geq d_D(\omega, E \sqcup \{K'\})$.

Since $d_{min}(E \sqcup_{\mu}^{d_D,f} \{K\}) = d_{min}(E \sqcup_{\mu}^{d_D,f} \{K'\})$ with (1), we obtain: $d_{min}(E \sqcup_{\mu}^{d_D,f} \{K'\}) \geq d_D(\omega, E \sqcup \{K'\})$. By definition of $min$ and since $\omega \models \mu$ (because $\omega \models E \bigtriangleup_{\mu}^{d_D,f} K$), we deduce that $\omega$ is a model of $E \bigtriangleup_{\mu}^{d_D,f} K'$. We can conclude that every model of $E \bigtriangleup_{\mu}^{d_D,f} K$ which is not a model of $K$ is a model of $E \bigtriangleup_{\mu}^{d_D,f} K'$. Hence: $[\neg K] \cap [E \bigtriangleup_{\mu}^{d_D,f} K] \subseteq [\neg K] \cap [E \bigtriangleup_{\mu}^{d_D,f} K']$.

– Finally, let $\omega$ be a model of $K$ which is a model of $E \bigtriangleup_{\mu}^{d_D,f} K'$: $\omega \models K \wedge (E \bigtriangleup_{\mu}^{d_D,f} K')$. Since $\omega \models K$, we have $d_D(\omega, K) = 0$ and this distance is minimal. Hence $d_D(\omega, K) \leq d_D(\omega, K')$. So:

$$d_D(\omega, E \sqcup \{K\}) \leq d_D(\omega, E \sqcup \{K'\}) \tag{3}$$

because the aggregation function is non-decreasing.

Since $\omega \models E \bigtriangleup_{\mu}^{d_D,f} K'$, we have $d_D(\omega, E \sqcup \{K'\}) = d_{min}(E \sqcup_{\mu}^{d_D,f} \{K'\})$. With (3), we get $d_D(\omega, E \sqcup \{K\} \leq d_{min}(E \sqcup_{\mu}^{d_D,f} \{K'\})$. Since $d_{min}(E \sqcup_{\mu}^{d_D,f} \{K\}) = d_{min}(E \sqcup_{\mu}^{d_D,f} \{K'\})$ with (1), we obtain $d_D(\omega, E \sqcup \{K\} \leq d_{min}(E \sqcup_{\mu}^{d_D,f} \{K\})$. By definition of $min$ and since $\omega \models \mu$ (because $\omega \models E \bigtriangleup_{\mu}^{d_D,f} K'$), we deduce that $\omega$ is a model of $E \bigtriangleup_{\mu}^{d_D,f} K$. We can conclude that every model of $E \bigtriangleup_{\mu}^{d_D,f} K'$ which is a model of $K$ is a model of $E \bigtriangleup_{\mu}^{d_D,f} K$. It follows that $[K] \cap [E \bigtriangleup_{\mu}^{d_D,f} K'] \subseteq [K] \cap [E \bigtriangleup_{\mu}^{d_D,f} K]$.

Since we can only increase the number of countermodels of $K$ in $E \bigtriangleup_{\mu}^{d_D,f} K'$ and decrease the number of models of $K$ in $E \bigtriangleup_{\mu}^{d_D,f} K'$, the proportion of models of $K$ in $E \bigtriangleup_{\mu}^{d_D,f} K'$ is smaller than in $E \bigtriangleup_{\mu}^{d_D,f} K$. This contradicts the assumption and shows that $\bigtriangleup_{\mu}^{d_D,f}$ is strategy-proof for $i_p$.

• Let $d$ be any distance. Provided that only two bases are to be merged, $\bigtriangleup_{\top}^{d,\Sigma}$ is strategy-proof for the indexes $i_{dw}$ and $i_{ds}$.

*In this proof, we first show that the merging of two bases is consistent with each base. Then, the property follows directly.*

Strategy-proofness for the two drastic indexes is a direct consequence of the following property:

**Lemma 1** *If $E = \{K_1, K_2\}$, then $\Delta_\top^{d,\Sigma}(E) \wedge K_1$ and $\Delta_\top^{d,\Sigma}(E) \wedge K_2$ are consistent.*

**Proof:** We show that $\Delta_\top^{d,\Sigma}(E) \wedge K_1$ is consistent (the remaining case is similar by symmetry). *Reductio ad absurdum.* Let us suppose that for two bases $K_1$ and $K_2$, $\Delta_\top^{d,\Sigma}(\{K_1, K_2\})$ is inconsistent with $K_1$. We can deduce that:

$$\exists \omega' \models \neg K_1, \forall \omega \models K_1, d(\omega, K_1 \triangle^{d,\Sigma} K_2) > d(\omega', K_1 \triangle^{d,\Sigma} K_2),$$

where $K_1 \triangle^{d,\Sigma} K_2$ is a light notation for $\Delta_\top^{d,\Sigma}(\{K_1, K_2\})$.

Since $\forall \omega \models K_1, d(\omega, K_1) = 0$, we get that $\exists \omega' \models \neg K_1, \forall \omega \models K_1, d(\omega, K_2) > d(\omega', K_1) + d(\omega', K_2)$. In particular, if we consider $\omega_1 \models K_1$ s.t. $d(\omega', \omega_1) = d(\omega', K_1)$ (such an $\omega_1$ exists by definition of $d(\omega', K_1)$), we have: $d(\omega_1, K_2) > d(\omega', \omega_1) + d(\omega', K_2)$. Similarly, if we consider $\omega_2 \models K_2$ s.t. $d(\omega', \omega_2) = d(\omega', K_2)$, we get:

$$d(\omega_1, K_2) > d(\omega', \omega_1) + d(\omega', \omega_2) \quad (*).$$

By definition of $d$, we have $\forall \omega \models K_2, d(\omega_1, K_2) \leq d(\omega_1, \omega)$; in particular, $d(\omega_1, K_2) \leq d(\omega_1, \omega_2)$. By transitivity of $\leq$, and with (*), we get $d(\omega_1, \omega_2) > d(\omega', \omega_1) + d(\omega', \omega_2)$. This contradicts the triangular inequality. $\square$

Let us now prove the main theorem:

<u>*Weak drastic index*</u>. For two bases $K_1$ and $K_2$, we always have $i_{dw}(K_1, K_1 \triangle K_2) = 1$, because $\Delta_\top^{d,\Sigma}(\{K_1, K_2\}) \wedge K_1$ is consistent (Lemma 1), so no manipulation is possible ($i_{dw}$ is maximal).

<u>*Strong drastic index*</u>. If $\Delta_\top^{d,\Sigma}$ is not strategy-proof, then we can find $K_1'$ s.t.:

$$i_{ds}(K_1, \Delta_\top^{d,\Sigma}(\{K_1, K_2\}) < i_{ds}(K_1, \Delta_\top^{d,\Sigma}(\{K_1', K_2\}).$$

For the strong drastic index, this means exactly that:

$$\Delta_\top^{d,\Sigma}(\{K_1, K_2\}) \not\models K_1 \tag{4}$$

and:

$$\Delta_\top^{d,\Sigma}(\{K_1', K_2\}) \models K_1. \tag{5}$$

Since $\Delta_\top^{d,\Sigma}(\{K_1', K_2\}) \wedge K_2$ is consistent (Lemma 1), we can find $\omega_2 \models K_2$ s.t. $\omega_2 \models \Delta_\top^{d,\Sigma}(\{K_1', K_2\})$. With (5), we can conclude that $\omega_2 \models K_1$ as well.

Since we have $\omega_2 \models K_1 \wedge K_2$, we can conclude that for every model $\omega$ of $\Delta_\top^{d,\Sigma}(\{K_1, K_2\})$, we have $d(\omega, \{K_1, K_2\}) = 0$. So $\forall \omega \models \Delta_\top^{d,\Sigma}(\{K_1, K_2\}), d(\omega, K_1) = d(\omega, K_2) = 0$. Hence $\forall \omega \models \Delta_\top^{d,\Sigma}(\{K_1, K_2\}), \omega \models K_1 \wedge K_2$. This contradicts (4), so no manipulation is possible.

- For any distance $d$, $\Delta_\mu^{d,\Sigma}$ is strategy-proof for the indexes $i_p$, $i_{dw}$ and $i_{ds}$ when the initial base $K$ is complete.

  *For the drastic indexes, the result is a consequence of Theorem 11, showing that if a manipulation occurs with an initial base $K$, then a manipulation with a complete base $K_\omega \models K$ is possible. If $K$ is complete, no such manipulation not possible.*

  *For the probabilistic index, the result is a consequence of the triangular inequality.*

  <u>Drastic indexes</u>. The property is a direct consequence of Theorem 11, showing that if $\Delta_\mu^{d,\Sigma}$ is manipulable for $i_{dw}$ and $i_{ds}$ by a base $K$, then it is manipulable by erosion. But manipulation by erosion is impossible whenever $K$ is complete.

  <u>Probabilistic index</u>. By *reductio ad absurdum*: let us suppose that there is an operator $\Delta_\mu^{d,\Sigma}$, where $d$ is any distance, that is manipulable for $i_p$ given a complete base $K = \{\omega_1\}$. So, there exists an integrity constraint $\mu$, a profile $E$, and a base $K'$ s.t.:

  $$i_p(\{\omega_1\}, \Delta_\mu^{d,\Sigma}(\{\omega_1\} \sqcup E)) < i_p(\{\omega_1\}, \Delta_\mu^{d,\Sigma}(\{K'\} \sqcup E)).$$

  If $i_p(\{\omega_1\}, \Delta_\mu^{d,\Sigma}(\{\omega_1\} \sqcup E)) = 0$, then $i_{dw}(\{\omega_1\}, \Delta_\mu^{d,\Sigma}(\{\omega_1\} \sqcup E)) = 0$ too. In that case, manipulation for $i_p$ implies manipulation for $i_{dw}$ and we have seen that no manipulation is possible for $i_{dw}$. As a consequence, we can suppose that $i_p(\{\omega_1\}, \Delta_\mu^{d,\Sigma}(\{\omega_1\} \sqcup E)) \neq 0$. Equivalently:

  $$\frac{\#(\{\omega_1\} \cap [E \triangle_\mu^\Sigma \{\omega_1\}])}{\#([E \triangle_\mu^\Sigma \{\omega_1\}])} \neq 0$$

  (where $E \triangle_\mu^\Sigma \{\omega_1\}$ is a light notation for $\Delta_\mu^{d,\Sigma}(\{\omega_1\} \sqcup E)$).

  This statement allows us to infer that $\omega_1$ is a model of $E \triangle_\mu^\Sigma \{\omega_1\}$. In order to increase $i_p(\{\omega_1\}, \Delta_\mu^{d,\Sigma}(K' \sqcup E))$, we have to reduce the number of models of $E \triangle_\mu^\Sigma K'$ compared to $E \triangle_\mu^\Sigma \{\omega_1\}$, without removing $\omega_1$ from $[E \triangle_\mu^\Sigma K']$. So we have to find $\omega_2 \neq \omega_1$ s.t. $\omega_2 \models E \triangle_\mu^\Sigma \{\omega_1\}$ and $\omega_2 \not\models E \triangle_\mu^\Sigma K'$. So, $\omega_2 \models \mu$ and we have $d(\omega_2, E \sqcup \{\omega_1\}) = d(\omega_1, E \sqcup \{\omega_1\})$ and: $d(\omega_2, E \sqcup \{K'\}) > d(\omega_1, E \sqcup \{K'\})$ (because $\omega_1$ is a model of both $E \triangle_\mu^\Sigma \{\omega_1\}$ and $E \triangle_\mu^\Sigma K'$). With the aggregation function $\Sigma$, we get: $d(\omega_2, \omega_1) + d(\omega_2, E) = d(\omega_1, E)$ and $d(\omega_2, K') + d(\omega_2, E) > d(\omega_1, K') + d(\omega_1, E)$.

  Replacing $d(\omega_1, E)$ by $d(\omega_2, \omega_1) + d(\omega_2, E)$, we obtain $d(\omega_2, K') + d(\omega_2, E) > d(\omega_1, K') + d(\omega_2, \omega_1) + d(\omega_2, E)$, so $d(\omega_2, K') > d(\omega_1, K') + d(\omega_2, \omega_1)$. If $\omega_1'$ is a model of $K'$ s.t. $d(\omega_1, K') = d(\omega_1, \omega_1')$, then we have $d(\omega_2, K') > d(\omega_1, \omega_1') + d(\omega_2, \omega_1)$. Furthermore by definition of *min*, we have $d(\omega_2, \omega_1') \geq d(\omega_2, K')$, so $d(\omega_2, \omega_1') > d(\omega_1, \omega_1') + d(\omega_2, \omega_1)$ which contradicts the triangular inequality.

  $\square$

### Theorem 3

- $\Delta_\mu^{d_H,\Sigma}$ is strategy-proof for $i_{dw}$ or $i_{ds}$ if and only if ($\mu \equiv \top$ *and* $\#(E) = 2$) *or* $K$ is complete.

- $\Delta_\mu^{d_H,\Sigma}$ is strategy-proof for $i_p$ if and only if $K$ is complete.

**Proof:** Theorem 2 entails straightforwardly the $\Leftarrow$ part of the proof, taking the Hamming distance $d_H$ for $d$.

For the $\Rightarrow$ part of the proof, we shall show by examples of manipulation that $\Delta_\mu^{d_H,\Sigma}$ is not strategy-proof in other cases.

- The first examples show that $\Delta_\mu^{d_H,\Sigma}$ is not strategy-proof for $i_{dw}$ or $i_{ds}$ if ($\mu \not\equiv \top$ or $\#(E) \neq 2$), and if $K$ is not complete.

  *Weak drastic index.*

  - $i_{dw}$ and $\mu \not\equiv \top$ ($K$ is not complete)

    We consider the constraint $\mu = a \vee b$ and the two bases $K_1$ and $K_2$ defined by their set of models: $[K_1] = \{00, 01\}$ and $[K_2] = \{10\}$. We have $[\Delta_\mu^{d_H,\Sigma}(\{K_1, K_2\})] = \{10\}$ and $i_{dw}(K_1, \Delta_\mu^{d_H,\Sigma}(\{K_1, K_2\})) = 0$. On the other hand, if the agent whose base is $K_1$ gives $K_1'$, with $[K_1'] = \{01\}$ instead of $K_1$, we obtain $[\Delta_\mu^{d_H,\Sigma}(\{K_1', K_2\})] = \{01, 10, 11\}$ and $i_{dw}(K_1, \Delta_\mu^{d_H,\Sigma}(\{K_1', K_2\})) = 1$. This example shows the manipulability of $\Delta_\mu^{d_H,\Sigma}$ if $\mu \not\equiv \top$, even if there are only two bases in the profile. Computations are detailed in Table 5. Interpretations that do not satisfy the constraint are shaded.

| $\omega$ | $d_H(\omega, K_1)$ | $d_H(\omega, K_1')$ | $d_H(\omega, K_2)$ | $\Delta_\mu^{d_H,\Sigma}(\{K_1, K_2\})$ | $\Delta_\mu^{d_H,\Sigma}(\{K_1', K_2\})$ |
|---|---|---|---|---|---|
| 00 | 0 | 1 | 1 | 1 | 2 |
| 01 | 0 | 0 | 2 | 2 | **2** |
| 10 | 1 | 2 | 0 | **1** | **2** |
| 11 | 1 | 1 | 1 | 2 | **2** |

Table 5: Manipulability of $\Delta_\mu^{d_H,\Sigma}$ for $i_{dw}$ with $\mu \not\equiv \top$.

  - $i_{dw}$ and $\#(E) \neq 2$ ($K$ is not complete)

    Let us consider the three bases $[K_1] = \{00, 10\}$, $[K_2] = \{01, 10, 11\}$ and $[K_3] = \{01\}$. Then $\Delta_\top^{d_H,\Sigma}(\{K_1, K_2, K_3\})$ has a unique model 01 and $i_{dw}(K_1, \Delta_\top^{d_H,\Sigma}(\{K_1, K_2, K_3\})) = 0$. If we consider now $[K_1'] = \{10\}$ instead of $K_1$, then $[\Delta_\top^{d_H,\Sigma}(\{K_1', K_2, K_3\})] = \{01, 10, 11\}$ and $i_{dw}(K_1, \Delta^{d_H,\Sigma}(\{K_1', K_2, K_3\})) = 1$. See Table 6.

| $\omega$ | $K_1$ | $K_1'$ | $K_2$ | $K_3$ | $\Delta_\top^{d_H,\Sigma}(\{K_1, K_2, K_3\})$ | $\Delta_\top^{d_H,\Sigma}(\{K_1', K_2, K_3\})$ |
|---|---|---|---|---|---|---|
| 00 | 0 | 1 | 1 | 1 | 2 | 3 |
| 01 | 1 | 2 | 0 | 0 | **1** | **2** |
| 10 | 0 | 0 | 0 | 2 | 2 | **2** |
| 11 | 1 | 1 | 0 | 1 | 2 | **2** |

Table 6: Manipulability of $\Delta_\top^{d_H,\Sigma}$ for $i_{dw}$ with $\#(E) \neq 2$.

  *Strong drastic index.*

78

– $i_{ds}$ and $\mu \not\equiv \top$ ($K$ is not complete)

We consider the constraint $\mu = (a \wedge b) \vee (a \wedge \neg b \wedge \neg c)$ and the two bases $K_1$ and $K_2$ defined by their sets of models: $[K_1] = \{000, 111\}$ and $[K_2] = \{000, 001\}$. We have $[\Delta_\mu^{d_H, \Sigma}(\{K_1, K_2\})] = \{111, 100\}$ and $i_{ds}(K_1, \Delta_\mu^{d_H, \Sigma}(\{K_1, K_2\})) = 0$. On the other hand, if the agent whose base is $K_1$ gives $K_1'$, with $[K_1'] = \{111\}$ instead of $K_1$, we obtain $[\Delta_\mu^{d_H, \Sigma}(\{K_1', K_2\})] = \{111\}$ and $i_{ds}(K_1, \Delta_\mu^{d_H, \Sigma}(\{K_1', K_2\})) = 1$. This example shows the manipulability of $\Delta_\mu^{d_H, \Sigma}$ for $i_{ds}$ if $\mu \not\equiv \top$, even if there are only two bases in the profile. Details of the computation are reported in Table 7.

| $\omega$ | $K_1$ | $K_1'$ | $K_2$ | $\Delta_\mu^{d_H, \Sigma}(\{K_1, K_2\})$ | $\Delta_\mu^{d_H, \Sigma}(\{K_1', K_2\})$ |
|---|---|---|---|---|---|
| 000 | 0 | 3 | 0 | 0 | 3 |
| 001 | 1 | 2 | 0 | 1 | 2 |
| 010 | 1 | 2 | 1 | 2 | 3 |
| 011 | 1 | 1 | 1 | 2 | 2 |
| 100 | 1 | 2 | 1 | **2** | 3 |
| 101 | 1 | 1 | 1 | 2 | 2 |
| 110 | 1 | 1 | 2 | 3 | 3 |
| 111 | 0 | 0 | 2 | **2** | **2** |

Table 7: Manipulability of $\Delta_\mu^{d_H, \Sigma}$ for $i_{ds}$ with $\mu \not\equiv \top$.

– $i_{ds}$ and $\#(E) \neq 2$ ($K$ is not complete)

Let us consider the three bases $[K_1] = \{000, 001, 111\}$, $[K_2] = \{110, 001\}$ and $[K_3] = \{110, 000\}$. Then $[\Delta_\top^{d_H, \Sigma}(\{K_1, K_2, K_3\})] = \{000, 001, 110\}$ and $i_{ds}(K_1, \Delta_\top^{d_H, \Sigma}(\{K_1, K_2, K_3\})) = 0$.

If we consider $[K_1'] = \{000, 001\}$ instead of $K_1$, then $[\Delta_\top^{d_H, \Sigma}(\{K_1, K_2, K_3\})] = \{000, 001\}$ and $i_{ds}(K_1, \Delta_\top^{d_H, \Sigma}(\{K_1', K_2, K_3\})) = 1$. See Table 8.

| $\omega$ | $K_1$ | $K_1'$ | $K_2$ | $K_3$ | $\Delta_\top^{d_H, \Sigma}(\{K_1, K_2, K_3\})$ | $\Delta_\top^{d_H, \Sigma}(\{K_1', K_2, K_3\})$ |
|---|---|---|---|---|---|---|
| 000 | 0 | 0 | 1 | 0 | **1** | **1** |
| 001 | 0 | 0 | 0 | 1 | **1** | **1** |
| 010 | 1 | 1 | 1 | 1 | 3 | 3 |
| 011 | 1 | 1 | 1 | 2 | 4 | 4 |
| 100 | 1 | 1 | 1 | 1 | 3 | 3 |
| 101 | 1 | 1 | 1 | 2 | 4 | 4 |
| 110 | 1 | 2 | 0 | 0 | **1** | 2 |
| 111 | 0 | 2 | 1 | 1 | 2 | 4 |

Table 8: Manipulability of $\Delta^{d_H, \Sigma}$ for $i_{ds}$ with $\#(E) \neq 2$.

• The following example shows that $\Delta_\mu^{d_H, \Sigma}$ is not strategy-proof for $i_p$ if $K$ is not complete. Table 9 shows the manipulability of $\Delta_\top^{d_H, \Sigma}$ for $i_p$ (even if there are only two bases in the profile and if $\mu \equiv \top$). Let us consider the two bases $K_1$ and $K_2$ defined by their sets of models: $[K_1] = \{000, 001, 010, 100\}$ and $[K_2] = \{110, 011, 101, 111\}$. We have $[\Delta_\top^{d_H,\Sigma}(\{K_1, K_2\})] = \{001, 010, 100, 110, 011, 101\}$ and $i_p(K_1, \Delta_\top^{d_H,\Sigma}(\{K_1, K_2\})) = \frac{1}{2}$. On the other hand, if the agent whose base is $K_1$ gives $K_1'$, with $[K_1'] = \{000\}$ instead of $K_1$, we obtain $[\Delta_\top^{d_H,\Sigma}(\{K_1', K_2\})] = \{000, 001, 010, 100, 110, 011, 101\}$ and $i_p(K_1, \Delta_\top^{d_H,\Sigma}(\{K_1', K_2\})) = \frac{4}{7}$.

| $\omega$ | $K_1$ | $K_1'$ | $K_2$ | $\Delta_\top^{d_H,\Sigma}(\{K_1, K_2\})$ | $\Delta_\top^{d_H,\Sigma}(\{K_1', K_2\})$ |
|---|---|---|---|---|---|
| 000 | 0 | 0 | 2 | 2 | **2** |
| 001 | 0 | 1 | 1 | **1** | **2** |
| 010 | 0 | 1 | 1 | **1** | **2** |
| 011 | 1 | 2 | 0 | **1** | **2** |
| 100 | 0 | 1 | 1 | **1** | **2** |
| 101 | 1 | 2 | 0 | **1** | **2** |
| 110 | 1 | 2 | 0 | **1** | **2** |
| 111 | 2 | 3 | 0 | 2 | 3 |

Table 9: Manipulability of $\Delta_\top^{d_H,\Sigma}$ for $i_p$ if $K$ is not complete.

$\square$

**Theorem 4**

- $\Delta_\mu^{d_H,GMax}$ is not strategy-proof for the satisfaction indexes $i_{dw}$ and $i_p$ (even if $\mu \equiv \top$, $K$ is complete and $\#(E) = 2$).

- $\Delta_\mu^{d_H,GMax}$ is strategy-proof for the satisfaction index $i_{ds}$ if and only if $\mu \equiv \top$, $K$ is complete and $\#(E) = 2$.

**Proof:**

- Table 10 shows the manipulability of $\Delta^{d_H,GMax}$ for the weak satisfaction index $i_{dw}$ even if $\mu \equiv \top$, $K$ is complete and $\#(E) = 2$. We consider $K_1$ s.t. $[K_1] = \{001\}$, $K_2$ with $[K_2] = \{111\}$, and $\mu \equiv \top$. We have $[\Delta_\mu^{d_H,GMax}(\{K_1, K_2\})] = \{011, 101\}$, so no model of $K_1$ belongs to $[\Delta_\mu^{d_H,GMax}(\{K_1, K_2\})]$ and $i_{dw}(K_1, \Delta_\mu^{d_H,GMax}(\{K_1, K_2\})) = 0$. If agent 1 gives $K_1'$ with $[K_1'] = \{000\}$ instead of $K_1$, then $[\Delta_\mu^{d_H,GMax}\{K_1', K_2\})] = \{001, 010, 011, 100, 101, 110\}$ and $i_{dw}(K_1, \Delta_\mu^{d_H,GMax}(\{K_1', K_2\})) = 1$.

| $\omega$ | $K_1$ | $K_1'$ | $K_2$ | $\Delta_\mu^{d_H,GMax}(\{K_1, K_2\})$ | $\Delta_\mu^{d_H,GMax}(\{K_1', K_2\})$ |
|---|---|---|---|---|---|
| 000 | 1 | 0 | 3 | (3, 1) | (3, 0) |
| 001 | 0 | 1 | 2 | (2, 0) | **(2, 1)** |
| 010 | 2 | 1 | 2 | (2, 2) | **(2, 1)** |
| 011 | 1 | 2 | 1 | **(1, 1)** | **(2, 1)** |
| 100 | 2 | 1 | 2 | (2, 1) | **(2, 1)** |
| 101 | 1 | 2 | 1 | **(1, 1)** | **(2, 1)** |
| 110 | 3 | 2 | 1 | (3, 1) | **(2, 1)** |
| 111 | 2 | 3 | 0 | (2, 0) | (3, 0) |

Table 10: Manipulability of $\Delta_\mu^{d_H,GMax}$ for $i_{dw}$.

Since manipulability for $i_{dw}$ holds, manipulability for $i_p$ holds as well (cf. Theorem 1).

- As to $i_{ds}$, we first show that $\Delta_\mu^{d_H,GMax}$ is strategy-proof for this index if $\mu \equiv \top$, $\#(E) = 2$ and $K$ is complete. Then, we give examples of manipulation if $\mu \not\equiv \top$, or $\#(E) \neq 2$, or $K$ is not complete.

  – $\Delta_\top^{d_H,GMax}$ is strategy-proof when $E = \{K_1, K_2\}$ and $\mu \equiv \top$, if $K_1$ is complete. We consider $E' = \{K_1', K_2\}$ with $K_1' = K_{\omega_1}'$ complete (thanks to the forthcoming Lemma 2, we know that if the operator is manipulable, it is manipulable for a complete base), and $\mu \equiv \top$. Let $\#(\mathcal{P}) = n$ and let $d(K_1', K_2) = m \leq n$. Then there exists a model $\omega_2$ of $K_2$ s.t. $d_H(K_{\omega_1}', \omega_2) = m$. By definition of the Hamming distance, $\omega_2$ can be generated from $\omega_1$ by flipping $m$ variables (since $K_{\omega_1}'$ and $\omega_2$ differ on the truth values of $m$ variables $x_1, \ldots, x_m$).

    If $m = 2k + 1$ ($m$ odd), then $d(\top, E') = (k+1, k)$; otherwise $m = 2k$ ($m$ even) and $d(\top, E') = (k, k)$. In the first case ($m$ odd), there exist at least two interpretations $\omega$ and $\omega'$ s.t. $d(\omega, E') = d(\omega', E') = d(\top, E')$ (for instance, $\omega$ is generated from $\omega_1$ by flipping $x_1, \ldots, x_k$ and $\omega'$ is generated from $\omega_2$ by flipping $x_{k+1}, \ldots, x_m$).

    A similar conclusion can be derived in the second case ($m$ even) as soon as $k \geq 1$.

    In these two cases, $\Delta_\top^{d_H,GMax}(E')$ has at least two models, hence we cannot have $\Delta_\top^{d_H,GMax}(E') \equiv K_1$ with $K_1$ complete: $E$ cannot be manipulated by $K_1$ for $i_{ds}$. The remaining case is $d(\top, E') = (0,0)$. It imposes that $K_{\omega_1}' \wedge K_2$ is consistent. Since $K_{\omega_1}'$ is complete, we have $\Delta_\top^{d_H,GMax}(E') \equiv K_{\omega_1}'$, hence no manipulation is possible for $i_{ds}$ (since $\Delta_\top^{d_H,GMax}(E') \equiv K_1$ if and only if $K_1 \equiv K_{\omega_1}'$ if and only if $\Delta_\top^{d_H,GMax}(\{K_1, K_2\}) \equiv K_1$).

  – For showing the manipulability for $i_{ds}$, we consider the following scenarios:

    * $\mu \not\equiv \top$, even if $\#(E) = 2$ and $K$ is complete.

      Let us consider $[K_1] = \{01\}, [K_2] = \{11\}, \mu = \neg a \wedge b$. Then $[\Delta_\mu^{d_H,GMax}(\{K_1, K_2\})] = \{01, 11\}$, and $i_{ds}(K_1, \Delta_\mu^{d_H,GMax}(\{K_1, K_2\})) = 0$. If agent 1 gives $K_1'$ with $[K_1'] = \{00\}$ instead of $K_1$, then the result is $[\Delta_\mu^{d_H,GMax}(\{K_1', K_2\})] = \{01\}$ and $i_{ds}(K_1, \Delta_\mu^{d_H,GMax}(\{K_1', K_2\})) = 1$. (see Table 11).

| $\omega$ | $K_1$ | $K_1'$ | $K_2$ | $\Delta_\mu^{d_H,GMax}(\{K_1, K_2\})$ | $\Delta_\mu^{d_H,GMax}(\{K_1', K_2\})$ |
|---|---|---|---|---|---|
| 00 | 1 | 0 | 2 | $(2,1)$ | $(2,0)$ |
| 01 | 0 | 1 | 1 | $(\mathbf{1}, \mathbf{0})$ | $(\mathbf{1}, \mathbf{1})$ |
| 10 | 2 | 1 | 1 | $(2,1)$ | $(1,1)$ |
| 11 | 1 | 2 | 0 | $(\mathbf{1}, \mathbf{0})$ | $(2,0)$ |

Table 11: Manipulability of $\Delta_\mu^{d_H,GMax}$ for $i_{ds}$ if $\mu \not\equiv \top$

    * $\#(E) \neq 2$, even if $\mu \equiv \top$ and $K$ is complete.

      Let us consider $[K_1] = \{01\}, [K_2] = \{11\}$, and $[K_3] = \{00, 01, 11\}$. Then $[\Delta_\top^{d_H,GMax}(\{K_1, K_2, K_3\})] = \{01, 11\}$, so $i_{ds}(K_1, \Delta_\top^{d_H,GMax}(\{K_1, K_2, K_3\})) =$

0. If agent 1 gives $K_1'$ with $[K_1'] = \{00\}$ instead of $K_1$, then $[\Delta_\top^{d_H,GMax}(\{K_1', K_2, K_3\})] = \{01\}$ and $i_{ds}(K_1, \Delta_\top^{d_H,GMax}(\{K_1', K_2, K_3\}) = 1$. (see Table 12).

| $\omega$ | $K_1$ | $K_1'$ | $K_2$ | $K_3$ | $\Delta_\mu^{d_H,GMax}(\{K_1,K_2,K_3\})$ | $\Delta_\mu^{d_H,GMax}(\{K_1',K_2,K_3\})$ |
|---|---|---|---|---|---|---|
| 00 | 1 | 0 | 2 | 0 | $(2,1,0)$ | $(2,0,0)$ |
| 01 | 0 | 1 | 1 | 0 | $(\mathbf{1,0,0})$ | $(\mathbf{1,1,0})$ |
| 10 | 2 | 1 | 1 | 1 | $(2,1,1)$ | $(1,1,1)$ |
| 11 | 1 | 2 | 0 | 0 | $(\mathbf{1,0,0})$ | $(2,0,0)$ |

Table 12: Manipulability of $\Delta_\top^{d_H,GMax}$ for $i_{ds}$ if $\#(E) \neq 2$.

* $K$ is not complete, even if $\mu \equiv \top$ and $\#(E) = 2$.
  The example given Table 13 shows that manipulation is possible if the initial base is not complete. Consider $[K_1] = \{01,10\}, [K_2] = \{11\}$, and $\mu \equiv \top$. Then $[\Delta_\mu^{d_H,GMax}(\{K_1, K_2\})] = \{01,10,11\}$, and $i_{ds}(K_1, \Delta_\mu^{d_H,GMax}(\{K_1, K_2\})) = 0$. If agent 1 gives $K_1'$ with $[K_1'] = \{00\}$ instead of $K_1$, then $[\Delta_\mu^{d_H,GMax}(\{K_1', K_2\})] = \{01,10\}$ and $i_{ds}(K_1, \Delta_\mu^{d_H,GMax}(\{K_1', K_2\})) = 1$.

| $\omega$ | $K_1$ | $K_1'$ | $K_2$ | $\Delta_\mu^{d_H,GMax}(\{K_1,K_2\})$ | $\Delta_\mu^{d_H,GMax}(\{K_1',K_2\})$ |
|---|---|---|---|---|---|
| 00 | 1 | 0 | 2 | $(2,1)$ | $(2,0)$ |
| 01 | 0 | 1 | 1 | $(\mathbf{1,0})$ | $(\mathbf{1,1})$ |
| 10 | 0 | 1 | 1 | $(1,0)$ | $(1,1)$ |
| 11 | 1 | 2 | 0 | $(\mathbf{1,0})$ | $(2,0)$ |

Table 13: Manipulability of $\Delta_\mu^{d_H,GMax}$ for $i_{ds}$ if $K$ is not complete.

$\square$

**Theorem 5**

- $\triangle_\mu^{C1}, \triangle_\mu^{C3}, \triangle_\mu^{C4},$ and $\triangle_\mu^{C5}$ are not strategy-proof for $i_p$ (even if $\mu \equiv \top$, $K$ is complete and $\#(E) = 2$).

- $\triangle_\mu^{C1}$ is strategy-proof for $i_{dw}$ and $i_{ds}$.

- $\triangle_\mu^{C3}$ is strategy-proof for $i_{dw}$ and $i_{ds}$ if and only if $\mu \equiv \top$.

- $\triangle_\mu^{C4}$ is not strategy-proof for $i_{dw}$ and $i_{ds}$ (even if $\mu \equiv \top$, $K$ is complete and $\#(E) = 2$).

- $\triangle_\mu^{C5}$ is strategy-proof for $i_{dw}$ if and only if $\mu \equiv \top$ or $K$ is complete, and is strategy-proof for $i_{ds}$ if and only if $\mu \equiv \top$.

**Proof:**

- We first give an example of manipulation of $\triangle_\mu^{C1}$ for $i_p$, with $\#(E) = 2$, a complete base $K_1$, and $\mu \equiv \top$.

  Consider $E = \{K_1, K_2\}$, with $K_1 = \{a \wedge b\}$ and $K_2 = \{\neg(a \wedge b)\}$. Then $\triangle_\top^{C1}(E) \equiv \top$, and $i_p(K_1, \triangle_\top^{C1}(E)) = \frac{1}{4}$. But if agent 1 gives $K_1' = \{a, b\}$ instead of $K_1$, then $\triangle_\top^{C1}(\{K_1', K_2\}) \equiv a \vee b$, and $i_p(K_1, \triangle_\top^{C1}(\{K_1', K_2\})) = \frac{1}{3}$. So $E$ is manipulable by $K_1$ for $i_p$. The same example holds for $\triangle_\mu^{C4}$. It remains to note now that $\triangle_\top^{C1} = \triangle_\top^{C3} = \triangle_\top^{C5}$ to conclude the first point of the proof.

- $\triangle_\mu^{C1}$ is strategy-proof for $i_{dw}$ and $i_{ds}$.

  <u>Weak drastic index</u>.

  For any $K \in E$ there are two cases:

  - $K \wedge \mu$ is consistent. Then there is at least one maximal consistent subset $M$ of $\bigcup_{K_i \in E} K_i$ which contains $\mu$ and all the formulas of $K$. So $\triangle_\mu^{C1}(E \sqcup \{K\}) \equiv M \vee R$ (where $R$ denotes the disjunction of the other maxcons) is consistent with $K \wedge \mu$. So $i_{dw}(K, \triangle_\mu^{C1}(E \sqcup \{K\})) = 1$ and no manipulation is possible.

  - $K \wedge \mu$ is not consistent. Since for any $K'$, we have $\triangle_\mu^{C1}(E \sqcup \{K'\})) \models \mu$, we also have $i_{dw}(K, \triangle_\mu^{C1}(E \sqcup \{K'\})) = 0$ and no manipulation is possible.

  <u>Strong drastic index</u>.

  By *reductio ad absurdum*. Assume that $\triangle_\mu^{C1}$ is not strategy-proof for $i_{ds}$. It means that

  $$\exists K \text{ s.t. } \triangle_\mu^{C1}(E \sqcup \{K\}) \not\models K, \tag{6}$$

  $$\exists K' \text{ s.t. } \triangle_\mu^{C1}(E \sqcup \{K'\}) \models K. \tag{7}$$

  From statement (7) we get that $\forall M \in \text{MAXCONS}(E \sqcup \{K'\}, \mu)$, $M \models K$. So if we consider $\triangle_\mu^{C1}(E \sqcup \{K'\} \sqcup \{K\})$, every $M' \in \text{MAXCONS}(E \sqcup \{K'\} \sqcup \{K\}, \mu)$ is of the form $M \cup \{K\}$, so $M' \models K$ and $\triangle_\mu^{C1}(E \sqcup \{K'\} \sqcup \{K\}) \models K$ (∗).

  From statement (6) we get that $\exists M \in \text{MAXCONS}(E \sqcup \{K\}, \mu)$, $M \not\models K$. Since $M$ is a maximal subset, it means that $M \wedge K$ is not consistent. So if we consider $\triangle_\mu^{C1}(E \sqcup \{K'\} \sqcup \{K\})$, then $M \subseteq M'$, with $M' \in \text{MAXCONS}(E \sqcup \{K'\} \sqcup \{K\}, \mu)$. So $M' \wedge K$ is not consistent. Hence $M' \not\models K$ and $\triangle_\mu^{C1}(E \sqcup \{K\} \sqcup \{K'\}) \not\models K$, which contradicts (∗).

- $\triangle_\mu^{C3}$ is strategy-proof for $i_{dw}$ and $i_{ds}$ if and only if $\mu \equiv \top$.

  <u>Weak drastic index</u>.

  Since $\triangle_\top^{C1} = \triangle_\top^{C3}$, it follows immediately from the above proof for $\triangle_\mu^{C1}$ that $\triangle_\top^{C3}$ is strategy-proof for $i_{dw}$.

  For showing that manipulation is possible for $\triangle_\mu^{C3}$ if $\mu \not\equiv \top$, even with two bases and a complete initial base $K_1$, consider the following example: let $K_1 = \{a \wedge b\}$, $K_2 = \{\neg a\}$, $\mu = \neg b$ and $K_1' = \{a\}$. We have $\Delta_\mu^{C3}(\{K_1, K_2\}) \equiv \neg a$, which is inconsistent with $K_1$. We also have $\Delta_\mu^{C3}(\{K_1', K_2\}) \equiv \top$, which is consistent with $K_1$.

_Strong drastic index_.

Since $\triangle_{\top}^{C1} = \triangle_{\top}^{C3}$, it follows immediately from the point above that $\triangle_{\top}^{C3}$ is strategy-proof for $i_{ds}$.

For showing that a manipulation is possible for $\triangle_{\mu}^{C3}$ for $i_{ds}$ if $\mu \not\equiv \top$, even with two bases and a complete initial base $K_1$, we consider the following example: let $K_1 = \{a \wedge b\}$, $K_2 = \{\neg a\}$, $\mu = \neg a \wedge \neg b$ and $K_1' = \{\neg a \wedge b\}$. We have $\Delta_{\mu}^{C3}(\{K_1, K_2\}) \equiv \neg a$, hence $\Delta_{\mu}^{C3}(\{K_1, K_2\}) \not\models K_1$. We also have $\Delta_{\mu}^{C3}(\{K_1', K_2\}) \equiv \bot$, hence $\Delta_{\mu}^{C3}(\{K_1', K_2\}) \models K_1$.

- $\triangle_{\mu}^{C4}$ is not strategy-proof for $i_{dw}$ and $i_{ds}$ (even if $\mu \equiv \top$, $K$ is complete and $\#(E) = 2$).

  _Weak drastic index_.

  For showing that manipulation is possible for $\triangle_{\mu}^{C4}$ with two bases, a complete initial base $K_1$ and $\mu \equiv \top$, we consider the following example: let $K_1 = \{a\}$, $K_2 = \{\neg a, \neg a \wedge \top\}$, $\mu = \top$ and $K_1' = \{a, a \wedge \top\}$. We have $\Delta_{\mu}^{C4}(\{K_1, K_2\}) \equiv \neg a$, hence $\Delta_{\mu}^{C4}(\{K_1, K_2\}) \wedge K_1$ is not consistent. We also have $\Delta_{\mu}^{C4}(\{K_1', K_2\}) \equiv \top$, hence $\Delta_{\mu}^{C4}(\{K_1', K_2\}) \wedge K_1$ is consistent.

  _Strong drastic index_.

  For showing that $\triangle_{\mu}^{C4}$ is not strategy-proof for $i_{ds}$ with two bases, a complete initial base $K_1$ and $\mu \equiv \top$, consider the following example: let $K_1 = \{a\}$, $K_2 = \{\neg a\}$, and $K_1' = \{a, a \wedge \top\}$. We have $\Delta_{\mu}^{C4}(\{K_1, K_2\}) \equiv \top$, hence $\Delta_{\mu}^{C4}(\{K_1, K_2\}) \not\models K_1$. We also have $\Delta_{\mu}^{C4}(\{K_1', K_2\}) \equiv a$, hence $\Delta_{\mu}^{C4}(\{K_1', K_2\}) \models K_1$.

- $\triangle_{\mu}^{C5}$ is strategy-proof for $i_{dw}$ if and only if $\mu \equiv \top$ or $K$ is complete, and is strategy-proof for $i_{ds}$ if and only if $\mu \equiv \top$.

  _Weak drastic index_.

  Since $\triangle_{\top}^{C1} = \triangle_{\top}^{C5}$, it follows from the above proof for $\triangle_{\mu}^{C1}$ that $\triangle_{\mu}^{C5}$ is strategy-proof for $i_{dw}$ if $\mu \equiv \top$.

  If the initial base $K_1$ is complete, $\triangle_{\mu}^{C5}$ is also strategy-proof. There are two cases:

  - $K_1 \models \mu$. Let $M = \{\phi \in \bigcup_{K \in E} K \mid K_1 \models \phi\}$. By construction, $M$ is an element of MAXCONS($\bigcup_{K \in E} K, \top$). Since $K_1 \models \mu$, $M$ is consistent with $\mu$ ($K_1$ is a model of each of them). Since we have both $K_1 \models M$ and $M \models \Delta_{\mu}^{C5}(E)$ (by definition of the operator), we also have $K_1 \models \Delta_{\mu}^{C5}(E)$. Hence $\Delta_{\mu}^{C5}(E) \wedge K_1$ is consistent, and this prevents $E$ from being manipulable by $K_1$ for $i_{dw}$ given $\Delta_{\mu}^{C5}$ and $\mu$.

  - $K_1 \models \neg\mu$. By definition of the operator, for any base $K_1'$ and any profile $E'$ (especially the profile obtained by removing $K_1$ from $E$), we have $\Delta_{\mu}^{C5}(\{K_1'\} \sqcup E')$ is consistent and $\Delta_{\mu}^{C5}(\{K_1'\} \sqcup E') \models \mu$. This implies that $\Delta_{\mu}^{C5}(\{K_1'\} \sqcup E') \wedge K_1$ is inconsistent, and no manipulation is possible for $i_{dw}$.

  For showing that manipulation is possible for $\triangle_{\mu}^{C5}$ if $\mu \not\equiv \top$ and if the initial base $K_1$ is not complete, we consider the following example:

let $K_1 = \{a\}$, $K_2 = \{b, \neg a\}$, $\mu = \neg a \vee \neg b$ and $K_1' = \{a \wedge \neg b\}$. We have $\Delta_\mu^{C5}(\{K_1, K_2\}) \equiv b \wedge \neg a$, hence $\Delta_\mu^{C5}(\{K_1, K_2\}) \wedge K_1$ is not consistent. We also have $\Delta_\mu^{C5}(\{K_1', K_2\}) \equiv (a \wedge \neg b) \vee (b \wedge \neg a)$, hence $\Delta_\mu^{C4}(\{K_1', K_2\}) \wedge K_1$ is consistent.

*Strong drastic index.*

Since $\triangle_\top^{C1} = \triangle_\top^{C5}$, it follows immediately from the above proof for $\triangle_\mu^{C1}$ that $\triangle_\mu^{C5}$ is strategy-proof for $i_{ds}$ if $\mu \equiv \top$.

For showing that manipulation is possible for $\triangle_\mu^{C5}$ if $\mu \not\equiv \top$, even with two bases and a complete initial base $K_1$, we consider the following example: let $K_1 = \{a \wedge b\}$, $K_2 = \{\neg b\}$, $\mu = a$ and $K_1' = \{a \wedge b, b \vee \neg a\}$. We have $\Delta_\mu^{C5}(\{K_1, K_2\}) \equiv a$, hence $\Delta_\mu^{C5}(\{K_1, K_2\}) \not\models K_1$. We also have $\Delta_\mu^{C5}(\{K_1', K_2\}) \equiv a \wedge b$. Hence $\Delta_\mu^{C5}(\{K_1', K_2\}) \models K_1$.

$\square$

**Theorem 6**

- $\triangle_\mu^{\widehat{C1}}$ *is strategy-proof for $i_{dw}$ and $i_{ds}$, and is strategy-proof for $i_p$ if and only if $\#(E) = 2$.*

- $\triangle_\mu^{\widehat{C3}}$ *is strategy-proof for $i_{dw}$ and $i_{ds}$ if and only if $\mu \equiv \top$, and is strategy-proof for $i_p$ if and only if $\#(E) = 2$ and $\mu \equiv \top$.*

- $\triangle_\mu^{\widehat{C4}}$ *is strategy-proof for $i_p$, $i_{dw}$ and $i_{ds}$.*

- $\triangle_\mu^{\widehat{C5}}$ *is strategy-proof for $i_{dw}$ if and only if $\#(E) = 2$ or $\mu \equiv \top$ or $K$ is complete. $\triangle_\mu^{\widehat{C5}}$ is strategy-proof for $i_{ds}$ if and only if $\#(E) = 2$ or $\mu \equiv \top$. Finally, $\triangle_\mu^{\widehat{C5}}$ is strategy-proof for $i_p$ if and only if $\#(E) = 2$.*

**Proof:**

- $\triangle_\mu^{\widehat{C1}}$ is strategy-proof for $i_{dw}$ and $i_{ds}$, and is strategy-proof for $i_p$ if and only if $\#(E) = 2$.

  *Drastic indexes.* The strategy-proofness of $\triangle_\mu^{\widehat{C1}}$ comes from the strategy-proofness of $\triangle_\mu^{C1}$ as reported in Theorem 5, because $\triangle_\mu^{\widehat{C1}}$ is a specialization of $\triangle_\mu^{C1}$. Furthermore, in the examples given in the proof of Theorem 5 concerning $\triangle_\mu^{C1}$, every base is a singleton, so these examples hold for $\triangle_\mu^{\widehat{C1}}$ too.

  *Probabilistic index.*

  *The proof for the probabilistic index and a profile $E$ s.t. $\#(E) = 2$ is based on the fact that the merging of two bases with $\triangle_\mu^{\widehat{C1}}$ is either the conjunction of the two bases, or their disjunction. In both cases, we shall show that no manipulation can occur.*

  For the $\Leftarrow$ part of the proof, suppose that $\#(E) = 2$. we shall show that $\triangle_\mu^{\widehat{C1}}$ and $\triangle_\mu^{\widehat{C5}}$ (we group here the two cases, because their proofs are similar) is strategy-proof for $i_p$. By case analysis:

– If $K_1$ is consistent with $\mu$, then there are two cases:

* $\triangle_\mu^{\widehat{C1}}(\{K_1, K_2\}) \equiv \triangle_\mu^{\widehat{C5}}(\{K_1, K_2\}) \equiv K_1 \wedge K_2 \wedge \mu$ if it is consistent.

* $\triangle_\mu^{\widehat{C1}}(\{K_1, K_2\}) \equiv \triangle_\mu^{\widehat{C5}}(\{K_1, K_2\}) \equiv (K_1 \vee K_2) \wedge \mu$ otherwise.

In the first case, $\triangle_\mu^{\widehat{C1}}(\{K_1, K_2\}) \models K_1$ and $\triangle_\mu^{\widehat{C5}}(\{K_1, K_2\}) \models K_1$ so $i_p$ takes its maximal value, and no manipulation is possible.

Let us consider the second case for $\triangle_\mu^{\widehat{C1}}$ (the case for $\triangle_\mu^{\widehat{C5}}$ is similar): we assume that $\triangle_\mu^{\widehat{C1}}(\{K_1, K_2\}) \equiv (K_1 \vee K_2) \wedge \mu$.

Since for every base $K_1'$, the definition of $\triangle_\mu^{\widehat{C1}}$ requires that we have $\triangle_\mu^{\widehat{C1}}(\{K_1', K_2\}) \models \mu$ and since we have $K_1 \wedge \mu \models \triangle_\mu^{\widehat{C1}}(\{K_1, K_2\})$, the following inequation holds for every base $K_1'$:

$$\#([K_1] \cap [\triangle_\mu^{\widehat{C1}}(\{K_1', K_2\})]) \leq \#([K_1] \cap [\triangle_\mu^{\widehat{C1}}(\{K_1, K_2\})]).$$

If $K_1' \wedge K_2 \wedge \mu$ is consistent, then $\triangle_\mu^{\widehat{C1}}(\{K_1', K_2\}) \equiv K_1' \wedge K_2 \wedge \mu$, hence $\#([K_1] \cap [\triangle_\mu^{\widehat{C1}}(\{K_1', K_2\})]) = 0$ is minimal since we have assumed that $K_1 \wedge K_2 \wedge \mu$ is inconsistent.

If $K_1' \wedge K_2 \wedge \mu$ is inconsistent, then there are two cases:

* $K_1' \wedge \mu$ is inconsistent and $K_2 \wedge \mu$ is inconsistent. We have $\triangle_\mu^{\widehat{C1}}(\{K_1', K_2\}) \equiv \mu$. Since we have assumed $K_1 \wedge \mu$ consistent, we also have $\triangle_\mu^{\widehat{C1}}(\{K_1, K_2\}) \equiv K_1 \wedge \mu$. Hence:

$$i_p(K_1, \triangle_\mu^{\widehat{C1}}(\{K_1', K_2\})) = \frac{\#([K_1] \cap [\mu])}{\#([\mu])},$$

and

$$i_p(K_1, \triangle_\mu^{\widehat{C1}}(\{K_1, K_2\})) = \frac{\#([K_1] \cap [K_1 \wedge \mu])}{\#([K_1 \wedge \mu])}.$$

Since the numerators of the two fractions coincide while $\#([K_1 \wedge \mu]) \leq \#([\mu])$, no manipulation is possible in this case.

* $K_1' \wedge \mu$ is consistent or $K_2 \wedge \mu$ is consistent. We have $\triangle_\mu^{\widehat{C1}}(\{K_1', K_2\}) \equiv (K_1' \vee K_2) \wedge \mu$. Since $K_1 \wedge K_2 \wedge \mu$ is inconsistent, we have

$$i_p(K_1, \triangle_\mu^{\widehat{C1}}(\{K_1', K_2\})) = \frac{\#([K_1 \wedge K_1' \wedge \mu])}{\#([(K_1' \vee K_2) \wedge \mu])}.$$

We also have:

$$i_p(K_1, \triangle_\mu^{\widehat{C1}}(\{K_1, K_2\})) = \frac{\#([K_1 \wedge \mu])}{\#([(K_1 \vee K_2) \wedge \mu])}.$$

Now, since $K_1' \wedge K_2 \wedge \mu$ is inconsistent, we have $\#([(K_1' \vee K_2) \wedge \mu]) = \#([K_1' \wedge \mu]) + \#([K_2 \wedge \mu])$. Similarly, since $K_1 \wedge K_2 \wedge \mu$ is inconsistent, we have $\#([(K_1 \vee K_2) \wedge \mu]) = \#([K_1 \wedge \mu]) + \#([K_2 \wedge \mu])$. Subsequently, suppose that we have $i_p(K_1, \triangle_\mu^{\widehat{C1}}(\{K_1', K_2\})) > i_p(K_1, \triangle_\mu^{\widehat{C1}}(\{K_1, K_2\}))$. This is the case if and only if $\#([K_1 \wedge K_1' \wedge \mu])(\#([K_1 \wedge \mu]) + \#([K_2 \wedge \mu])) > \#([K_1 \wedge \mu])(\#([K_1' \wedge \mu]) + \#([K_2 \wedge \mu]))$.

If we note $a = \#([K_1 \wedge K_1' \wedge \mu])$ and $b = \#([K_2 \wedge \mu])$, then there exist two natural integers $a'$ and $a''$ such that $\#([K_1 \wedge \mu]) = a + a'$ and $\#([K_1' \wedge \mu]) = a + a''$. Replacing in the previous inequation, it comes:

$$a(a + a' + b) > (a + a')(a + a'' + b)$$

which simplifies to $0 > aa'' + a'a'' + a'c$, which is impossible. Hence no manipulation is possible in this case as well.

— If $K_1$ is not consistent with $\mu$, then, since $\forall E', \triangle_\mu^{\widehat{C1}}(E') \models \mu$ and $\triangle_\mu^{\widehat{C5}}(E') \models \mu$, we have: $\forall E', i_p(K_1, \triangle_\mu^{\widehat{C1}}(E')) = 0$ and $i_p(K_1, \triangle_\mu^{\widehat{C5}}(E')) = 0$, so no manipulation is possible.

For the $\Rightarrow$ part of the proof, the following example shows that strategy-proofness for $i_p$ does not hold any longer for $\triangle_\mu^{\widehat{C1}}$ or $\triangle_\mu^{\widehat{C5}}$ when $\#(E) \neq 2$, even if the initial base is complete and $\mu \equiv \top$. We consider $K_1 = \{a \wedge b\}, K_2 = \{a \wedge b\}$ and $K_3 = \{\neg a\}$, with the integrity constraint $\mu \equiv \top$. There are two maximal consistent sets in MAXCONS($\{K_1, K_2, K_3\}, \mu$): $M_1 = \{a \wedge b, \top\}$ and $M_2 = \{\neg a, \top\}$. Hence $\triangle_\mu^{\widehat{C1}}(\{K_1, K_2, K_3\}) \equiv (a \wedge b) \vee (\neg a)$. We get $i_p(K_1, \triangle_\mu^{\widehat{C1}}(\{K_1, K_2, K_3\})) = \frac{1}{3}$. If agent 1 gives $K_1' = \{\neg a \wedge b\}$ instead of $K_1$, then there are two maximal consistent sets $M_1' = \{a \wedge b, \top\}, M_2' = \{\neg a \wedge b, \neg a, \top\}$ in MAXCONS($\{K_1', K_2, K_3\}, \mu$), so $\triangle_\mu^{\widehat{C1}}(\{K_1', K_2, K_3\}) \equiv (a \wedge b) \vee (\neg a \wedge b)$. We get $i_p(K_1, \triangle_\mu^{\widehat{C1}}(\{K_1', K_2, K_3\})) = \frac{1}{2}$, so this is an example of manipulation of $\triangle_\top^{\widehat{C1}}$ for $i_p$.

Since $\triangle_\top^{\widehat{C1}} = \triangle_\top^{\widehat{C5}}$, this example shows the manipulability of $\triangle_\mu^{\widehat{C5}}$ as well.

- $\triangle_\mu^{\widehat{C3}}$ is strategy-proof for $i_{dw}$ and $i_{ds}$ if and only if $\mu \equiv \top$, and is strategy-proof for $i_p$ if and only if $\#(E) = 2$ and $\mu \equiv \top$.

*Drastic indexes.*

Since $\triangle_\top^{C3}$ is strategy-proof for $i_{dw}$ and $i_{ds}$, its specialization $\triangle_\mu^{\widehat{C3}}$ is also strategy-proof for $i_{dw}$ and $i_{ds}$.

In the example showing the manipulation for $\triangle_\mu^{C3}$ with two bases and a complete initial base $K_1$, every base is a singleton, so this example holds for $\triangle_\mu^{\widehat{C3}}$ too.

*Probabilistic index.*

Since $\triangle_\top^{C1} = \triangle_\top^{C3}$, it follows immediately from the above proof for $\triangle_\mu^{\widehat{C1}}$ that $\triangle_\top^{\widehat{C3}}$ is strategy-proof for $i_p$ when two bases are considered, and it is not the case when three agents or more are taken into account.

For showing that manipulation is possible for $\triangle_\mu^{\widehat{C3}}$ with two bases and a complete initial base $K_1$, when the integrity constraint $\mu$ is not true, consider the following example: let $K_1 = \{a \wedge b\}$, $K_2 = \{\neg a\}$, $\mu = \neg b$ and $K_1' = \{a\}$. We have $\triangle_\mu^{\widehat{C3}}(\{K_1, K_2\}) \equiv$ $\neg a$, which is inconsistent with $K_1$, so $i_p(K_1, \triangle_\mu^{\widehat{C3}}(\{K_1, K_2\})) = 0$. We also have $\triangle_\mu^{\widehat{C3}}(\{K_1', K_2\}) \equiv \top$, which is consistent with $K_1$, so $i_p(K_1, \triangle_\mu^{\widehat{C3}}(\{K_1', K_2\})) > 0$.

- $\triangle_\mu^{\widehat{C4}}$ is strategy-proof for $i_p$, $i_{dw}$ and $i_{ds}$. This is a direct consequence of Remark 1 and Theorem 2.

- $\triangle_\mu^{\widehat{C5}}$ is strategy-proof for $i_{dw}$ if and only if $\#(E) = 2$ or $\mu \equiv \top$ or $K$ is complete. $\triangle_\mu^{\widehat{C5}}$ is strategy-proof for $i_{ds}$ if and only if $\#(E) = 2$ or $\mu \equiv \top$. Finally, $\triangle_\mu^{\widehat{C5}}$ is strategy-proof for $i_p$ if and only if $\#(E) = 2$.

_Weak drastic index._

As $\triangle_\mu^{\widehat{C5}}$ is strategy-proof for $i_p$ with two bases, it is also strategy-proof for $i_{dw}$ in that case.

Since $\triangle_\top^{C5}$ is strategy-proof for $i_{dw}$, we have that $\triangle_\top^{\widehat{C5}}$ is also strategy-proof for $i_{dw}$. Finally, as no manipulation is possible for $\triangle_\mu^{C5}$ when the initial base $K_1$ is complete (Theorem 5), no manipulation exists for $\triangle_\mu^{\widehat{C5}}$ in this cas.

Contrastingly, a manipulation example exists for $\triangle_\mu^{\widehat{C5}}$ if $\#(E) \neq 2$ and $\mu \not\equiv \top$ and the initial base $K_1$ not complete. We consider the three (singleton) bases: $K_1 = \{b\}$, $K_2 = \{\neg a\}$, $K_3 = \{a \wedge \neg b\}$, and $\mu = a$. We have $\triangle_\mu^{\widehat{C5}}(\{K_1, K_2, K_3\}) \equiv a \wedge \neg b$, so $i_{dw}(K_1, \triangle_\mu^{\widehat{C5}}(\{K_1, K_2, K_3\})) = 0$. And with $K_1' = \{b \wedge a\}$, we have $\triangle_\mu^{\widehat{C5}}(\{K_1', K_2, K_3\}) \equiv a$, and $i_{dw}(K_1, \triangle_\mu^{\widehat{C5}}(\{K_1', K_2, K_3\})) = 1$.

_Strong drastic index._

As $\triangle_\top^{C5}$ is strategy-proof for $i_{ds}$, $\triangle_\top^{\widehat{C5}}$ is also strategy-proof for $i_{ds}$.

As $\triangle_\mu^{C5}$ is strategy-proof for $i_p$ when there are two bases in $E$, $\triangle_\mu^{\widehat{C5}}$ is also strategy-proof for $i_{ds}$ in this case, since for any profile $E$, $\triangle_\mu^{\widehat{C5}}(E)$ is consistent.

A manipulation example exists for $\triangle_\mu^{\widehat{C5}}$ if $\#(E) \neq 2$ and $\mu \not\equiv \top$, even if the initial base $K_1$ is complete: $K_1 = \{a \wedge b\}$, $K_2 = \{a \wedge b\}$, $K_3 = \{\neg a \vee \neg b\}$, and $\mu = a$. There are two maximal consistent sets in MAXCONS$(\{K_1, K_2, K_3\}, \top)$: $M_1 = \{a \wedge b\}$ and $M_2 = \{\neg a \vee \neg b\}$. Hence $\triangle_\mu^{\widehat{C5}}(\{K_1, K_2, K_3\}) \equiv (a \wedge b) \vee (a \wedge \neg b)$. We get $i_{ds}(K_1, \triangle_\mu^{\widehat{C5}}(\{K_1, K_2, K_3\})) = 0$. With $K_1' = \{\neg a \wedge \neg b\}$, there are two maximal consistent sets in MAXCONS$(\{K_1, K_2, K_3\}, \top)$: $M_1' = \{a \wedge b\}$ and $M_2' = \{\neg a \wedge \neg b\}$. Hence $\triangle_\mu^{\widehat{C5}}(\{K_1', K_2, K_3\}) \equiv (a \wedge b)$, and $i_{ds}(K_1, \triangle_\mu^{\widehat{C5}}(\{K_1', K_2, K_3\})) = 1$.

_Probabilistic index._

The proof for $\triangle_\mu^{\widehat{C5}}$ is similar to the proof for $\triangle_\mu^{\widehat{C1}}$.

$\square$

**Theorem 7** _The strategy-proofness results reported in Table 14 hold, under the restriction that each base is complete ($f$ stands for any aggregation function, and $d$ for any distance)._

*f is any aggregation function, d is any distance,* **sp** *means "strategy-proof",* $\overline{sp}$ *means "non strategy-proof" even if* $\#(E) = 2$ *and* $\mu \equiv \top$, $\overline{sp}^*$ *means "non strategy-proof" even if either* $\#(E) = 2$ *or* $\mu \equiv \top$, *but "strategy-proof" if both* $\#(E) = 2$ *and* $\mu \equiv \top$. *Finally,* $\overline{sp}^{\top}$ *means "non strategy-proof" even if* $\#(E) = 2$, *but "strategy-proof" whenever* $\mu \equiv \top$.

| $\Delta$ | $i_p$ | $i_{dw}$ | $i_{ds}$ |
|---|---|---|---|
| $\Delta_\mu^{d_D,f}$ | **sp** | **sp** | **sp** |
| $\Delta_\mu^{d,\Sigma}$ | **sp** | **sp** | **sp** |
| $\Delta_\mu^{d_H,GMax}$ | $\overline{sp}$ | $\overline{sp}$ | $\overline{sp}^*$ |
| $\triangle_\mu^{C1}$ | $\overline{sp}$ | **sp** | **sp** |
| $\triangle_\mu^{C3}$ | $\overline{sp}$ | $\overline{sp}^{\top}$ | $\overline{sp}^{\top}$ |
| $\triangle_\mu^{C4}$ | $\overline{sp}$ | $\overline{sp}$ | $\overline{sp}$ |
| $\triangle_\mu^{C5}$ | $\overline{sp}$ | **sp** | $\overline{sp}^{\top}$ |
| $\triangle_\mu^{C1}$ | **sp** | **sp** | **sp** |
| $\triangle_\mu^{C3}$ | **sp** | **sp** | **sp** |
| $\triangle_\mu^{C4}$ | **sp** | **sp** | **sp** |
| $\triangle_\mu^{\widehat{C5}}$ | **sp** | **sp** | **sp** |

Table 14: Strategy-proofness: complete bases.

**Proof:** The first line of the table ($\Delta_\mu^{d_D,f}$) is a direct consequence of Theorem 2.

The second line of the table ($\Delta_\mu^{d,\Sigma}$) is a direct consequence of Theorem 2.

The third line of the table ($\Delta_\mu^{d_H,GMax}$) comes from the proof of Theorem 4.

The first column of the fourth line ($\triangle_\mu^{C1}$ and $i_p$) comes from the following example. Let $K_1 = \{a, b\}$, $K_2 = \{\neg a, \neg b\}$, $K_1' = \{a \wedge b\}$ and $\mu = \top$. We have $\triangle_\mu^{C1}(\{K_1, K_2\}) \equiv \top$, hence $i_p(K_1, \triangle_\mu^{C1}(\{K_1, K_2\})) = \frac{1}{4}$, while we have $\triangle_\mu^{C1}(\{K_1', K_2\}) \equiv (a \wedge b) \vee (\neg a \wedge \neg b)$, showing that $i_p(K_1, \triangle_\mu^{C1}(\{K_1', K_2\})) = \frac{1}{2}$. The rightmost columns of the fourth line ($\triangle_\mu^{C1}$ and $i_{dw}$, $i_{ds}$) come directly from Theorem 5.

The first column of the fifth line ($\triangle_\mu^{C3}$ and $i_p$) comes from the first column of the fourth line given that the example is s.t. $\mu \equiv \top$ (in that case both operators coincide). Similarly for the second and the third columns ($\triangle_\mu^{C3}$ and $i_{dw}$, $i_{ds}$) in the case $\mu \equiv \top$ ($\triangle_\top^{C3}$ coincides with $\triangle_\top^{C1}$). In the remaining case, $\triangle_\mu^{C3}$ is not strategy-proof for $i_{dw}$ even if $\#(E) = 2$ as the following example shows: take $E = \{K_1, K_2\}$ with $K_1 = \{a \wedge b \wedge c\}$, $K_2 = \{\neg a \wedge \neg b, c\}$ and $\mu = \neg b$; we have $\triangle_\mu^{C3}(E) \equiv \neg a \wedge \neg b \wedge c$, which is inconsistent with $K_1$; if the agent gives $K_1' = \{a, b \wedge \neg c\}$ instead of $K_1$, we obtain $\triangle_\mu^{C3}(\{K_1', K_2\}) \equiv (a \wedge c) \vee (\neg a \wedge \neg b \wedge c)$, which is consistent with $K_1$. Finally, $\triangle_\mu^{C3}$ is not strategy-proof for $i_{ds}$ even if $\#(E) = 2$ when $\mu \neq \top$; let $K_1 = \{a, \neg b\}$, $K_2 = \{\neg a, b\}$, $K_1' = \{a \wedge \neg b\}$ and $\mu = \neg b$. We have $\triangle_\mu^{C3}(\{K_1, K_2\}) \equiv (a \wedge \neg b) \vee (\neg a \wedge \neg b)$, hence $i_{ds}(K_1, \triangle_\mu^{C3}(\{K_1, K_2\})) = 0$, while we have $\triangle_\mu^{C3}(\{K_1', K_2\}) \equiv a \wedge \neg b$, showing that $i_{ds}(K_1, \triangle_\mu^{C3}(\{K_1', K_2\})) = 1$.

The sixth line ($\triangle_\mu^{C4}$) comes from the proof of Theorem 5.

The first column of the seventh line ($\triangle_\mu^{C5}$ and $i_p$) comes from the first column of the fourth line given that the example is s.t. $\mu \equiv \top$ (in that case both operators coincide). The second column ($\triangle_\mu^{C5}$ and $i_{dw}$) comes directly from Theorem 5. The third column ($\triangle_\mu^{C5}$ and $i_{ds}$) in the case $\mu \equiv \top$ comes from the third column of the fourth line ($\triangle_\top^{C5}$ coincides with $\triangle_\top^{C1}$). Finally, $\triangle_\mu^{C5}$ is not strategy-proof for $i_{ds}$ even if $\#(E) = 2$ when $\mu \not\equiv \top$; let $K_1 = \{a, \neg b\}$, $K_2 = \{\neg a, b\}$, $K_1' = \{a \wedge \neg b\}$ and $\mu = \neg b$. We have $\triangle_\mu^{C5}(\{K_1, K_2\}) \equiv (a \wedge \neg b) \vee (\neg a \wedge \neg b)$, hence $i_{ds}(K_1, \triangle_\mu^{C5}(\{K_1, K_2\})) = 0$, while we have $\triangle_\mu^{C5}(\{K_1', K_2\}) \equiv a \wedge \neg b$, showing that $i_{ds}(K_1, \triangle_\mu^{C5}(\{K_1', K_2\})) = 1$.

Finally, it remains to consider the $\triangle_\mu^{\widehat{C}}$ operators. As to $\triangle_\mu^{\widehat{C4}}$, we know from Theorem 6 that it is strategy-proof for $i_p$ (hence for $i_{dw}$ and $i_{ds}$) since $\triangle_\mu^{\widehat{C4}}$ is strategy-proof for $i_p$ when all bases are singletons. Let us focus on $\triangle_\mu^{\widehat{C1}}$, $\triangle_\mu^{\widehat{C3}}$ and $\triangle_\mu^{\widehat{C5}}$. Since each base is complete and can be assumed to be a singleton without loss of generality, we have

$$\triangle_\mu^{\widehat{C1}}(\{K_1, \ldots, K_n\}) \equiv \triangle_\mu^{\widehat{C5}}(\{K_1, \ldots, K_n\}) \equiv (\bigvee_{i=1}^n K_i) \wedge \mu \text{ if consistent,}$$

$$\triangle_\mu^{\widehat{C1}}(\{K_1, \ldots, K_n\}) \equiv \triangle_\mu^{\widehat{C5}}(\{K_1, \ldots, K_n\}) \equiv \mu \text{ otherwise.}$$

We also have:

$$\triangle_\mu^{\widehat{C3}}(\{K_1, \ldots, K_n\}) \equiv (\bigvee_{i=1}^n K_i) \wedge \mu \text{ if consistent,}$$

$$\triangle_\mu^{\widehat{C3}}(\{K_1, \ldots, K_n\}) \equiv \bot \text{ otherwise.}$$

Let $\triangle_\mu^{\widehat{C}}$ be any operator among $\triangle_\mu^{\widehat{C1}}$, $\triangle_\mu^{\widehat{C3}}$ and $\triangle_\mu^{\widehat{C5}}$. There are two cases:

- $\triangle_\mu^{\widehat{C}}$ consistent. There are again two cases:

    - $K_1 \models \neg\mu$. Since for every profile $E'$ we have $\triangle_\mu^{\widehat{C1}}(E') \models \mu$ and $\triangle_\mu^{\widehat{C5}}(E') \models \mu$, we also have $\triangle_\mu^{\widehat{C1}}(E') \wedge K_1$ inconsistent and $\triangle_\mu^{\widehat{C5}}(E') \wedge K_1$ inconsistent, showing that no manipulation is possible for $i_p$, hence for $i_{dw}$ and $i_{ds}$. In the specific case we consider (all bases are singletons and are complete), we also have for every profile $E'$, $\triangle_\mu^{\widehat{C3}}(E') \models \mu$ since each base from $E'$ that is kept as an element from MAXCONS$(E', \top)$ must satisfy $\mu$ and if no base is kept, $\triangle_\mu^{\widehat{C3}}(E')$ is inconsistent, hence $\triangle_\mu^{\widehat{C3}}(E') \models \mu$ trivially holds. The previous argument can be used to show that no manipulation is possible with $\triangle_\mu^{\widehat{C3}}$ for $i_p$, hence for $i_{dw}$ (and $i_{ds}$ under the assumption $\triangle_\mu^{\widehat{C3}}(E)$ consistent).

    - $K_1 \models \mu$. We necessarily have $\#([K_1] \cap [\triangle_\mu^{\widehat{C}}(\{K_1, \ldots, K_n\})]) = 1$. By *reductio ad absurdum*. A manipulation for $i_p$ is possible only if we can find a complete base $K_1'$ s.t. (1) $K_1 \models \triangle_\mu^{\widehat{C}}(\{K_1', \ldots, K_n\})$ and (2) $\#([\triangle_\mu^{\widehat{C}}(\{K_1', \ldots, K_n\})]) < \#([\triangle_\mu^{\widehat{C}}(\{K_1, \ldots, K_n\})])$. (2) requires that $K_1' \models \neg\mu$. (1) imposes that $K_1 \models K_1' \vee K_2 \vee \ldots \vee K_n$. Since $K_1 \models \mu$ while $K_1' \models \neg\mu$, we have $K_1 \not\equiv K_1'$. Subsequently, there exists $K_j$ ($j \in 2, \ldots, n$) s.t. $K_1 \equiv K_j$. Since $K_j$ is a model of

$\triangle_\mu^{\widehat{C}}(\{K_1', \ldots, K_n\})$, inequation (2) cannot be satisfied. Hence, $\triangle_\mu^{\widehat{C}}$ is strategy-proof for $i_{dw}$, hence for $i_{dw}$. Since $\triangle_\mu^{\widehat{C}}(E)$ is assumed consistent, no manipulation is possible for $i_{ds}$.

- $\triangle_\mu^{\widehat{C}}(E)$ inconsistent. This is only possible for $\triangle_\mu^{\widehat{C}} = \triangle_\mu^{\widehat{C3}}$ and requires that each $K_i$ ($i \in 1, \ldots, n$) is s.t. $K_i \models \neg\mu$. Since $\triangle_\mu^{\widehat{C3}}(E)$ is inconsistent, we have $i_p(K_1, \triangle_\mu^{\widehat{C3}}(E)) = 0$. Since for every $K_1'$ complete, $K_1$ is not a model of $\triangle_\mu^{\widehat{C3}}(\{K_1', \ldots, K_n\})$, we have $i_p(K_1, \triangle_\mu^{\widehat{C3}}(\triangle_\mu^{C3}(\{K_1', \ldots, K_n\}))) = 0$ as well. This shows that $\triangle_\mu^{\widehat{C3}}$ is strategy-proof for $i_p$, hence for $i_{dw}$. Finally, when $\triangle_\mu^{\widehat{C3}}(E)$ is inconsistent, we have $\triangle_\mu^{\widehat{C3}}(E) \models K_1$, showing that no manipulation is possible for $i_{ds}$ as well.

□

**Theorem 8** *None of $\triangle_\mu^{d_D, \Sigma}$, $\triangle_\mu^{d_D, G_{Max}}$, $\triangle_\mu^{d_H, \Sigma}$ or $\triangle_\mu^{d_H, G_{Max}}$ is strategy-proof for $i_{Dalal}$, even in the restricted case $E$ consists of two complete bases.*

**Proof:**

- $\triangle_\mu^{d_D, \Sigma} = \triangle_\mu^{d_D, G_{Max}}$. Let us consider $[K_1] = \{000\}$, $[K_2] = \{110\}$ and $\mu = a \wedge b \wedge c$ where $\mathcal{P} = \{a, b, c\}$. We have $[\triangle_\mu^{d_D, \Sigma}(\{K_1, K_2\})] = \{110\}$ and $i_{Dalal}(K_1, \triangle_\mu^{d_D, \Sigma}(\{K_1, K_2\})) = 1 - \frac{2}{3}$. With $[K_1'] = \{001\}$, we get $[\triangle_\mu^{d_D, \Sigma}(\{K_1', K_2\})] = \{110, 001\}$ and $i_{Dalal}(K_1, \triangle_\mu^{d_D, \Sigma}(\{K_1', K_2\})) = 1 - \frac{1}{3}$, showing the manipulation (details are reported in Table 15).

| $\omega$ | $K_1$ | $K_1'$ | $K_2$ | $\triangle_\mu^{d_D, \Sigma}(\{K_1, K_2\})$ | $\triangle_\mu^{d_D, \Sigma}(\{K_1', K_2\})$ |
|---|---|---|---|---|---|
| 000 | 0 | 1 | 1 | 2 | 2 |
| 001 | 1 | 0 | 1 | 2 | **1** |
| 010 | 1 | 1 | 1 | 2 | 2 |
| 011 | 1 | 1 | 1 | 2 | 2 |
| 100 | 1 | 1 | 1 | 2 | 2 |
| 101 | 1 | 1 | 1 | 2 | 2 |
| 110 | 1 | 1 | 0 | **1** | **1** |
| 111 | 1 | 1 | 1 | 2 | 2 |

Table 15: Manipulation of $\triangle_\mu^{d_D, \Sigma}$ for $i_{Dalal}$ with two complete bases.

- $\triangle_\mu^{d_H, \Sigma}$. Let us consider $[K_1] = \{000\}$, $[K_2] = \{110\}$ and $\mu = \neg((\neg a \wedge \neg b \wedge \neg c) \vee (\neg a \wedge b \wedge \neg c) \vee (a \wedge \neg b \wedge \neg c))$ where $\mathcal{P} = \{a, b, c\}$. We have $[\triangle_\mu^{d_H, \Sigma}(\{K_1, K_2\})] = \{110\}$ and $i_{Dalal}(K_1, \triangle_\mu^{d_H, \Sigma}(\{K_1, K_2\})) = 1 - \frac{2}{3}$. With $[K_1'] = \{001\}$, we get $[\triangle_\mu^{d_H, \Sigma}(\{K_1', K_2\})] = \{110, 001, 011, 111\}$ and $i_{Dalal}(K_1, \triangle_\mu^{d_H, \Sigma}(\{K_1', K_2\})) = 1 - \frac{1}{3}$, showing the manipulation (details are reported in Table 16).

- $\triangle_\mu^{d_H, G_{Max}}$. Let us consider $[K_1] = \{0001\}$, $[K_2] = \{0111\}$ and $\mu = (\neg a \wedge \neg b \wedge \neg c \wedge \neg d) \vee (\neg a \wedge b \wedge c) \vee (a \wedge \neg b \wedge \neg c) \vee (a \wedge \neg b \wedge c \wedge \neg d) \vee (a \wedge b \wedge \neg c \wedge \neg d) \vee (a \wedge b \wedge c \wedge d)$ where $\mathcal{P} = \{a, b, c, d\}$. We have $[\triangle_\mu^{d_H, G_{Max}}(\{K_1, K_2\})] = \{0111\}$ and

| $\omega$ | $K_1$ | $K_1'$ | $K_2$ | $\Delta_\mu^{d_H,\Sigma}(\{K_1,K_2\})$ | $\Delta_\mu^{d_H,\Sigma}(\{K_1',K_2\})$ |
|---|---|---|---|---|---|
| 000 | 0 | 1 | 2 | 2 | 3 |
| 001 | 1 | 0 | 3 | 4 | **3** |
| 010 | 1 | 2 | 1 | 2 | 3 |
| 011 | 2 | 1 | 2 | 4 | **3** |
| 100 | 1 | 2 | 1 | 2 | 3 |
| 101 | 2 | 1 | 2 | 4 | **3** |
| 110 | 2 | 3 | 0 | **2** | **3** |
| 111 | 3 | 2 | 1 | 4 | **3** |

Table 16: Manipulation of $\Delta_\mu^{d_H,\Sigma}$ for $i_{Dalal}$ with two complete bases.

$i_{Dalal}(K_1, \Delta_\mu^{d_H,G_{Max}}(\{K_1,K_2\})) = 1 - \frac{2}{4}$. With $[K_1'] = \{1000\}$, we get $[\Delta_\mu^{d_H,G_{Max}}(\{K_1', K_2\})] = \{0000,0110,1001,1010,1100,1111\}$ and $i_{Dalal}(K_1, \Delta_\mu^{d_H,G_{Max}}(\{K_1',K_2\})) = 1 - \frac{1}{4}$, showing the manipulation (details are reported in Table 17).

| $\omega$ | $K_1$ | $K_1'$ | $K_2$ | $\Delta_\mu^{d_H,G_{Max}}(\{K_1,K_2\})$ | $\Delta_\mu^{d_H,G_{Max}}(\{K_1',K_2\})$ |
|---|---|---|---|---|---|
| 0000 | 1 | 1 | 3 | (3,1) | (**3,1**) |
| 0001 | 0 | 2 | 2 | (2,0) | (2,2) |
| 0010 | 2 | 2 | 2 | (2,2) | (2,2) |
| 0011 | 1 | 3 | 1 | (1,1) | (3,1) |
| 0100 | 2 | 2 | 2 | (2,2) | (2,2) |
| 0101 | 1 | 3 | 1 | (1,1) | (3,1) |
| 0110 | 3 | 3 | 1 | (3,1) | (**3,1**) |
| 0111 | 2 | 4 | 0 | (2,0) | (4,0) |
| 1000 | 2 | 0 | 4 | (4,2) | (4,0) |
| 1001 | 1 | 1 | 3 | (3,1) | (**3,1**) |
| 1010 | 3 | 1 | 3 | (3,3) | (**3,1**) |
| 1011 | 2 | 2 | 2 | (2,2) | (2,2) |
| 1100 | 3 | 1 | 3 | (3,3) | (**3,1**) |
| 1101 | 2 | 2 | 2 | (2,2) | (2,2) |
| 1110 | 4 | 2 | 2 | (4,2) | (2,2) |
| 1111 | 3 | 3 | 1 | (3,1) | (**3,1**) |

Table 17: Manipulation of $\Delta_\mu^{d_H,G_{Max}}$ for $i_{Dalal}$ with two complete bases.

$\square$

**Theorem 9** *None of the $\triangle_\mu^{\widehat{C}}$ operators (hence, none of the $\triangle_\mu^C$ operators) is strategy-proof for $i_{Dalal}$, even in the restricted case $E$ consists of two complete bases.*

**Proof:** Let us consider the complete bases $K_1 = \{a \wedge b\}$ and $K_2 = \{\neg a \wedge \neg b\}$, with $\mu = \neg(a \wedge b)$. We have $\mathrm{MAXCONS}(\{K_1, K_2\}, \mu) = \{\{\neg a \wedge \neg b, \neg(a \wedge b)\}\} = \mathrm{MAXCONS}_{card}(\{K_1, K_2\}, \mu)$ and $\mathrm{MAXCONS}(\{K_1, K_2\}, \top) = \{\{\neg a \wedge \neg b, \top\}, \{a \wedge b, \top\}\}$, hence $\triangle_\mu^{\widehat{C1}}(\{K_1, K_2\}) \equiv \triangle_\mu^{\widehat{C4}}(\{K_1, K_2\}) \equiv \triangle_\mu^{\widehat{C5}}(\{K_1, K_2\}) \equiv \neg a \wedge \neg b$.
We get $i_{Dalal}(K_1, \triangle_\mu^{\widehat{C}}(\{K_1, K_2\})) = 1 - \frac{2}{2} = 0$.
With $K_1' = \{\neg a \wedge b\}$, we have $\mathrm{MAXCONS}(\{K_1', K_2\}, \mu) = \{\{\neg a \wedge \neg b, \neg(a \wedge b)\}, \{\neg a \wedge b, \neg(a \wedge b)\}\} = \mathrm{MAXCONS}_{card}(\{K_1', K_2\}, \mu)$ and $\mathrm{MAXCONS}(\{K_1', K_2\}, \top) = \{\{\neg a \wedge \neg b, \top\}, \{\neg a \wedge$

$b, \top\}\}$, hence $\triangle_\mu^{\widehat{C1}}(\{K_1', K_2\}) \equiv \triangle_\mu^{\widehat{C3}}(\{K_1', K_2\}) \equiv \triangle_\mu^{\widehat{C4}}(\{K_1', K_2\}) \equiv \triangle_\mu^{\widehat{C5}}(\{K_1', K_2\}) \equiv (\neg a \wedge \neg b) \vee (\neg a \wedge b) \equiv \neg a$.

Thus $i_{Dalal}(K_1, \triangle_\mu^{\widehat{C}}(\{K_1', K_2\})) = 1 - \frac{1}{2}$, showing the manipulation. $\qquad\square$

**Theorem 10** *Let $d$ be a pseudo-distance and let $f$ be an aggregation function. $\Delta_\mu^{d,f}$ is dilation strategy-proof for the indexes $i_p$, $i_{dw}$ and $i_{ds}$.*

**Proof:**     *The idea is that if the untruthful base $K'$ contains all the models of the "true" base $K$, then the merged base when $K'$ is provided contains at most the same models of $K$ as those appearing in the merged base when $K$ is reported, and more countermodels of $K$. So no manipulation is possible.*

From Theorem 1, it is sufficient to show that $\Delta_\mu^{d,f}$ is strategy-proof for $i_p$. By *reductio ad absurdo*. Let us suppose that there is an operator $\Delta_\mu^{d,f}$, where $d$ and $f$ are respectively a pseudo-distance and an aggregation function, which is manipulable by dilation for $i_p$. Under this assumption, we can find an integrity constraint $\mu$, a profile $E$, two bases $K$ and $K'$ with $K \models K'$, s.t. $i_p(K, \Delta_\mu^{d,f}(\{K\} \sqcup E)) < i_p(K, \Delta_\mu^{d,f}(\{K'\} \sqcup E))$. Using the light notation $E \triangle_\mu K$ instead of $\Delta_\mu^{d,f}(\{K\} \sqcup E)$, we have:

$$\frac{\#([K] \cap [E \triangle_\mu K])}{\#([E \triangle_\mu K])} < \frac{\#([K] \cap [E \triangle_\mu K'])}{\#([E \triangle_\mu K'])}.$$

Since $K \models K'$, for any pseudo-distance $d$, we have $\forall \omega \in \mathcal{W}, d(\omega, K) \geq d(\omega, K')$. So, for any aggregation function $f$ (that satisfies non-decreasingness):

$$\forall \omega \in \mathcal{W}, d(\omega, E \sqcup \{K\}) \geq d(\omega, E \sqcup \{K'\}). \tag{8}$$

Let us note $d_{min}(E \sqcup_\mu \{K\}) = min(\{d(\omega, E \sqcup \{K\}) \mid \omega \models \mu\}, \leq)$. With (8), we can immediately infer: $d_{min}(E \sqcup_\mu \{K\}) \geq d_{min}(E \sqcup_\mu \{K'\})$. Two cases have to be considered:

- $d_{\min}(E \sqcup_\mu \{K\}) > d_{min}(E \sqcup_\mu \{K'\})$ (*).

  If $\omega_1$ is a model of $K \wedge \mu$ then, since $K \models K'$, $d(\omega_1, K) = d(\omega_1, K') = 0$, so $d(\omega_1, E \sqcup \{K\}) = d(\omega_1, E \sqcup \{K'\})$ . If furthermore $\omega_1$ is a model of $E \triangle_\mu K'$, then $d(\omega_1, E \sqcup \{K'\}) = d_{min}(E \sqcup_\mu \{K'\})$, and $d(\omega_1, E \sqcup \{K\}) = d_{min}(E \sqcup_\mu \{K'\})$. By definition of $min$, we have $d(\omega_1, E \sqcup \{K\}) \geq d_{min}(E \sqcup_\mu \{K\})$, because $\omega_1 \models \mu$. So we can conclude that $d_{min}(E \sqcup_\mu \{K\}) \leq d_{min}(E \sqcup_\mu \{K'\})$, but this contradicts (*). So, no model of $K \wedge \mu$ is a model of $E \triangle_\mu K'$. Consequently, $i_p(K, E \triangle_\mu K') = 0$ and is minimal, and this prevents from any manipulation for $\Delta_\mu^{d,f}$. So, we can exclude case (*).

- $d_{min}(E \sqcup_\mu \{K\}) = d_{min}(E \sqcup_\mu \{K'\})$ (**).

  If $\omega$ is a model of $E \triangle_\mu K$, then we have both $\omega \models \mu$ and $d(\omega, E \sqcup \{K\}) = d_{min}(E \sqcup_\mu \{K\})$. So, $d(\omega, E \sqcup \{K\}) = d_{min}(E \sqcup_\mu \{K'\})$ with the equation (**). Furthermore, with the inequation (8), we infer that $d(\omega, E \sqcup \{K\}) \geq d(\omega, E \sqcup \{K'\})$. Hence, $d(\omega, E \sqcup \{K'\}) \leq d_{min}(E \sqcup_\mu \{K\})$. Since $\omega$ is a model of $\mu$, we can finally infer that $\omega$ is a model of $E \triangle_\mu K'$ as well. So we have that any model of $E \triangle_\mu K$ is a model of $E \triangle_\mu K'$, and then:

$$\#([E \triangle_\mu K]) \leq \#([E \triangle_\mu K']). \tag{9}$$

We can deduce as well that any model of $E \triangle_\mu K$ which is a model of $K \wedge \mu$ is a model of $E \triangle_\mu K'$ (and of $K$), so:

$$\#([K] \cap [E \triangle_\mu K]) \leq \#([K] \cap [E \triangle_\mu K']).$$

Furthermore, if $\omega_1 \models K \wedge \mu$ is a model of $E \triangle_\mu K'$, then we have both:

- $d(\omega_1, E \sqcup \{K'\}) = d_{min}(E \sqcup_\mu \{K'\}) = d_{min}(E \sqcup_\mu \{K\})$ with (**), and
- $d(\omega_1, E \sqcup \{K\}) = d(\omega_1, E \sqcup \{K'\})$ because $K \models K'$: since $d(\omega_1, K) = 0$, we have $d(\omega_1, K') = 0$ too.

We obtain: $d(\omega_1, E \sqcup \{K\}) = d_{min}(E \sqcup_\mu \{K\})$ and $\omega_1 \models \mu$, so $\omega_1$ is a model of $E \triangle_\mu K$. Then we can state $\#([K] \cap [E \triangle_\mu K]) \geq \#([K] \cap [E \triangle_\mu K'])$. So we get:

$$\#([K] \cap [E \triangle_\mu K]) = \#([K] \cap [E \triangle_\mu K']). \tag{10}$$

With (9) and (10), we get immediately that:

$$\frac{\#([K] \cap [E \triangle_\mu K])}{\#([E \triangle_\mu K])} \geq \frac{\#([K] \cap \#([E \triangle_\mu K'])}{\#([E \triangle_\mu K'])}. \tag{11}$$

As a consequence, $i_p(K, \Delta_\mu^{d,f}(\{K\} \sqcup E) \geq i_p(K, \Delta_\mu^{d,f}(\{K'\} \sqcup E))$. This inequation shows that $\Delta_\mu^{d,f}$ is not manipulable for $i_p$, which contradicts the assumption. So case (**) has to be excluded as well, and this concludes the proof.

$\square$

**Theorem 11** *Let $d$ be any distance. If $\Delta_\mu^{d,\Sigma}$ is not strategy-proof for the index $i_{dw}$ (resp. $i_{ds}$), then it is not erosion strategy-proof for $i_{dw}$ (resp. $i_{ds}$).*

**Proof:** We first need the following lemma:

**Lemma 2** *Let $d$ be a pseudo-distance and let $f$ be an aggregation function. If a profile $E$ is manipulable by $K$ for $i_{dw}$ (resp. $i_{ds}$) given $\Delta_\mu^{d,f}$ and $\mu$, then one can find a complete base $K'$ – the base the agent gives instead of her true base $K$ – s.t. $i_{dw}(K, \Delta_\mu(E \sqcup \{K'\})) > i_{dw}(K, \Delta_\mu(E \sqcup \{K\}))$ (resp. $i_{ds}(K, \Delta_\mu(E \sqcup \{K'\})) > i_{ds}(K, \Delta_\mu(E \sqcup \{K\})))$.*

**Proof:** *This lemma is mainly a consequence of the definition of the distance between an interpretation and a base $K'$, as the minimal distance between an interpretation and a model $\omega''$ of the base; the complete base whose unique model is $\omega''$ allows as well a manipulation*

<u>*Weak drastic index*</u>.

We suppose that $\Delta_\mu^{d,f}$ is manipulable for $i_{dw}$, i.e., we can find an integrity constraint $\mu$, a profile $E = \{K_1, \ldots, K_n\}$, and two bases $K$ and $K'$ s.t.:

$$i_{dw}(K, \Delta_\mu^{d,f}(\{K\} \sqcup E)) < i_{dw}(K, \Delta_\mu^{d,f}(\{K'\} \sqcup E)). \tag{12}$$

This is equivalent to: $i_{dw}(K, E \triangle_\mu K) = 0$ and $i_{dw}(K, E \triangle_\mu K') = 1$, where $\Delta_\mu^{d,f}(\{K\} \sqcup E)$ is noted $E \triangle_\mu K$ for simplifying notations.

Statement (13) states that $i_{dw}(K, E \triangle_\mu K) = 0$: no model of $K \wedge \mu$ is a model of $E \triangle_\mu K$; statement (14) states that $i_{dw}(K, E \triangle_\mu K') = 1$: there is as least one model of $K \wedge \mu$ in $E \triangle_\mu K'$:

$$\forall \omega \models K \wedge \mu, \exists \omega' \models (\neg K) \wedge \mu, d(\omega', E \sqcup \{K\}) < d(\omega, E \sqcup \{K\}). \tag{13}$$

$$\exists \omega_1 \models K \wedge \mu, \forall \omega \models \mu, d(\omega_1, E \sqcup \{K'\}) \leq d(\omega, E \sqcup \{K'\}). \tag{14}$$

Since in (13) the choice of $\omega'$ can be made apart from $\omega$, (13) is equivalent to:

$$\exists \omega' \models (\neg K) \wedge \mu, \forall \omega \models K \wedge \mu, d(\omega', E \sqcup \{K\}) < d(\omega, E \sqcup \{K\}). \tag{15}$$

Let $\omega'' \models K'$ s.t. $d(\omega_1, K') = d(\omega_1, \omega'')$. We consider the complete base $K''$ s.t. $[K''] = \{\omega''\}$. we shall show in the rest of the proof that $\Delta_\mu^{d,f}$ is manipulable with that base. If the agent whose beliefs/goals are $K$ gives $K''$ as a base instead of $K$, then, since $d(\omega_1, K'') = d(\omega_1, K')$, we have:

$$d(\omega_1, E \sqcup \{K''\}) = d(\omega_1, E \sqcup \{K'\}), \tag{16}$$

and then:

$$\forall \omega \models \mu, d(\omega_1, E \sqcup \{K''\}) \leq d(\omega, E \sqcup \{K'\}), \tag{17}$$

with (14) and (16).

Furthermore, since the aggregation function $f$ is non-decreasing (by definition) and since $K'' \models K'$, we have $\forall \omega \models \mu, d(\omega, E \sqcup \{K'\}) \leq d(\omega, E \sqcup \{K''\})$, so we get immediately with (17):

$$\forall \omega \models \mu : d(\omega_1, E \sqcup \{K''\}) \leq d(\omega, E \sqcup \{K''\}). \tag{18}$$

So $\omega_1$ is a model of $\Delta_\mu^{d,f}(E \sqcup \{K''\})$, and we have $i_{dw}(K, \Delta_\mu^{d,f}(\{K''\} \sqcup E)) = 1$ and $i_{dw}(K, \Delta_\mu^{d,f}(\{K\} \sqcup E)) < i_{dw}(K, \Delta_\mu^{d,f}(\{K''\} \sqcup E))$. This shows that $\Delta_\mu^{d,f}$ is manipulable for a complete base $K''$.

<u>Strong drastic index.</u> Let us assume that an operator $\Delta_\mu^{d,f}$, where $d$ is any pseudo-distance and $f$ any aggregation function, is manipulable for the strong drastic index $i_{ds}$. Then we can find an integrity constraint $\mu$, a profile $E$ and bases $K$ and $K'$ s.t. $i_{ds}(K, \Delta_\mu^{d,f}(E \sqcup \{K\}) < i_{ds}(K, \Delta_\mu^{d,f}(E \sqcup \{K'\})$. This implies that $i_{ds}(K, \Delta_\mu^{d,f}(E \sqcup \{K\}) = 0$, and $i_{ds}(K, \Delta_\mu^{d,f}(E \sqcup \{K'\}) = 1$. This means, by definition of the index, that $\Delta_\mu^{d,f}(E \sqcup \{K\}) \not\models K$, and $\Delta_\mu^{d,f}(E \sqcup \{K'\}) \models K$.

Given a model $\omega_1$ of $\Delta_\mu^{d,f}(E \sqcup \{K'\})$ and a model $\omega_2$ of $K'$ s.t. $d(\omega_1, K') = d(\omega_1, \omega_2)$, we define $K'' = \{\omega_2\}$. Then we have $d(\omega_1, K'') = d(\omega_1, K')$, and: $d(\omega_1, E \sqcup \{K''\}) = d(\omega_1, E \sqcup \{K'\})$.

Let us note $d_{min}(E \sqcup_\mu^{d,f} \{K'\}) = min(\{d(\omega, E \sqcup \{K'\}) \mid \omega \models \mu\}, \leq)$. Since $\omega_1$ is a model of $\Delta_\mu^{d,f}(E \sqcup \{K'\})$, we have $d(\omega_1, E \sqcup \{K'\}) = d_{min}(E \sqcup_\mu^{d,f} \{K'\})$. Hence:

$$d(\omega_1, E \sqcup \{K''\}) = d_{min}(E \sqcup_\mu^{d,f} \{K'\}). \tag{19}$$

By definition of $min$ and since $\omega_1 \models \mu$, we also have: $d(\omega_1, E \sqcup \{K''\}) \geq d_{min}(E \sqcup_\mu^{d,f} \{K''\})$. So we get:

$$d_{min}(E \sqcup_\mu^{d,f} \{K'\}) \geq d_{min}(E \sqcup_\mu^{d,f} \{K''\}). \tag{20}$$

On the other hand, since $K'' \models K'$, we have: $\forall \omega \in \mathcal{W}, d(\omega, K') \leq d(\omega, K'')$. Since the aggregation function $f$ is non-decreasing, we get:

$$\forall \omega \in \mathcal{W}, d(\omega, E \sqcup \{K'\}) \leq d(\omega, E \sqcup \{K''\}), \tag{21}$$

and then $d_{min}(E \sqcup_\mu^{d,f} \{K'\}) \leq d_{min}(E \sqcup_\mu^{d,f} \{K''\})$. With (20) we get $d_{min}(E \sqcup_\mu^{d,f} \{K'\}) = d_{min}(E \sqcup_\mu^{d,f} \{K''\})$. Then, with (19), we obtain $d(\omega_1, E \sqcup \{K''\}) = d_{min}(E \sqcup_\mu^{d,f} \{K''\})$. Since $\omega_1 \models \mu$, we have that $\omega_1$ is a model of $\Delta_\mu^{d,f}(E \sqcup \{K''\})$ too.

Let $\omega$ be a model of $\Delta_\mu^{d,f}(E \sqcup \{K''\})$. Then, $\omega \models \mu$ and $d(\omega, E \sqcup \{K''\}) = d_{min}(E \sqcup_\mu^{d,f} \{K''\})$. Then, since $d_{min}(E \sqcup_\mu^{d,f} \{K'\}) = d_{min}(E \sqcup_\mu^{d,f} \{K''\})$, we have $d(\omega, E \sqcup \{K''\}) = d_{min}(E \sqcup_\mu^{d,f} \{K'\})$. With (21), we get $d(\omega, E \sqcup \{K'\}) \leq d_{min}(E \sqcup_\mu^{d,f} \{K'\})$.

Then, by definition of $min$ we have $d(\omega, E \sqcup \{K'\}) = d_{min}(E \sqcup_\mu^{d,f} \{K'\})$.

This implies that $\omega$ is a model of $\Delta_\mu^{d,f}(E \sqcup \{K'\})$ too (because $\omega \models \mu$), so we can write: $\Delta_\mu^{d,f}(E \sqcup \{K''\}) \models \Delta_\mu^{d,f}(E \sqcup \{K'\})$. Since we have $\Delta_\mu^{d,f}(E \sqcup \{K'\}) \models K$, we can infer that $\Delta_\mu^{d,f}(E \sqcup \{K''\}) \models K$, and then $i_{ds}(K, \Delta_\mu^{d,f}(E \sqcup \{K''\}) = 1$.

We get then a manipulation for $i_{ds}$ with a complete base $K''$, and this completes the proof of the lemma. $\square$

Let us now give the proof of the main theorem:

<u>Weak drastic index</u>. By *reductio ad absurdum*. We assume that $\Delta_\mu^{d,\Sigma}$ is a manipulable operator and that it is not manipulable by erosion. Then we can find an integrity constraint $\mu$, a profile $E$, and two bases $K$ and $K'$ with $K' \not\models K$ s.t.:

$$i_{dw}(K, \Delta_\mu^{d,\Sigma}(E \sqcup \{K\})) < i_{dw}(K, \Delta_\mu^{d,\Sigma}(E \sqcup \{K'\})).$$

Lemma 2 shows that that we can assume $[K'] = \{\omega_1'\}$ complete; it comes:

$$i_{dw}(K, \Delta_\mu^{d,\Sigma}(E \sqcup \{K\})) < i_{dw}(K, \Delta_\mu^{d,\Sigma}(E \sqcup \{\omega_1'\})).$$

This implies that:

$$i_{dw}(K, \Delta_\mu^{d,\Sigma}(E \sqcup \{K\}) = 0 \tag{22}$$

$$i_{dw}(K, \Delta_\mu^{d,\Sigma}(E \sqcup \{\omega_1'\}) = 1.$$

This means there is no model of $K \wedge \mu$ which is a model of $\Delta_\mu^{d,\Sigma}(E \sqcup \{K\})$, and that there is at least one model $\omega_1$ of $K \wedge \mu$ which is a model of $\Delta_\mu^{d,\Sigma}(E \sqcup \{\omega_1'\})$. We can express those facts by two statements:

$$\forall \omega \models K \wedge \mu, \exists \omega' \models \neg K \wedge \mu, d(\omega', E \sqcup \{K\}) < d(\omega, E \sqcup \{K\})$$

and:

$$\exists \omega_1 \models K \wedge \mu, \forall \omega \models \mu, d(\omega_1, E \sqcup \{\omega_1'\}) \leq d(\omega, E \sqcup \{\omega_1'\}).$$

Hence:

$$\exists \omega_1 \models K \wedge \mu, \forall \omega \models \mu, d(\omega_1, \omega_1') + d(\omega_1, E) \leq d(\omega, \omega_1') + d(\omega, E) \tag{23}$$

Let us now define a new base $K'' = \{\omega_1\}$. Since we have supposed the operator not manipulable by erosion and since $K'' \models K$, we can then state that it is strategy-proof for $i_{dw}$ with $K''$: $i_{dw}(K, \Delta_\mu^{d,\Sigma}(E \sqcup \{K\})) \geq i_{dw}(K, \Delta_\mu^{d,\Sigma}(E \sqcup \{K''\}))$. This implies that:

- either $i_{dw}(K, \Delta_\mu^{d,\Sigma}(E \sqcup \{K\})) = 1$,

- or $i_{dw}(K, \Delta_\mu^{d,\Sigma}(E \sqcup \{K\})) = i_{dw}(K, \Delta_\mu^{d,\Sigma}(E \sqcup \{K''\})) = 0$.

From equation (22), we can infer that $i_{dw}(K, \Delta_\mu^{d,\Sigma}(E \sqcup \{K''\})) = 0$, and we have $\forall \omega \models K \wedge \mu, \exists \omega' \models \neg K \wedge \mu, d(\omega', E \sqcup \{K''\}) < d(\omega, E \sqcup \{K''\})$.

Since $K'' = \{\omega_1\}$ and the choice of $\omega'$ can be made independently of $\omega_1$, we have $\exists \omega_2 \models (\neg K) \wedge \mu, \forall \omega \models K \wedge \mu, d(\omega_2, E \sqcup \{\omega_1\}) < d(\omega, E \sqcup \{\omega_1\})$, that is:

$$\exists \omega_2 \models \neg K \wedge \mu, \forall \omega \models K \wedge \mu, d(\omega_2, \omega_1) + d(\omega_2, E) < d(\omega, \omega_1) + d(\omega, E).$$

In particular, this statement holds for $\omega = \omega_1$, because $\omega_1 \models K \wedge \mu$. Hence:

$$d(\omega_2, \omega_1) + d(\omega_2, E) < d(\omega_1, \omega_1) + d(\omega_1, E).$$

Since $d(\omega_1, \omega_1) = 0$, we obtain finally:

$$d(\omega_2, \omega_1) + d(\omega_2, E) < d(\omega_1, E). \tag{24}$$

On the other hand, since $\omega_2 \models \mu$, with (23), we get:

$$d(\omega_1, \omega_1') + d(\omega_1, E) \leq d(\omega_2, \omega_1') + d(\omega_2, E). \tag{25}$$

Summing (24) and (25) side by side, we get:

$$d(\omega_2, \omega_1) + d(\omega_2, E) + d(\omega_1, \omega_1') + d(\omega_1, E) < d(\omega_1, E) + d(\omega_2, \omega_1') + d(\omega_2, E).$$

Simplifying by $d(\omega_1, E)$ and $d(\omega_2, E)$, we obtain:

$$d(\omega_2, \omega_1) + d(\omega_1, \omega_1') < d(\omega_2, \omega_1').$$

This contradicts the triangular inequality. So, if manipulation is possible, then it is possible by erosion with a complete base $K'' = \{\omega_1\}$.

_Strong drastic index._ Let us assume that $\Delta_\mu^{d,\Sigma}$ is manipulable for $i_{ds}$. So we can find a profile $E$, an integrity constraint $\mu$ and two bases $K$ and $K'$ s.t.:

$$i_{ds}(K, \Delta_\mu^{d,\Sigma}(E \sqcup \{K\})) < i_{ds}(K, \Delta_\mu^{d,\Sigma}(E \sqcup \{K'\})).$$

This implies that:

$$i_{ds}(K, \Delta_\mu^{d,\Sigma}(E \sqcup \{K\}) = 0 \tag{26}$$

and

$$i_{ds}(K, \Delta_\mu^{d,\Sigma}(E \sqcup \{K'\}) = 1.$$

This means that there is at least one model of $\Delta_\mu^{d,\Sigma}(E \sqcup \{K\})$ which is not a model of $K$, and that every model of $\Delta_\mu^{d,\Sigma}(E \sqcup \{K'\})$ is a model of $K \wedge \mu$. Let us consider a model $\omega_1$ of $\Delta_\mu^{d,\Sigma}(E \sqcup \{K'\})$. We can write that $\omega_1 \models K \wedge \mu$ and:

$$\forall \omega \models \mu, d(\omega_1, E \sqcup \{K'\}) \leq d(\omega, E \sqcup \{K'\}).$$

So we have:

$$\forall \omega \models \mu, d(\omega_1, K') + d(\omega_1, E) \leq d(\omega, K') + d(\omega, E).$$

Let us define $K'' = \{\omega_1\}$ and show that there is manipulation with this base. Let us assume that we can find $\omega'' \models \Delta_\mu^{d,\Sigma}(E \sqcup \{K''\})$ s.t. $\omega'' \not\models K$. Since $\omega'' \models \Delta_\mu^{d,\Sigma}(E \sqcup \{\omega_1\})$, we have $\omega'' \models \mu$ and

$$d(\omega'', E \sqcup \{\omega_1\}) \leq d(\omega_1, E \sqcup \{\omega_1\}),$$

and then:

$$d(\omega'', \omega_1) + d(\omega'', E) \leq d(\omega_1, E)$$

(as $d(\omega_1, \omega_1) = 0$).

We know that $\omega'' \not\models K$ and $\omega'' \models \mu$, so we can infer that $\omega'' \not\models \Delta_\mu^{d,\Sigma}(E \sqcup \{K'\})$ (else we should have $\omega'' \models K$). Since $\omega_1$ is a model of $\Delta_\mu^{d,\Sigma}(E \sqcup \{K'\})$, we have:

$$d(\omega'', E \sqcup \{K'\}) > d(\omega_1, E \sqcup \{K'\}).$$

Equivalently:

$$d(\omega'', K') + d(\omega'', E) > d(\omega_1, K') + d(\omega_1, E).$$

Since $d(\omega_1, E) \geq d(\omega'', \omega_1) + d(\omega'', E)$, we obtain:

$$d(\omega'', K') + d(\omega'', E) > d(\omega_1, K') + d(\omega'', \omega_1) + d(\omega'', E).$$

Simplifying this equation by $d(\omega'', E)$, we get:

$$d(\omega'', K') > d(\omega_1, K') + d(\omega'', \omega_1).$$

If $\omega_2$ is a model of $K'$ s.t. $d(\omega_1, K') = d(\omega_1, \omega_2)$, we have:

$$d(\omega'', K') > d(\omega_1, \omega_2) + d(\omega'', \omega_1),$$

and finally, since $d(\omega", \omega_2) \geq d(\omega", K')$ by definition of $min$, we get:

$$d(\omega", \omega_2) > d(\omega_1, \omega_2) + d(\omega", \omega_1).$$

This contradicts the triangular inequality.

We have shown that every model of $\Delta_\mu^{d,\Sigma}(E \sqcup \{K''\})$ is a model of $K$. Hence, $i_{ds}(K, \Delta_\mu^{d,\Sigma}(E \sqcup \{K''\}) = 1$. Since $i_{ds}(K, \Delta_\mu^{d,\Sigma}(E \sqcup \{K\}) = 0$ with (26), we have:

$$i_{ds}(K, \Delta_\mu^{d,\Sigma}(E \sqcup \{K\})) < i_{ds}(K, \Delta_\mu^{d,\Sigma}(E \sqcup \{K''\})),$$

and a manipulation by erosion with a complete base is possible.

<div style="text-align: right;">□</div>

**Corollary 12** *A profile $E$ is manipulable by $K$ for $i_{dw}$ (resp. $i_{ds}$) given $\Delta_\mu^{d,\Sigma}$ and $\mu$ if and only if the manipulation is possible using a complete base $K_\omega \models K$, i.e., there exists $K_\omega \models K$ s.t. $i_{dw}(K, \Delta_\mu^{d,\Sigma}(E \sqcup \{K_\omega\})) > i_{dw}(K, \Delta_\mu^{d,\Sigma}(E \sqcup \{K\}))$ (resp. $i_{ds}(K, \Delta_\mu^{d,\Sigma}(E \sqcup \{K_\omega\})) > i_{ds}(K, \Delta_\mu^{d,\Sigma}(E \sqcup \{K\}))$).*

**Proof:** ($\Rightarrow$): This is a consequence of Theorem 11 and of Lemma 2, which enable to state that if $K$ is manipulable for $i = i_{dw}$ or $i = i_{ds}$ given $\Delta_\mu^{d,\Sigma}$ and $E$, then one can find a base $K' = \{\omega'\} \models K$ complete such that $i(K, \Delta_\mu^{d,f}(E \sqcup \{\omega'\})) > i(K, \Delta_\mu^{d,f}(E \sqcup \{K\}))$.

($\Leftarrow$): If $K$ is strategy-proof for $i = i_{dw}$ or $i = i_{ds}$ given $\Delta_\mu^{d,\Sigma}$ and $E$, then $\forall [K'] \subseteq \mathcal{W}, i(K, \Delta_\mu^{d,\Sigma}(E \sqcup \{K'\}) \leq i(K, \Delta_\mu^{d,\Sigma}(E \sqcup \{K\}))$. This is in particular true when $K'$ is reduced to a singleton.

<div style="text-align: right;">□</div>

**Theorem 14** *$\Delta_{max}$, $\Delta_{min_1}$, $\Delta_{min_2}$, and $\Delta_\Sigma$ are strategy-proof for $i_{dw}$, $i_{ds}$ and $i_p$.*

**Proof:** $\Delta_{max}$ (or equivalently $\Delta_{min_2}$) is strategy-proof for $i_p$. Indeed, if $\bigwedge E$ is consistent, then all the models of the merged base are models of $K_1 \in E$, thus the satisfaction of $K_1$ is maximal for $i_p$. If $\bigwedge E$ is not consistent, then the merged base is valid. Assume that agent 1 reports $K_1'$ instead of $K_1$. If $K_1' \wedge \bigwedge \{K_2, \ldots, K_n\}$ is consistent, then no model of the resulting merged base is a model of $K_1$. In the case $K_1' \wedge \bigwedge \{K_2, \ldots, K_n\}$ is inconsistent, then the resulting merged base is still valid. From Theorem 1, $\Delta_{max}$ (or equivalently $\Delta_{min_2}$) is also strategy-proof for the two drastic indexes.

$\Delta_{min_1}$ is also strategy-proof for $i_p$. The case when $\bigwedge E$ is consistent is as above. If $\bigwedge E$ is not consistent, the result of the merging is $\bigvee E$ and all the models of $K_1$ are models of the merged base. So, in order to increase the value of the $i_p$ index, it is not possible to increase the number of models of $K_1$ in the result of the merging. Hence one needs to decrease the number of countermodels of $K_1$ in the merged base. But we shall show that it is not possible: assume that agent 1 reports $K_1'$ instead of $K_1$. If $K_1' \wedge \bigwedge \{K_2, \ldots, K_n\}$ is consistent, then no model of the resulting merged base is a model of $K_1$ (as $\bigwedge E$ is not consistent).

Then $K'_1 \wedge \bigwedge\{K_2, \ldots, K_n\}$ is not consistent, and we have:

$$i_p(K_1, \Delta_{min_1}(K_1 \sqcup \{K_2, \ldots, K_n\})) = \frac{\#([K_1])}{\#([K_1 \vee \bigvee\{K_2, \ldots, K_n\}])}$$

and

$$i_p(K_1, \Delta_{min_1}(K'_1 \sqcup \{K_2, \ldots, K_n\})) = \frac{\#([K_1 \wedge (K'_1 \vee \bigvee\{K_2, \ldots, K_n\})])}{\#([K'_1 \vee \bigvee\{K_2, \ldots, K_n\}])}.$$

The numerator of $i_p(K_1, \Delta_{min_1}(K_1 \sqcup \{K_2, \ldots, K_n\}))$ is maximal, so in order to increase $i_p(K_1, \Delta_{min_1}(K'_1 \sqcup \{K_2, \ldots, K_n\}))$, we have to decrease the denominator of $i_p(K_1, \Delta_{min_1}(K'_1 \sqcup \{K_2, \ldots, K_n\}))$, $\#([K'_1 \vee \bigvee\{K_2, \ldots, K_n\}])$.

We can write the following equality:

$$\#([K'_1 \vee \bigvee\{K_2, \ldots, K_n\}]) = \#([K'_1 \wedge K_1 \wedge \neg(\bigvee\{K_2, \ldots, K_n\})]) +$$
$$\#([K'_1 \wedge \neg K_1 \wedge \neg(\bigvee\{K_2, \ldots, K_n\})]) + \#([\bigvee\{K_2, \ldots, K_n\}]).$$

In this sum, $\#([\bigvee\{K_2, \ldots, K_n\}])$ cannot be changed. We have to decrease the two other terms of this sum: $\#([K'_1 \wedge \neg K_1 \wedge \neg(\bigvee\{K_2, \ldots, K_n\})])$ is minimal if $K'_1$ is such that $\#([K'_1 \wedge \neg K_1 \wedge \neg(\bigvee\{K_2, \ldots, K_n\})]) = 0$. In the following, we suppose then that $\#([K'_1 \wedge \neg K_1 \wedge \neg(\bigvee\{K_2, \ldots, K_n\})]) = 0$.

For the first term of the sum, $\#([K'_1 \wedge K_1 \wedge \neg(\bigvee\{K_2, \ldots, K_n\})])$, we can write:

$$\#([K'_1 \wedge K_1 \wedge \neg(\bigvee\{K_2, \ldots, K_n\})]) =$$
$$\#([K_1 \wedge \neg(\bigvee\{K_2, \ldots, K_n\})]) - \#([K_1 \wedge \neg K'_1 \wedge \neg(\bigvee\{K_2, \ldots, K_n\})])$$

So we obtain for the probabilistic index:

$$i_p(K_1, \Delta_{min_1}(K'_1 \sqcup \{K_2, \ldots, K_n\})) =$$
$$\frac{\#([K_1]) - \#([K_1 \wedge \neg K'_1 \wedge \neg(\bigvee\{K_2, \ldots, K_n\})])}{\#([K_1 \wedge \neg(\bigvee\{K_2, \ldots, K_n\})]) - \#([K_1 \wedge \neg K'_1 \wedge \neg(\bigvee\{K_2, \ldots, K_n\})]) + \#([\bigvee\{K_2, \ldots, K_n\}])}.$$

Now remark that if $k$ is an integer and if $a \leq b$, then $\frac{a-k}{b-k} \leq \frac{a}{b}$. If we subtract the same integer $\#([K_1 \wedge \neg K'_1 \wedge \neg(\bigvee\{K_2, \ldots, K_n\})])$ from the numerator and the denominator, then we get the following inequation:

$$\frac{\#([K_1]) - \#([K_1 \wedge \neg K'_1 \wedge \neg(\bigvee\{K_2, \ldots, K_n\})])}{\#([K_1 \wedge \neg(\bigvee\{K_2, \ldots, K_n\})]) - \#([K_1 \wedge \neg K'_1 \wedge \neg(\bigvee\{K_2, \ldots, K_n\})]) + \#([\bigvee\{K_2, \ldots, K_n\}])} \leq$$
$$\frac{\#([K_1])}{\#([K_1 \wedge \neg(\bigvee\{K_2, \ldots, K_n\})]) + \#([\bigvee\{K_2, \ldots, K_n\}])}.$$

So:

$$i_p(K_1, \Delta_{min_1}(K'_1 \sqcup \{K_2, \ldots, K_n\})) \leq \frac{\#([K_1])}{\#([K_1 \wedge \neg(\bigvee\{K_2, \ldots, K_n\})]) + \#([\bigvee\{K_2, \ldots, K_n\}])}$$

then

$$i_p(K_1, \Delta_{min_1}(K_1' \sqcup \{K_2, \ldots, K_n\})) \leq i_p(K_1, \Delta_{min_1}(K_1 \sqcup \{K_2, \ldots, K_n\})).$$

No manipulation is possible, and $\Delta_{min_1}$ is strategy-proof for $i_p$. From Theorem 1, $\Delta_{min_1}$ is also strategy-proof for the two drastic indexes.

Finally, $\Delta_\Sigma$ is strategy-proof for the three indexes (see Theorem 2). □

**Theorem 15** $\Delta$ *satisfies the* **(IP)** *property if and only if for every profile $E$ and every pair of bases $K$ and $K'$:*

- $K \wedge \neg\Delta(E \sqcup \{K\}) \models \neg\Delta(E \sqcup \{K'\})$, *and*

- $\neg K \wedge \Delta(E \sqcup \{K\}) \models \Delta(E \sqcup \{K'\})$.

**Proof:**     Suppose that a merging operator $\Delta = Bel(\kappa_\Delta)$) does not satisfy the **(IP)** property. Then there is a profile $E$, an agent $i$, an OCF $\kappa$, an interpretation $\omega \in \mathcal{W}$ s.t.

$$|\kappa_\Delta(E)(\omega) - \kappa_i(\omega)| > |\kappa_\Delta(rep(E, \{i\}, \kappa)(\omega) - \kappa_i(\omega)|$$

where $rep(E, \{i\}, \kappa)$ is the profile identical to $E$ except that the OCF $\kappa_i$ is replaced by $\kappa$. In the following we note $\Delta(E \sqcup \{K\})$ for $Bel(\kappa_\Delta(E))$: it is the initial merged base when agent $i$ reports her true base $K \equiv Bel(\kappa_i)$, and we note $\Delta(E \sqcup \{K'\})$ for $Bel(\kappa_\Delta(rep(E, \{i\}, \kappa)))$: it is the merged base obtained by replacing the base $K$ of agent $i$ by another $K' \equiv Bel(\kappa)$.

Focusing on two-strata OCFs, this inequation entails that $|\kappa_\Delta(E)(\omega) - \kappa_i(\omega)| = 1$ and $|\kappa_\Delta(rep(E, \{i\}, \kappa)(\omega) - \kappa_i(\omega)| = 0$. Since $|\kappa_\Delta(rep(E, \{i\}, \kappa)(\omega) - \kappa_i(\omega)| = 0$, $\omega$ is a model of both $\Delta(E \sqcup \{K'\})$ and of $K$ (*), or $\omega$ is a countermodel of both $\Delta(E \sqcup \{K'\})$ and of $K$ (**).

To get $|\kappa_\Delta(E)(\omega) - \kappa_i(\omega)| = 1$, there are also two cases:

- either $\kappa_\Delta(E)(\omega) = 1$ and $\kappa_i(\omega) = 0$: then $\omega$ is a model of $K$ and a countermodel of $\Delta(E \sqcup \{K\})$. Since $\omega$ is a model of $K$, with (*), we know that $\omega$ is a model of $\Delta(E \sqcup \{K'\})$. Then $\omega$ is a model of $K$, of $\neg(\Delta(E \sqcup \{K\}))$ and not of $\neg\Delta(E \sqcup \{K'\})$: it entails that

  $$K \wedge \neg\Delta(E \sqcup \{K\}) \not\models \neg\Delta(E \sqcup \{K'\}).$$

- or $\kappa_\Delta(E)(\omega) = 0$ and $\kappa_i(\omega) = 1$: then $\omega$ is a countermodel of $K$ and a model of $\Delta(E \sqcup \{K\})$. Since $\omega$ is a countermodel of $K$, with (**), we know that $\omega$ is a countermodel of $\Delta(E \sqcup \{K'\})$. Then $\omega$ is a model of $\neg K$, of $\Delta(E \sqcup \{K\})$ and not of $\Delta(E \sqcup \{K'\})$: it entails that

  $$\neg K \wedge \Delta(E \sqcup \{K\}) \not\models \Delta(E \sqcup \{K'\}).$$

Hence it is not the case that both

- $K \wedge \neg\Delta(E \sqcup \{K\}) \models \neg\Delta(E \sqcup \{K'\})$, and

- $\neg K \wedge \Delta(E \sqcup \{K\}) \models \Delta(E \sqcup \{K'\})$.

are satisfied.

As to the converse, suppose that there is a profile $E$, an agent $i$ with a base $K$, and another base $K'$ s.t.:

$$K \wedge \neg\Delta(E \sqcup \{K\}) \not\models \neg\Delta(E \sqcup \{K'\})$$

or

$$\neg K \wedge \Delta(E \sqcup \{K\}) \not\models \Delta(E \sqcup \{K'\}).$$

In the first case, there is a model $\omega$ of $K \wedge \neg\Delta(E \sqcup \{K\})$ which is a countermodel of $\neg\Delta(E \sqcup \{K'\})$ :

$$\kappa_\Delta(E)(\omega) = 1, \kappa_i(\omega) = 0, \kappa_\Delta(rep(E, \{i\}, \kappa)(\omega) = 0.$$

So

$$|\kappa_\Delta(E)(\omega) - \kappa_i(\omega)| > |\kappa_\Delta(rep(E, \{i\}, \kappa)(\omega) - \kappa_i(\omega)|.$$

In the second case, there is a model $\omega$ of $\neg K \wedge \Delta(E \sqcup \{K\})$ which is a countermodel of $\Delta(E \sqcup \{K'\})$:

$$\kappa_\Delta(E)(\omega) = 0, \kappa_i(\omega) = 1, \kappa_\Delta(rep(E, \{i\}, \kappa)(\omega) = 1.$$

So

$$|\kappa_\Delta(E)(\omega) - \kappa_i(\omega)| > |\kappa_\Delta(rep(E, \{i\}, \kappa)(\omega) - \kappa_i(\omega)|.$$

In both cases, $\Delta$ does not satisfy the **(IP)** property. $\qquad\square$

**Theorem 16** *Let* $i_{wip}(K, K_\Delta) = \frac{1}{\#([K \oplus K_\Delta]) + 1}$. $\Delta$ *satisfies the* **(WIP)** *property if and only if it is strategy-proof for* $i_{wip}$.

**Proof:**  By definition $K \oplus K_\Delta \equiv (\neg K \wedge K_\Delta) \vee (K \wedge \neg K_\Delta)$ So we can write:

$$i_{wip}(K, K_\Delta) = \frac{1}{\#([(\neg K \wedge K_\Delta) \vee (K \wedge \neg K_\Delta)]) + 1}$$

if and only if

$$i_{wip}(K, K_\Delta) = \frac{1}{\#([\neg K \wedge K_\Delta]) + \#([K \wedge \neg K_\Delta]) + 1}.$$

Suppose that a merging operator $\Delta = Bel(\kappa_\Delta))$ does not satisfy the **(WIP)** property. Then there is a profile $E$, an agent $i$, an OCF $\kappa$ s.t.

$$\Sigma_{\omega \in \mathcal{W}}|\kappa_\Delta(E)(\omega) - \kappa_i(\omega)| > \Sigma_{\omega \in \mathcal{W}}|\kappa_\Delta(rep(E, \{i\}, \kappa)(\omega) - \kappa_i(\omega)| \quad (*)$$

In the following we note $\Delta(E \sqcup \{K\})$ for $Bel(\kappa_\Delta(E))$, that is the initial merged base when agent $i$ reports her true base $K \equiv Bel(\kappa_i)$, and we note $\Delta(E \sqcup \{K'\})$ for $Bel(\kappa_\Delta(rep(E, \{i\}, \kappa)))$, i.e., the merged base obtained by replacing the base $K$ of the agent $i$ by another $K' \equiv Bel(\kappa)$.

In the two-strata case, $|\kappa_\Delta(E)(\omega) - \kappa_i(\omega)|$ is equal to 0 or 1. In fact, $|\kappa_\Delta(E)(\omega) - \kappa_i(\omega)| = 1$ if and only if:

- either $\kappa_\Delta(E)(\omega) = 1$ and $\kappa_i(\omega) = 0$: this is equivalent to $\omega \models K \wedge \neg\Delta(E \sqcup \{K\})$.

- or $\kappa_\Delta(E)(\omega) = 0$ and $\kappa_i(\omega) = 1$: this is equivalent to $\omega \models \neg K \wedge \Delta(E \sqcup \{K\})$.

We deduce the following equation:

$$\Sigma_{\omega \in \mathcal{W}} |\kappa_\Delta(E)(\omega) - \kappa_i(\omega)| = \#([K \wedge \neg\Delta(E \sqcup \{K\})]) + \#([\neg K \wedge \Delta(E \sqcup \{K\})]).$$

Similarly, we get:

$$\Sigma_{\omega \in \mathcal{W}} |\kappa_\Delta(rep(E, \{i\}, \kappa)(\omega) - \kappa_i(\omega)| = \#([K \wedge \neg\Delta(E \sqcup \{K'\})]) + \#([\neg K \wedge \Delta(E \sqcup \{K'\})])$$

The inequation (*) is then equivalent to:

$$\#([K \wedge \neg\Delta(E \sqcup \{K\})]) + \#([\neg K \wedge \Delta(E \sqcup \{K\})]) > \#([K \wedge \neg\Delta(E \sqcup \{K'\})]) + \#[\neg K \wedge \Delta(E \sqcup \{K'\})])$$

which is equivalent to

$$\frac{1}{\#([K \wedge \neg\Delta(E \sqcup \{K\})]) + \#([\neg K \wedge \Delta(E \sqcup \{K\})]) + 1} <$$
$$\frac{1}{\#([K \wedge \neg\Delta(E \sqcup \{K'\})]) + \#([\neg K \wedge \Delta(E \sqcup \{K'\})]) + 1}$$

which is equivalent to

$$i_{wip}(K, \Delta(E \sqcup \{K\})) < i_{wip}(K, \Delta(E \sqcup \{K'\}))$$

which is equivalent to the fact that $\Delta$ is not strategy-proof for $i_{wip}$.

$\square$